\pdfoutput=1
\documentclass[11pt,oneside,openany]{dissertation}

\newif\iftth
\tthfalse

\newif\ifht
\htfalse

\usepackage{tabularx}

\newif\iftwoside
\twosidefalse

\newif\ifdraft
\draftfalse

\newif\ifsingle
\singlefalse

\usepackage{verbatim}
\usepackage{tikz}
\usetikzlibrary{matrix,arrows}
\usepackage{listings}
\usepackage{wrapfig}

\usepackage{setspace}
\usepackage{amsmath}
\usepackage{amssymb}
\usepackage{amsthm}
\usepackage{fancyvrb}

\usepackage{url}

\newcommand{\urlBiBTeX}[1]{\url{#1}}

\iftth
\newcommand{\xspace}{\ }
\else
\usepackage{xspace}
\fi

\ifht
  \usepackage[html,2]{tex4ht}
\else
\fi

\newif\ifwww

\iftth
\wwwtrue
\fi

\ifht
\wwwtrue
\fi

\usepackage{cite}

\ifdraft
\else
\fi

\iftwoside
\else
\fi

\DefineVerbatimEnvironment{figurecodeverbatim}%
  {Verbatim}%
  {fontfamily=tt,%
   fontsize=\small,%
   commandchars=\\\{\},%
   formatcom=\def\{{\symbol{123}}\def\}{\symbol{125}}\def\\{\symbol{92}},%
   listparameters=\setlength{\topsep}{0pt}%
                  \setlength{\partopsep}{0pt}%
                  \setlength{\parskip}{0pt}%
   }

\DefineVerbatimEnvironment{quotecodeverbatim}%
  {Verbatim}%
  {fontfamily=tt,%
   fontsize=\small,%
   commandchars=\\\{\},%
   formatcom=\def\{{\symbol{123}}\def\}{\symbol{125}}\def\\{\symbol{92}}%
   }

\newcommand{\fancyfloatrule}{\ifwww\else{\noindent\hrulefill\par}\fi}
\newcommand{\fancyfloatsize}{\small}

  {\begin{figure}[#1]%
   \fancyfloatsize%
   \fancyfloatrule%
  }
  {\fancyfloatrule%
   \end{figure}}

\newcommand*{\Month}{%
  \ifcase\month \or
  January\or February\or March\or April\or May\or June\or
  July\or August\or September\or October\or November\or
  December\fi \xspace
}

\newcommand*{\Year}{\number\year\xspace}

\usepackage{calc}

 {\begin{list}{}{%
  \settowidth{\labelwidth}{\textsf{#1:}}%
  \setlength{\itemsep}{0pt}%
  \setlength{\parsep}{0pt}%
  \setlength{\leftmargin}{\labelwidth+\labelsep}}}%
  {\end{list}}

\ifdraft
\else
\fi

\begin{document}

\frontmatter

\thispagestyle{empty}

\mbox{}
\vskip1in

\begin{center}

{
{\Large
\textbf{
Synthesis from Formal Partial Abstractions
}}
\\[40pt]
A Dissertation\\
Presented to\\
the faculty of the School of Engineering and Applied Science \\
University of Virginia\\[20pt]
in partial fulfillment\\
of the requirements for the degree\\[15pt]
Doctor of Philosophy\\
Computer Science\\
by\\[25pt]
{\large\bf Hamid Bagheri} \\[30pt]
May\\ 
2013
}

\end{center}

\newpage
\thispagestyle{empty}
\mbox{}

\newpage
\section*{\centering Approval Sheet}

\thispagestyle{empty}
\begin{center}

This dissertation is submitted in partial fulfillment of the
requirements for the degree of

Doctor of Philosophy (Computer Science)

\vskip20pt
\rule{3in}{0.01in} \\
\vspace{-0.25cm}Hamid Bagheri \\[18pt]
\end{center}
This dissertation has been read and approved by the Examining Committee: \\[5pt]
\begin{center}
\rule{3in}{0.01in}\\
\vspace{-0.25cm}Kevin Sullivan (Advisor) \\[18pt]
\rule{3in}{0.01in} \\
\vspace{-0.25cm}Mary Lou Soffa (Chair) \\[18pt]
\rule{3in}{0.01in}\\
\vspace{-0.25cm}Westley Weimer\\[18pt]
\rule{3in}{0.01in}\\
\vspace{-0.25cm}William G. Griswold\\[18pt]
\rule{3in}{0.01in}\\
\vspace{-0.25cm}John Lach\\[18pt]

\end{center}
Accepted for the School of Engineering and Applied Science: \\[5pt]
\begin{center}
\rule{3in}{0.01in} \\
\vspace{-0.25cm}Dean, School of Engineering and Applied Science \\[9pt]

\Month\\ 
\Year
\end{center}

\newpage
\thispagestyle{empty}
\mbox{}

\chapter*{Abstract}

Developing complex software systems is costly, time-consuming and error-prone. Model-driven development (MDD) promises to improve software productivity, timeliness, quality and cost through the transformation of abstract application models to code-level implementations. However, it remains unreasonably difficult to build the modeling languages and translators required for software synthesis. This difficulty, in turns, limits the applicability of MDD, and makes it hard to achieve reliability in MDD tools.
This dissertation research seeks to reduce the cost, broaden the applicability, and increase the quality of model-driven development systems by embedding modeling languages within established formal languages and by using the analyzers provided with such languages for synthesis purposes to reduce the need for hand coding of translators. 
This dissertation, in particular, explores the proposed approach using relational logic as expressed in Alloy as the general specification language, and the Alloy Analyzer as the general-purpose analyzer. Synthesis is thus driven by finite-domain constraint satisfaction.
One important aspect of this work is its focus on partial specifications of particular aspects of the system, such as application architectures and target platforms, and synthesis of partial code bases from such specifications.
Contributions of this work include novel insights, methods and tools for (1) synthesizing architectural models from abstract application models; (2) synthesizing partial, platform-specific application frameworks from application architectures; and (3) synthesizing object-relational mapping tradeoff spaces and database schemas for database-backed object-oriented applications. 

\newpage
\thispagestyle{empty}
\mbox{}

\chapter*{Acknowledgments}

\newcommand{\skiiip}{\vskip5pt}

I would like to express my sincere gratitude and appreciation to my advisor, Professor Kevin Sullivan, for his invaluable guidance, mentorship, encouragement, and enthusiasm. His insight has made me a more effective researcher. I greatly appreciate all the effort he has put into mentoring me. Thank you Kevin!

Special thanks to my PhD committee, Professors May Lou Soffa, Westley Weimer, William Griswold, and John Lach for their support, guidance and helpful suggestions. Their guidance has served me well and I owe them my heartfelt appreciation.

I would also like to thank my qualifying exam committee, Professors Sang Son, John Knight, and Jack Davidson who provided constructive feedback 
on the first steps of this research.

I thank my friends at UVa, Tamal Saha, Ge Gao, Yuanyuan Song, Chong Tang, Xi Wang, Wei Wang, Runjie Zhang, Juhi Ranjan, Tanima Dey, Chris Gregg, John Hott, Taniya Siddiqua, Shahriar Nirjon, and Chih-hao Shen, who made my graduate studies a wonderful academic experience. Thanks are also certainly due to my great Iranian friends in Charlottesville.

I would like to express my immense gratitude to my parents, for their encouragement and advice. Their support throughout the years has been unwavering despite the physical distance that has separated us during my PhD study. They encouraged my intellectual curiosity from a young age. Throughout my life, they have always ensured that every opportunity is available to me. Last but certainly not least, special thanks go out to my wonderful wife, Zeinab, for her love, support and constant encouragement.

\newpage
\thispagestyle{empty}
\mbox{}

\ifht
\else
\tableofcontents
\fi

\newpage
\thispagestyle{empty}
\mbox{}

\ifdraft
\else
\listoffigures
\fi

\newpage
\thispagestyle{empty}
\mbox{}

\mainmatter

\pdfoutput=1
\chapter{Introduction}

Developing complex software systems is costly, time-consuming, and error-prone, in large part due to wide conceptual gaps between application descriptions and code implementation spaces~\cite{france_model-driven_2007}. This difficulty has led to the emergence of model-driven development (MDD), the main goal of which is to span these gaps through the transformation of abstract, domain-oriented application models to code-level implementations. Model-driven development is a software development paradigm centered around the extensive use of abstract models and top-down synthesis of implementations from such models using model transformation and code generation techniques~\cite{schmidt_MDE_2006}.
MDD holds out the promise of improved software productivity, timeliness, quality and cost~\cite{france_model-driven_2007}.
There are numerous success stories for model driven development~\cite{friedman_matlab/simulink_2006,bringmann_model-based_2008,li_study_2005,aulagnier_soc/sopc_2009}. It is, for example, being used in the automotive, aerospace, electronic and communications industries.


To enable the use of a MDD approach, engineers first develop a domain-specific language (DSL).
These languages are typically specified in terms of metamodeling languages provided by MDD frameworks.
Taking a DSL specification as input, some MDD frameworks automatically produce a model-editor for the DSL under consideration~\cite{bezivin_model_2006}.
Software engineers then implement model transformations for each DSL. Automated code generation in model-driven development is achieved through such model transformations. A model transformation maps input models in the DSL to outputs in a target language. Domain experts model the system in the DSL, which is then automatically transformed to the target language.
By shifting the focus of development from source code to model abstractions, MDD aims to reduce development cost, improve productivity and quality, and facilitate the engineering of complex systems.

\section{Problem}

The problems that this dissertation addresses are threefold:
First, in the current state of the art in MDD, it remains unreasonably difficult to build modeling languages and translators that are needed for software synthesis~\cite{white_improving_2009}.
Designing language syntax and semantics for DSLs that support modeling of complete applications in a particular domain is difficult~\cite{Edwards_XTEAM_2008}.
Mernik et al. say, ``DSL development is hard, requiring both domain knowledge and language development expertise~\cite{mernik_when_2005}".
Moreover, model transformers are complex, and thus difficult to develop and maintain~\cite{Edwards_XTEAM_2008}. 
It is widely recognized that developing code generation model interpreters is a very difficult problem~\cite{Edwards_XTEAM_2008,Santos_JSS_2010}.
Second, the commitment to heavyweight modeling languages limits the applicability of model driven development.
Ambler says, ``Although the [Model Driven Architecture] MDA\footnote{Model Driven Architecture (MDA) is a registered trademark of the Object Management Group (OMG) for model driven development.} is a very wonderful idea I suspect that it will succeed in only a very small percentage of organizations~\cite{scott_ambler_examining_MDA}.''
This commitment also drives up the cost of language and translator design and implementation.
Third, complexity of modeling languages and transformation systems increases the difficulty of achieving reliability in MDD tools.  Among other researchers, Karsai and his colleagues say, ``Writing translators by hand\ldots is the most time consuming and error prone phase~\cite{Karsai_et_al_2003}''.

\section{Goals}
The long term goal of my research is to reduce the cost, broaden the applicability, and increase the quality of model-driven development systems.
The specific goal of this research is to show that relational logic specification languages and formal analyzers that accompany them provide a basis for a style of specification-driven model driven development that has significant and demonstrable potential to address these problems.

This research aims to:
\begin{itemize}
  \item reduce the costs of developing modeling languages by enabling rapid and declarative development of narrow DSLs and platform models;
  \item reduce the costs of developing translators by reducing the need for traditional hand coding of translators using specification-driven transformation systems;
  \item increase the reliability of MDD by enabling formal analysis and verification of mapping specifications;
  \item broaden the applicability of MDD by facilitating lightweight development and application of MDD tools and methods within larger software development processes.
\end{itemize}

\section{Current state of the art}

There are related efforts that have made some progress in addressing the problems identified in model driven software development.
This section describes some notable attempts in this area, and discusses shortcomings in the current state of the art.

Edwards and Medvidovic~\cite{Edwards_XTEAM_2008} developed an MDD framework, called eXtensible Tool-chain for Evaluation of Architectural Models (XTEAM), to reduce the burden of manually developing domain-specific analyzers and code generators for each target platform. Their approach is based on separating domain-independent model interpretation logic from domain-specific model interpretation logic, where the former 
is modularized in an extensible transformation tool, called a model interpreter framework (MIF).
For example, for software architecture synthesis and analysis, the authors developed a generic metamodeling language, called abstract component technology (ACT). Upon ACT, they then constructed a model interpreter framework that provides an extensible infrastructure for implementing architectural analysis and synthesis. Developers can further extend the MIF to support synthesis for particular architecture description languages (ADLs) and target platforms.

This work shows that the burden of manual development of model interpreters can be mitigated to an extent through the design of extensible transformation tools. But it also has some shortcomings. 
First, each extensible transformation tool is specialized for a particular domain. Producing such an extensible transformer for each particular domain is costly. For example, while the MIF the authors developed for software architecture analysis and synthesis is independent of any particular ADL and target platform, it is specialized for the domain of software architecture.
Second, manually programming particular transformation components that extend MIFs is still required.

Along the same line, Santos et al. \cite{Santos_JSS_2010} proposed an aspect-oriented approach, implemented in a prototype tool called ALFAMA. 
Using ALFAMA, instead of defining the domain-specific language and the related code generator for each platform, developers extend the platform with an additional aspect-oriented modeling layer, called a domain specific modeling (DSM) layer. The aim of this aspectual layer is to encode both the specialization aspects that modularize platform extension points as well as the meta-model of the custom language specialized for the given platform. This additional layer records the information required to generate platform- and application-specific code from application models.

Given these DSM layers as input, ALFAMA then automatically extracts domain specific language meta-models and can generate application-specific code for instances of those meta-models. However, these productivity gains entail a significant additional effort involved in the development of DSM layers expressed in an aspect-oriented programming language. Moreover, the domain specific languages extracted from such layers are tightly coupled with the given platform rather than just being designed based on the application domain.


Malavolta et al.~\cite{Dually_2010} proposed a model-driven framework, called DUALLy, that 
automates generation of model transformations among instances of various DSLs.
The basic idea is that they defined a minimal but extensible, set of core architectural concepts in a meta-model called {$A_0$}. The {$A_0$} is designated as a bridge among various custom languages. For each custom language, DUALLy users then need to specify mappings, called semantic links, between the metamodel of the given custom language and {$A_0$}. As such, code generation model interpreters developed for any particular custom language is then accessible in other domain specific languages.
DUALLy thus relieves the burden of developing transformation systems for each custom language through interoperability among domain specific architecture languages. However, two steps transformations---from source to {$A_0$} and then to the target for which code generation is available---complicates the transformation process which in turn reduces the traceability from implementation artifacts to input models.

\section{Approach}
To achieve the aforementioned goals, this dissertation introduces an approach, called \emph{automated synthesis from formal partial abstractions}.
This approach combines four key elements.
First, it uses general-purpose, semantically well defined relational specification languages for modeling, rather than ad hoc or semantically unclear languages, e.g., UML or many of domain-specific languages.
Second, it uses general-purpose relational logic model finders for synthesis, rather than custom-built model transformers. 
While this approach can, in principal, accommodate a range of relational logic formalisms and tools,
in this dissertation, I have used Alloy as a specification language~\cite{daniel_jackson_alloy:lightweight_2002}, and the the Alloy Analyzer as a model finder.
Third, the approach targets selected aspects of a system for modeling and synthesis, rather than trying to impose a top-down MDD approach to whole system development.
For example, instead of generating complete enterprise system applications, using this approach, one can develop partial modeling and synthesis capabilities that target only synthesis of the database schema elements of a system.

\begin{figure*}[tbh]
\centering
\includegraphics[width=\textwidth]{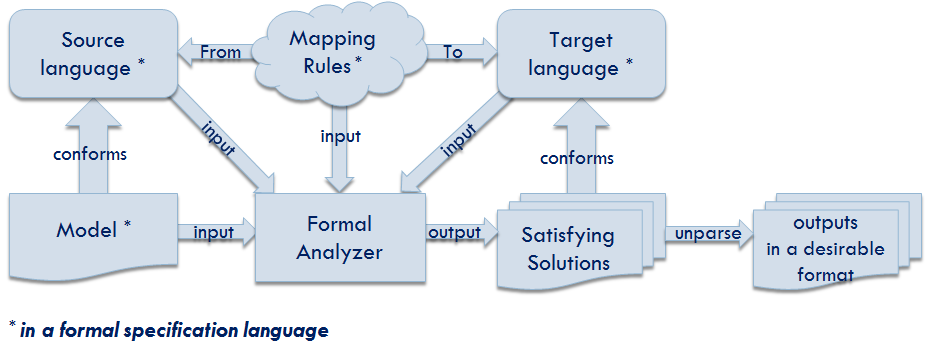}
\caption{\label{fig:approach}High-level view of the approach.}
\end{figure*}

Figure~\ref{fig:approach} represents a high-level view of the approach.
One starts by specifying source and target languages in relational logic.
These specifications define element types, and how they are related and constrained to constitute valid expressions in these respective domains. This task is carried out by the MDD system designers.
Application engineers then use the system by defining models in a given source language.

For example, consider an architectural style~\cite{shaw_software_1996} as a source domain specific language, where an architectural model is defined as an instance model of architectural style specifications.
The user then expresses mapping rules as additional predicates in relational logic. The mapping rules define the relationships required to hold between elements of the source domain and those of the target domain.

Given source and target languages specifications, mapping rules, and an application model as inputs, a relational logic model finder then automatically transforms the given model instance into one (or more) target model instance(s).
The transformation process involves the synthesis of solutions that satisfy the constraints of source and target languages specifications, the particular model, as well as mapping constraints.
The generated models are relational logic structures that encode the desired solutions. These models are then unparsed into a desirable form, such as human-readable architecture descriptions or executable code.


I believe that this approach has significant potential to achieve the aforementioned goals for the following reasons.
First, relational logic is a highly expressive notation, particularly for specifying structural properties of systems~\cite{daniel_jackson_alloy:lightweight_2002}, and it has been shown that Alloy is appropriate for formally defining domain specific languages~\cite{kelsen_lightweight_2008}.
It is also well suited to formalizing mapping rules between such domains. Altogether, highly expressive relational logic enables quick development of declarative specifications of MDD concepts.
Moreover, formalizing these concepts using an analyzable specification language enables automatic synthesis of resulting models as satisfying solutions, without having to write any code other than specifications.

Second, the automatic formal analyzer facilitates the process of verifying the correctness of specifications and mapping rules, which in turn, makes model driven software development more reliable.
By expressing essential properties intended to follow from the specifications and mapping rules, one can use automated analysis to check them. In my own experience, I have conducted formal validation of object-relational mapping rules in terms of developing a set of assertions.

Finally, 
abstracting from application details and focusing on particular aspects of the system relieves the synthesizer of responsibility for designing full-spectrum modeling languages and transformation systems, 
while facilitating use of formal analyzers for partial synthesis. One can more easily develop little languages for model-driven development and associated transformation systems. 
This, in turn, broadens the applicability of model-driven development to many aspects of software development.

\section{Example}
To make my approach clearer, I present an illustrative example.
The example is to synthesize relational schema for an application object model.

\begin{figure}
\lstset{ %
basicstyle=\scriptsize,       
numbers=left,                   
numberstyle=\tiny,      
stepnumber=1,                   
numbersep=7pt,
backgroundcolor=\color{white},  
showspaces=false,               
showstringspaces=false,         
showtabs=false,                 
frame=bottomline, 
tabsize=2,                    
captionpos=b,                   
breaklines=true,                
breakatwhitespace=false,        
numberbychapter=false,
xleftmargin=1.5em,
morekeywords={module,sig,abstract,extends,one,some,set,open,pred,all,in,no}
}
\lstinputlisting[caption=Part of the formal specification of a language in Alloy.,numberbychapter=false,label=objectModel]
{Pics/ch1/object-model.als}
\vspace{-0.6cm}
\end{figure}

The first step is to formally specify source and target languages. 
In this case, we need languages for expressing source object models and target relational schemas, in a manner that also allows for the expression of mapping rules and efficient model finding.

Listing~\ref{objectModel} partially represents the formal specification of a source language in Alloy.
This code snippet defines the Class construct as a signature in the Alloy language. Signature is a fundamental construct in Alloy, and represents the basic types of elements and the relationships between them.
Each Class signature has a set of fields: attrSet, id, parent and isAbstract.
The \emph{attrSet} field in each Class signature specifies the set of attributes of the class. The \emph{id} field represents the identifier of the class. The inheritance relationship is then represented by the \emph{parent} relation. Finally, the \emph{isAbstract} field denotes whether the class under consideration is abstract or not. 

\begin{figure}
\lstset{ %
basicstyle=\scriptsize,       
numbers=left,                   
numberstyle=\scriptsize ,      
stepnumber=1,                   
numbersep=5pt,                  
backgroundcolor=\color{white},  
showspaces=false,               
showstringspaces=false,         
showtabs=false,                 
frame=bottomline,                    
tabsize=2,                    
captionpos=b,                   
breaklines=true,                
breakatwhitespace=false,        
numberbychapter=false,
xleftmargin=1.8em,
morekeywords={module,sig,abstract,extends,one,set,open,pred,all,in,none}
}
\begin{lstlisting}[caption=An example of a mapping rule specified in the Alloy language,numberbychapter=false,label=mapping]
(c.isAbstract=No)&&(c.^parent != none) =>{
    all a:Attribute | a in c.^parent.attrSet =>{
        one f:Field|f.fAssociate=a && 
        f in (c.~tAssociate.fields) }	
}
\end{lstlisting}
\vspace{-0.6cm}
\end{figure}

\begin{wrapfigure}{r}{0.5\textwidth}
\vspace{-25pt}
  \begin{center}
    \includegraphics[width=0.53\textwidth]{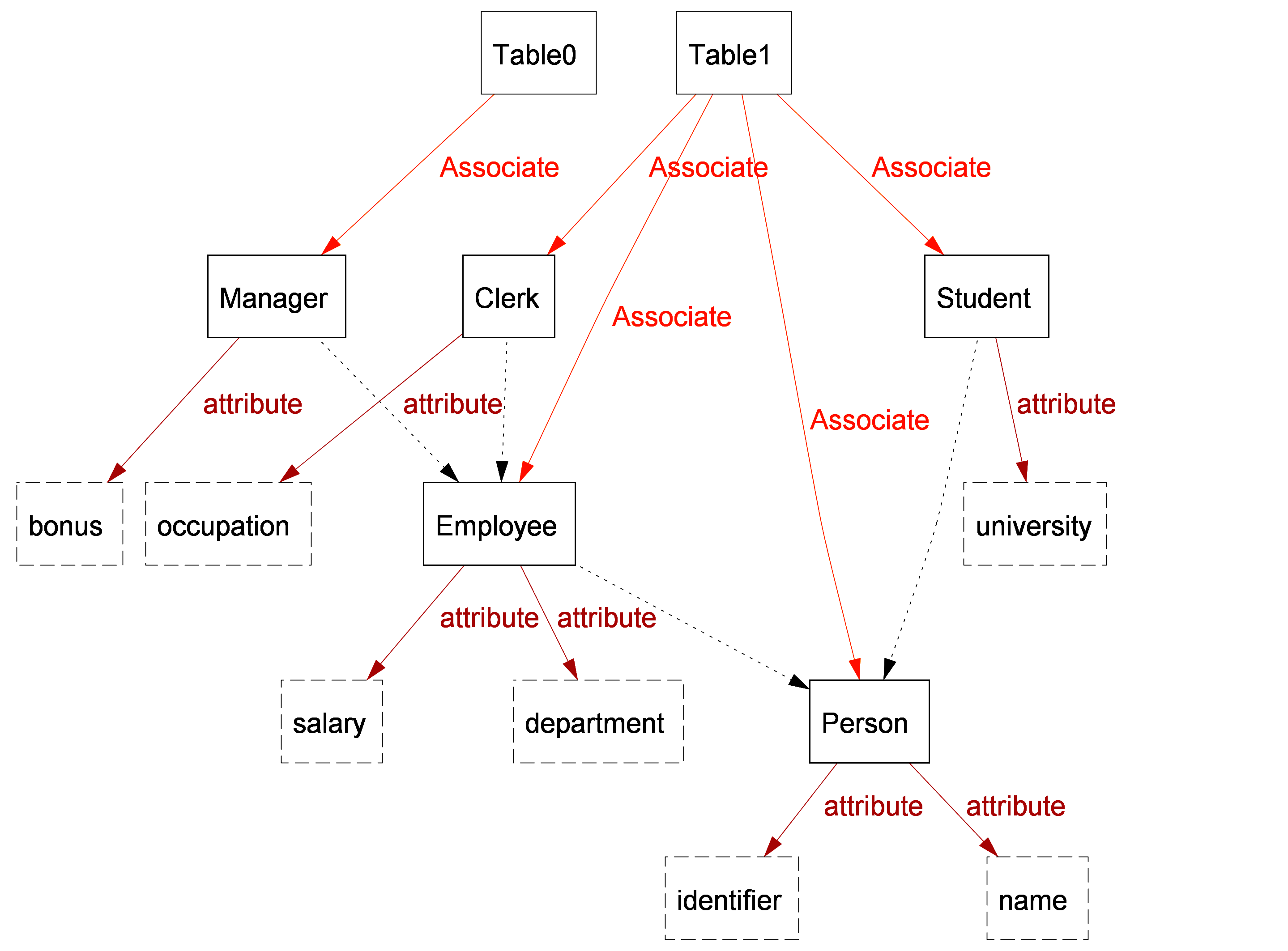}
  \end{center}
  \vspace{-25pt}
  \caption{\label{fig:generatedInstance}A generated instance of an object-relational mapping. 
  }
\vspace{-0.1cm}
\end{wrapfigure}

After specifying the abstractions involved in the source and destination models, the next step is to express mapping rules as additional predicates that relate elements of the source models to the constructs in the destination.
In the case of this example, we can use object-relational mapping (ORM) strategies~\cite{luca_cabibbo_managing_2005,keller_mapping_1997} as mapping rules and formalize them using relational logic. 
Without going into the details of these ORM strategies, which are discussed in Chapter 5, Listing~\ref{mapping} represents an example of a mapping rule specified in the Alloy language.
\emph{tAssociate} is a relation from a table to its associated classes, and \emph{fAssociate} is a relation from a field in the relational schema to its associated construct in the object model. The mapping expression thus states that the table encompasses relational fields corresponding to all inherited attributes of the given class, should the class not be abstract.

Given the specifications and mapping constraints along with a formal specification of the application object model, the Alloy Analyzer computes a set of relational database schemas and associated object-relational mappings.
Figure~\ref{fig:generatedInstance} illustrates a generated instance of an object-relational mapping, in which the object model is represented by two tables in the relational database.
Having computed satisfying solutions, we can then unparse these solutions from low-level Alloy objects to SQL counterparts.

\section{Hypotheses}
The main claim of this thesis is that the approach I have described promises to reduce the costs of building narrowly targeted model-driven development languages and systems, to provide a formal basis for quality assurance, and to broaden the applicability of model driven development methods.
More specifically, this dissertation makes the following claims for this approach:

\begin{itemize}
  \item the approach substantially eliminates the need for traditional hand coding of translators, replacing traditional coding with specification-driven synthesis; 
  \item the approach supports the reuse of formal specifications of source and target languages as well as mapping rules between them; 
  \item the approach can be used to synthesize a variety of artifacts, ranging from code targeted widely-used industrial software platforms to architectural models; 
  \item the approach enables the application of lightweight formal validation techniques~\cite{daniel_jackson_alloy:lightweight_2002} during the development of MDD systems; 
  \item the approach can be employed at various stages of the software development lifecycle. 
\end{itemize}

\section{Evaluation}


The overall approach I have taken to evaluate the claims of this dissertation is based on experimental systems methods. I do not believe it is possible to assess the potential of the proposed approach analytically. Rather, experience applying it to, and assessing it in the context of a diversity of applications is needed. To test my claims, I have selected three distinctly different applications, developed MDD infrastructure for each of them, and assessed the results in several key dimensions, including its potential to support  the reuse of formal specifications of MDD concepts---source and target languages as well as mapping rules---to synthesize a variety of software artifacts, to formally validate correctness of mapping rules, and in reducing the need for traditional hand coding of translators.

\begin{figure*}[tbh]
\centering
\includegraphics[width=4.5in]{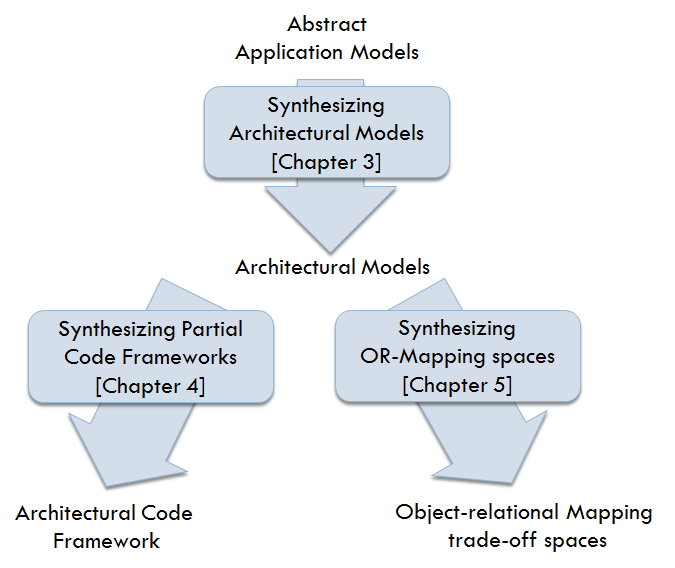}
\caption{\label{fig:researchComponents}Three components of the thesis research}
\end{figure*}

The first application is the mapping of high-level application models into style-specific application architectures. The second is the mapping of application architectures into platform-specific code frameworks, and the third is the synthesis of object-relational mapping (ORM) tradeoff spaces for object-oriented application architectures. For each of these problems, I design and conduct a separate experimental system, collect the results of these experiments, and interpret the resulting data in relation to the hypotheses.
Figure~\ref{fig:researchComponents} shows three high-level research components of this dissertation.

\subsection{Synthesizing architectural models from abstract application models}
The first experiment with my approach focuses on formal synthesis of software architectures. Today, the success of architecture development depends heavily on the experience of human architects, and this manual process is unreliable, costly and labor-intensive. As our understanding of architectures grows, we can systematize and eventually formalize and automate synthesis of architectures.  Architectural styles~\cite{shaw_software_1996} are the results of earlier efforts to systematize successful architectures in terms of constraints on architectural elements and their patterns of composition. I develop a formally-precise, automated technique, called Monarch~\cite{bagheri_monarch_2010}, for mapping application models to software architectures in selected architectural styles.

There is a fundamental separation of concerns underlying this approach:
given an application model that expresses abstract application structure independent of architectural style,
we can choose a compatible architectural style and then map the application model to architectural models in this style.
I formulate the mapping problem as one of finding satisfying solutions to a specification that conjoins an application model of a given application type---a formal specification of a family of application---with an architectural style specification, and with rules for mapping application models of the given type to architectural models in the given style. Precisely, four pieces of formal specifications are conjoined in the process of mapping an application model to architectural models in a given style: (1) an application type, in which the application model is defined; (2) an application model; (3) an architectural style specification; and (4) mapping predicates that specify the relationships required to hold between an application of the given type and an architecture in the given style. A formal analyzer then computes satisfying solutions to the conjoined specification, yielding the synthesized architectures.

\subsection{Synthesizing code frameworks from application architectures}
The second experiment with my approach focuses on architecture-driven development of applications on software platforms. In this dissertation, I use the term `platform' to broadly refer to middleware and software frameworks. Examples include the Restlet~\cite{restlet} framework for implementing RESTful web applications, the HornetQ~\cite{hornetq} middleware that supports reliable distributed messaging, and the CouchDB~\cite{couchdb} for data persistence.
Developing applications that use such complex platforms 
is hard and error-prone. 
I develop a declarative technique, called Pol~\cite{Pol_2012}, for constraint-based synthesis of partial code frameworks without the need for hand-coded transformation systems. Rather, synthesis is driven by formal, partial specifications of target platforms and application architectures, and by code fragments encoding platform usage patterns.

Using Pol, one specifies both application architecture and target platform models in a relational logic specification language.
The user then expresses mapping rules from architectural elements to platforms constructs as additional predicates.
The approach then involves the synthesis of a platform-specific implementation model that satisfies the constraints of both style and platforms specifications, and that does so in a particular manner described by additional mapping constraints.
The next step is to map such a formally synthesized implementation model to code.
Code is synthesized based on the matching of design fragments, which are platform usage code patterns, to elements of the implementation model. Binding elements that relate code patterns to implementation model elements are specified declaratively in a relational specification language, which in turn enables the use of a formal analyzer to automate the binding process.



\subsection{Synthesizing ORM tradeoff spaces from object-oriented architectures}
The third experiment of my research focuses on effective persistence layers for object-oriented application architectures. Developing object-relational (OR) mappings that achieve desirable quality attributes for object-oriented applications is difficult, tedious, and costly~\cite{philippi_model_2005}. While a wealth of research has been performed~\cite{luca_cabibbo_managing_2005,ireland_understanding_2009,keller_mapping_1997} on bridging application architecture and databases to address the object-relational impedance mismatch problem, little has been done on automated support for derivation of OR mapping specifications. I develop a novel technique for exhaustive, formal synthesis of large spaces of such mappings, and classifying individual mappings in these spaces into multidimensional quality equivalence classes.

The basic idea is to formalize object-relational mapping (ORM) strategies using relational logic. These mapping strategies are previously informally defined in the literature~\cite{philippi_model_2005,keller_mapping_1997,luca_cabibbo_managing_2005}.
This formal definition of ORM strategies enables the use of a relational model finder to generate space of possible mappings for each application object model.
The approach takes as inputs a formal specification of an object model and optional class-specific mapping strategies for those
classes that the user wants mapped in a specific manner. Employing a constraint-solving technique, it then generates the space of possible mapping candidates subject to the user specified constraints. It uses six metrics 
for each candidate, and classifies results into quality equivalence classes.
Given that quality characteristics are usually conflicting, there is generally no single optimum solution but there are several pareto-optimal choices representing best trade-offs.
The approach implementation presents pareto-optimal solutions to the user along with the measures of its quality attributes. The user then selects one of these candidates according to his or her tradeoff preferences.

\section{Contributions}

The overall contribution of this work is an approach to addressing fundamental problems in the current state of the art in model driven development, including its cost, reliability, and breadth of applicability. In more detail, this work makes several component contributions.


The first contribution of this research is a novel approach to enable rapid and reliable model-driven software development by substantially eliminating the need for  traditional hand coding of transformation systems, using relational logic model finders driven by partial formal specifications as an alternative mechanism. More precisely,
this dissertation demonstrates that analyzable formal specification languages can be leveraged for encoding general MDD abstractions, which in turn admits the use of constraint solvers for model synthesis.
Validation to date flows from experience of applying these ideas in three key dimensions of software synthesis: synthesizing architectural models, 
synthesizing partial code frameworks, 
and synthesizing object-relational mapping tradeoff spaces for database-centric applications.


Second, this research work, and particularly the first experiment, contributes a formally precise approach to separate architectural style design decisions from application-specific decisions in a way that  supports formal refinement of application models into software architecture in light of separate choices of architectural style. I developed the concepts of application type and architectural map as key constructs needed, in addition to that of architectural style, to achieve such a separation of concerns.

Third, the work on specification-driven synthesis of architectural code frameworks contributes an evolutionary development approach combining specification-driven synthesis of architectural code frameworks for multi-platform based applications and manual extension of these frameworks to complete applications.
This work 
leads to the notion that future developers might work with \emph{hybrid} code bases comprising both traditional source code as well as formal models and code that is synthesized from them, evolving in ways that include ongoing refactoring between imperative code and declarative specifications.


Finally, the research work on synthesis of tradeoff space for OR mappings, backed with formally precise definition of object-relational mapping strategies, contributes a novel approach that aims to deliver the quality of expert-hand-crafted mappings and the productivity benefits of automated techniques.
This work, among other things, contributes a novel formal technique to search-based software engineering (SBSE)~\cite{harman_search_2010}. 
Success in applying such a formally-specified tradeoff analysis technique, can open a new path to applying \emph{formal SBSE} in other disciplines. 

\section{Outline}
The rest of this dissertation is organized as follows:
Chapter 2 presents the background of this work and puts it in context with related efforts.
Chapters 3--5 present the three main experiments I have conducted.
Chapter 3 focuses on applying a formally precise approach into an automated development of architectural models from abstract application models. 
Chapter 4 discusses synthesizing partial code frameworks from application architectures. Specifically, it illustrates that with modest and principled development of code fragments capturing idiosyncratic use of given platforms in given applications, the proposed approach can map architectural descriptions to object-oriented application frameworks that use a range of modern software platforms and standards.
Chapter 5 focuses on formal synthesis of object-relational mapping tradeoff spaces for object-oriented application architectures. This chapter discusses how this form of formal synthesis of tradeoff spaces can create valuable opportunities for novel forms of trade space analysis.
Chapter 6 evaluates this work as a whole, and discusses its research impact, novelty, and shortcomings. 
Chapter 7 discusses potential directions for future work, and introduces the notion of bottom-up model-driven development towards the possibility of a re-conception of model-driven engineering.
Finally, Chapter 8 concludes this dissertation. 


\newpage
\thispagestyle{empty}
\mbox{}

\chapter{Background and Related Work}

In this chapter, I first discuss background and research efforts related to 
using constraint solvers for code synthesis. I then present work that is related to the application of my synthesis approach in three particular domains aforementioned in Chapter 1.
This includes work on (1) separation of application and architectural style concerns, (2) formalization of architectural styles, (3) filling the gap between software architecture and implementation, (4) software platforms, (5) partial code synthesis, (6) object relational mappings and (7) derivation of database-centric implementations from formal specifications.

\section{Constraint Solving for Code Synthesis}

Constraint solving for code synthesis is an emerging class of techniques aimed at reducing the code synthesis problem to that of solving a logical formula by means of an off-the-shelf constraint solver, such as a SAT or SMT Solver.

Sketeching~\cite{armando_solar-lezama_program_2008} is a synthesis technique in which programmers partially define the control structure of the program with holes, leaving the details unspecified. This technique uses an unoptimized program as correctness specification. Given these partial programs along with correctness specification as inputs, a synthesizer---developed upon a SAT-based constraint solver---is then used to complete the low-level details to complete the sketch by ensuring that no assertions are violated for any inputs. This work shares with mine the common insight on both using partial models and synthesis based on constraint solving. However, my work focuses on automating mapping from practically meaningful abstract application models to useful, platform-specific implementations, as is done in MDD.

Along the same line, Srivastava et al.~\cite{srivastava_program_2010} developed a proof-theoretic synthesis, in which the user provides the relation between inputs and outputs of a program in the form of logical specifications, specifications of the program control structure as a looping template, a set of program expressions, and allowed stack space for the program to be synthesized. It then generates a constraint system such that solutions to that set of constraints lead to the specified program. They have shown feasibility of their approach by synthesizing program implementations for several algorithms form logical specifications.

While these research efforts mainly focus on low-level details of programs, this dissertation, by contrast, focuses on the end-to-end transformation as is done in model-driven development. Specifically, this work uses analyzable formal specification languages for encoding general MDE abstractions, which in turn enables utilizing constraint solvers to provide state-of-the-art formal methods for model synthesis. It thus relieves the tedium and errors associated with manually developing transformation systems.

\section{Separation of Application and Architectural Style Concerns}
Researchers and practitioners have long separated application descriptions from choices of architectural form. Parnas's 1972 paper make this distinction~\cite{parnas_criteria_1972}, showing how one application, key word in context (KWIC), could be mapped to two distinct architectures: one based on a functional decomposition and one on information hiding. In analogous work published in 2009, Taylor, et al.~\cite{taylor_software_2009} described a lunar landing control system and showed how it could be realized in a wide variety of architectural styles. In many research efforts (e.g.~\cite{dewayne_e._perry_foundations_1992,erich_gamma_design_1995,garlan_evolution_styles_2009,shaw_software_1996}) I found the same basic schema: an application is described, often informally; a choice of architectural style is made; an architecture consistent with the style is exhibited, and relative costs and benefits of the results are compared in selected dimensions. The problem is that the mapping process has remained implicit, informal, and not itself subject to rigorous investigation or explicit representation and analysis. In most work to date, the focus is on \emph{ex post} analysis of the relevant properties of the resulting architectures (e.g., performance). Key application properties are inadequately addressed until it is too late.
One of the aims of the work presented here is to move consideration of all essential application properties to the application description level, and to precisely determine a \emph{priori} the checkable conditions under which particular
architectural mappings are allowed. This change in perspective is significant.

\section{Formalization of Architectural Styles}
The notion of architectural styles has been present since the identification of software architecture as a discipline within software engineering~\cite{dewayne_e._perry_foundations_1992,shaw_software_1996}.

A variety of approaches have been proposed to model and analyze architectural styles.
In this context, Alloy has been applied by numerous researchers to formal work in software architecture~\cite{ioannis_georgiadis_self-organising_2002,jung_soo_kim_analyzing_Journal,ian_warren_automated_2006,stephen_wong_scalable_2008}.
Among others,
Wong et al.~\cite{stephen_wong_scalable_2008} proposed an approach based on the Alloy language for modeling and verification of complex systems that exploit multi-style structures. Warren et al.~\cite{ian_warren_automated_2006} similarly proposed an approach to specify an intended configuration using Alloy in order to check consistency with structural architectural constraints before performing an architectural evolution.
My work differs in its focus on separating application description from style choices.
Furthermore, I use Alloy not only to check the consistency of a given description against the rules of a style, but also to synthesize spaces of architectural models consistent with given styles.

The idea that a platform induces an architectural style, and that systems using a platform are required to conform to those styles, was first considered by Di Nitto and Rosenblum \cite{nitto_exploiting_1999}, and then by others~\cite{baresi_style-based_2006,sousa_formal_2001,sullivan_analysis_1999,Hou_SCL_2006}. My work maintains a separation between architectural styles and platform constraints, while connecting them through implementation mappings.

Along the same line, Sousa and Garlan~\cite{sousa_formal_2001}, using Wright, modeled structural and behavioral constraints imposed by the Enterprise JavaBeans framework. Hou and Hoover~\cite{Hou_SCL_2006} also proposed a specification language for framework constraints (SCL) that provides framework developers with means to define the structural constraints for using a framework, and to check the compliance of framework instantiations with respect to  framework constraints. These approaches use formal modeling to detect architectural mismatch. My approach prevents mismatches from arising in synthesized code.

\section{Architecture--Implementation Mappings}
A long line of research has focused on filling the gap between software architecture and implementation.
ArchJava~\cite{aldrich_archjava:_2002} and Archface~\cite{ubayashi_archface:_2010} are among notable attempts in this context. ArchJava provides capabilities of architecture description languages within Java programming language. Archface, by leveraging implementation-level concepts borrowed from \emph{aspect-oriented programming}~\cite{kiczales_aspect-oriented_1997}, similarly provides a means to describe and enforce component interactions. However, these research efforts do not explicitly support architectural styles or application synthesis. Moreover, my work goes beyond these approaches by separating not only architectural details from implementation but also application properties from architectural choices, and then spanning the gaps by means of automated declarative mapping specifications and automated solvers.

Medvidovic et al.~\cite{nenad_medvidovic_role_2003} focused on relating the modeling facilities of architectures and the implementation abilities of middleware platforms, and proposed an approach for using middleware to implement architectural connectors. Malek et al.~\cite{malek_style-aware_2005} extended this work by proposing that using application frameworks supporting architectural styles is a crucial approach to bridging the gap between architecture description and implementation.
They suggested that the lack of support by traditional middleware platforms for architectural abstractions leads to uncertainty about the consistency of implemented systems with their software architectures, making such traditional platforms a poor fit to architecture-driven development. This dissertation presents an architecture-driven approach to improve framework usability and to reduce architectural drift by explicit definition and automation of architecture-to-platform mappings.

\section{Software Platforms}
During the past several years, a considerable number of approaches have been proposed to improve usability of software platforms. 
These approaches fall into two main categories: pattern-oriented techniques and formal approaches.
The first one focuses on using patterns as a way of documenting conventional solutions to how applications can use platform hot-spots~\cite{Fairbanks_DesignFragments_2006,Supporting-Framework_2009}. The idea of documenting software platforms as patterns is first introduced by Ralph Johnson~\cite{johnson_frameworks_1992}. While pattern-oriented techniques are not aimed at documenting all possible uses of platforms, they are targeted at reducing the burden of platform instantiation activities by recording selected well-known usage examples in terms of platform usage code patterns.

The second category encompasses work on formal specification of the information needed by platform users, including what is provided by a platform and what is required to use it, in a way that enables consistency analysis of platform instances. Although formal verification of platform-based applications has a long tradition, automated support for the derivation of platform instances has received little attention~\cite{Hou_SCL_2006,sousa_formal_2001}. 
This dissertation presents an approach that leverages the substantial advancements in both categories, design fragments~\cite{Fairbanks_DesignFragments_2006} to record platform usage code patterns and logical representation of platforms as well as architecture-implementation mappings,
to substantially automate the synthesis of platform-specific code implementations.

\section{Partial Code Synthesis}
Recent research efforts~\cite{zheng-2012,damien_cassou_leveraging_2011,Pol_2012} recognize the potential benefits of the emerging class of partial code synthesis in different domains, where partial models generate partial code which are then combined with hand-written code to constitute the application.
Among others, recent work of Zheng and Taylor~\cite{zheng-2012} on {\em 1.x-way  mapping} shares our commitment to partial synthesis from architectural models. Their work differs in supporting synthesis from both structural and behavioral specifications, whereas to date my work has focused on structural models. They also emphasize a {\em deep separation} model that puts synthesized code in classes separate from hand-crafted code. 
Zheng and Taylor also emphasize incremental regeneration of code from changes to architectural models, and automated notification of developers when code changes are needed.

By contrast, my work leverages formal methods in a way that they do not attempt. My work avoids the need for any hand-crafted synthesizer, and focuses on synthesizing code targeting complex modern application platforms. These two bodies of work appear to be complementary.
This dissertation also suggests the insight that partial synthesis techniques can provide a way of addressing the learning challenge that developers face when attempting to realize the MDD~\cite{anthony_finkelstein_bottom_2012}.

\section{Object-Relational Mapping}
A large body of work has focused on  object-relational mapping strategies and their impacts to address the impedance mismatch
problem~\cite{luca_cabibbo_managing_2005,ireland_understanding_2009,keller_mapping_1997,philippi_model_2005}. Among others,
Philippi~\cite{philippi_model_2005} categorized the mapping strategies in a set of pre-defined quality trade-off levels, which are used to develop a model driven approach for the generation of OR mappings. Cabibbo and Carosi~\cite{luca_cabibbo_managing_2005} also discussed more complex mapping strategies for inheritance hierarchies, in which the various strategies can be applied independently to different parts of a multi-level hierarchy.
I propose a novel approach in this area by formalizing ORM strategies previously informally described in some of these research efforts, and thereby automating generation of OR mappings for each application object model.

Drago et al.~\cite{drago_quality_2011} also considered OR mapping strategy as one of the variation points in their work on feedback provisioning. They extended the QVT-relations language with annotations for describing design variation points, and provided a feedback-driven backtracking capability to enable engineers to explore the design space.
While this work is concerned with the performance implications of choices of per-inheritance-hierarchy OR mapping strategies, it does not really attack the problem that I am addressing, namely the automated and exhaustive synthesis of the equivalence classes of mixed OR mapping specifications.

\section{Deriving Database-centric Implementations from Formal Specifications}
Krishnamurthi et al.~\cite{krishnamurthi_alchemy:_2008_short} proposed an approach to refine Alloy specifications into PLT Scheme implementations with special focus on persistent databases. Cunha and Pacheco~\cite{cunha_mapping_2009} are similarly focused on translating a subset of Alloy into the corresponding relational database operations. These research efforts share with mine an emphasis on using formal methods. However, my work differs in its focus on separating application description from other independent design decisions, such as choices of OR mapping strategies. Furthermore, I use Alloy not only to specify the object model, but also to model the spaces of OR mappings consistent with both a given object model and a choice of mapping strategy, and to automate the mapping process.

\newpage
\thispagestyle{empty}
\mbox{}

\chapter{Synthesizing Architectural Models from Abstract Application Models}

In this chapter, I present Monarch.
This is one of the three experiments the I have conducted in the context of this dissertation.
The specific aim of this system is to automate the costly and unreliable process of refining abstract application models into software architectures in selected architectural styles. To this end, I develop a formally precise and automated approach for synthesis of architectural models from formal specification of application models and choices of architectural style.
This work introduces the application type---a formal specification of a family of application---as a source modeling language, and uses style specifications as target languages.
This work, among other things, provides evidence in support of 
reducing the costs of developing modeling languages.
It supports rapid and declarative development of narrowly targeted application types.
It enables the reuse of formal specifications of target domain specifications, including previously published specifications of architectural styles as well as mapping rules.
In support of the claim of broadening the applicability of MDD in various stages of the software development lifecycle, this work demonstrates a model-driven approach for development of style-specific software architectures from high-level application structures.
Finally, 
this work incorporates benefits of formal specifications of architectural styles to reliably synthesize style-specific architectural models.

This chapter is organized as follows.
Section~\ref{problem_1} presents motivation and research problems.
Section~\ref{approach_1} provides a high-level introduction to the approach.
Section~\ref{framework} details the~\emph{Monarch} Framework I have developed to support automated development of software architecture.
In Section~\ref{Case Study_1}, I use a well-known and widely used example of the Lunar Lander application~\cite{taylor_software_2009} to demonstrate the key steps in the mapping process.
Section~\ref{Evaluation_1} reports data from the experimental testing of the approach.
Finally, Section~\ref{conclusion_1} concludes this chapter. 

\section{Motivation and Research Problem}
\label{problem_1}

Software architecture is an essential means for managing complexity and meeting demanding requirements in developing complex software systems~\cite{shaw_software_1996,taylor_software_2009}. Architectural styles systematize successful architectural design practices in terms of constraints on architectural elements and their composition into systems~\cite{shaw_software_1996}.

Developing a sound and appropriate architecture, however, remains a significant and intellectually challenging activity. To develop architectural models effectively one must understand both the application domain in question and the discipline of software architecture. However, these bodies of knowledge are typically held by different people. Domain experts better understand requirements and specifications, while architects understand architectural styles, their implications, and techniques for mapping abstract application models to architectural models in given styles.

The required communication and coordination, and the manual mapping of application models to architectures, are costly and error-prone activities.
This has called for an approach to close this arduous line of actions by enabling domain experts to model application requirements more directly, while automating key parts of the work done by human software architects.

\section{Approach}
\label{approach_1}

To address this problem, this work contributes a formally precise approach to separating application-specific decisions from architectural style design decisions and then using these separate decisions as inputs to an automated synthesizer. The key to this separation is a means of reconciliation---an `architectural map'---connecting application descriptions to realizations in particular architectural styles. Architectural map, in this work, defines mapping rules required by the overall approach.
Making these maps central complements past work on architectural styles with new attention to how style choices combine with application models to yield architectures.

Researchers and practitioners have long recognized that {\em application models} can be considered largely separately from {\em architectural style}. As long ago as 1972, for example, Parnas~\cite{parnas_criteria_1972} described a {\em key word in context (KWIC)} system and consequences of choosing to design it in one of two styles: functional decomposition or information hiding. As recently as 2009, Taylor et al.~\cite{taylor_software_2009} similarly described a lunar landing embedded control system and showed how programs implementing it could be realized in many different styles.

The same basic idea recurs in many works~\cite{robert_deline_avoiding_1999,dewayne_e._perry_foundations_1992,erich_gamma_design_1995,garlan_evolution_styles_2009,shaw_software_1996}: a choice of application model independent of architectural style can be combined with a separate choice of architectural style giving rise to one or more architectures for the application in the given style. Much of this work is concerned with the implications of style choice in terms of evolvability, reusability and other {\em quality attributes.} Research in this area has provided insights and technology that help designers to make good architectural choices.

Work in this broad area has also led to many practically important advances, e.g., object-oriented design patterns~\cite{erich_gamma_design_1995}, which abstract styles from successful instances, and the REST architectural style~\cite{roy_t._fielding_principled_2000}, which systematizes architectural constraints that are important for producing scalable web-based, data-oriented systems.
Yet while we already have a theory of architectural styles, and while we understand that we should seek to separate applications and architectural style, we have little theory for separating application models and architectural styles in ways that provide precise definition of the mappings that combine them to produce style-specific architectures.
Rather, architects gain tacit knowledge of these mappings through study and practice, and they implement them manually. Such work is labor-intensive and expensive, and, once completed, prohibitively hard to change.
Moreover, failure to separate these concerns complicates reasoning about application properties, and retargeting of applications to varying architectural styles and platforms.
This work seeks to automate such mappings, making architectural style more of a separate and easily changeable variable in design.


The remainder of this section introduces the theoretical framework for the notion of \emph{architectural map} to separating application models from architectural styles and automating the associated mappings, and then presents the correspondence of the proposed theory and the formal structure of model-driven development.

\subsection{Theoretical Framework}
\label{Formalizatio}
This section makes explicit and elaborates the notion that an {\em architectural map} combines an application model, $m,$ of a given application type, with a specification, $s$, of a given architectural style, to produce a set of architectural models, $\{a_i\},$ for application $m$ in style $s$.
These architectural models refine the application model while complying with the rules implied by the architectural style.

\[\{a_i: ArchModel \} =  ArchMap(m: AppType, s: ArchStyle)\]

$ArchMap$ is the principal object of this research study. It captures architectural knowledge that we seek to formalize and automate. Putting it at the center begins to balance attention to architectural styles, with attention to how style choices combine with application descriptions to yield architectures. Knowledge of this mapping is crucial to expertise in software design.
Given an application description, the experienced designer knows both what architectural style to pick, and how to map an application description of the given kind to an architectural description in the chosen style.
Clearly $map$ is a complicated object. In some sense, it embodies knowledge of how to realize different types of applications in different styles. We need a way to study it in pieces.

One contribution of this work is an approach to doing this: I decompose $map$ by treating it as a function polymorphic in both application type and architectural style. I then investigate it for specific pairs.
Thus study requires to make explicit a notion of {\em application type}. Application descriptions come in a variety forms. Examples include {\em composition of functions} (which is in essence how Parnas characterized KWIC~\cite{parnas_criteria_1972}, for
example), or {\em state machine}, or {\em sense-compute-control}~\cite{taylor_software_2009}.
Each of these {\em application types} provides a vocabulary and structuring mechanisms for organizing application descriptions prior to the choice of architectural style for the system implementation. An architectural map in essence converts the structure and content of such a description into a form consistent with a given architectural style choice.
Specifically, I view \emph{ArchMap} as parameterized by type ($AppType$) and style ($ArchStyle$), and develop separate mappings for each compatible $AppType/ArchStyle$ pair. Compatibility captures the idea that not every architectural style is appropriate for every application type.



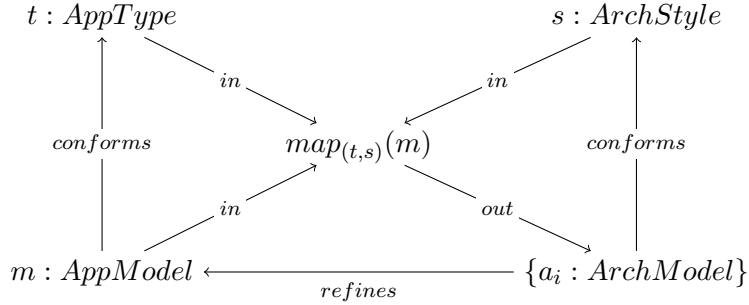
\begin{figure}[tbh]
\centering
\begin{tikzpicture}[description/.style={fill=white,inner sep=2pt}]
\matrix (m) [matrix of math nodes, row sep=3em,
column sep=2.5em, text height=1.5ex, text depth=0.25ex]
{ t: AppType &  & s: ArchStyle \\
  & map_{(t,s)}(m) & & \\
  m: AppModel & & \{ a_i: ArchModel \} \\};
\path[->,font=\scriptsize]
(m-3-1) edge node[description] {$conforms$} (m-1-1)
(m-1-1) edge node[description] {$in$} (m-2-2)
(m-1-3) edge node[description] {$in$} (m-2-2)
(m-3-1) edge node[description] {$in$} (m-2-2)
(m-3-3) edge node[auto] {$refines$} (m-3-1)
(m-3-3) edge node[description] {$conforms$} (m-1-3)
(m-2-2) edge node[description] {$out$} (m-3-3)
;
\end{tikzpicture}
\caption{\label{fig:category-theory-illustration}Key entities and relations in \emph{architectural maps}.}
\end{figure}

Figure~\ref{fig:category-theory-illustration} represents the fundamental elements of this model on the basis of the architectural maps and their relationships: (1) $\{a_i\}$, a set of architectural models (architectures) derived by the processes I describe in this chapter; (2) $s$, an architectural style specification; (3) {\em conforms}, a relation encoding the conformance of an architectural model, $a_i$, to an architectural style, $s$; (4) $m$, an application model; (5) $t$, an application type; (6) a (second) {\em conforms} relation, encoding the conformance of  $m$ to  $t$; (7) $map_{(t,s)}$, a map parameterized by $t$ and $s$ that takes application model, $m,$ to the set of architectural models, $\{a_i\};$ (8) a {\em refines} relation encoding the property that each such $a_i$ refines the application model, $m$. Given input parameters, $t, s,$ and $m$, our map yields a set of architectures because, in general, multiple architectures in a given style satisfy the required conformance and refinement constraints.

With these terms in hand, I can now say more precisely what I mean by {\em architectural style as a separate design variable.} For a given application model and type, one can select among compatible styles and maps and automatically synthesize architectures in these styles. To the extent that the essence of modularity is a decoupling of design parameters, the approach realizes a new form of modularity: it modularizes architectural style.

\subsection{Model-based Automation}

As software systems become larger and more complex, there is an ever greater need to employ higher levels of abstraction in application development. Model-driven development is centered around abstract, domain-specific models and transformations of abstract models into the constructs of specific underlying platforms. To be more precise, MDD is rooted in a mapping that takes a {\em platform-independent model}, $p$, and a {\em platform definition model,} $s$, to a {\em platform-specific model,} $i$. That is, $i: PSM = map(p: PIM, s: PDM).$
In my approach, application models are mapped to software architectural models, or targets, by way of choices of a software architectural style. $\{a_i: ArchModel \} \in ArchMap(m: AppType, s: ArchStyle).$

\begin{figure}[bth]
\begin{center}
\begin{tabular}{|c|c|}
  \hline
  {\bf Model-driven Development} &  {\bf Our Theory} \\  \hline \hline
Meta-model & Application type \\   \hline

Platform Independent Model (PIM) & Application Model \\   \hline

Model Transformation & Architectural Mapping  \\   \hline
Platform Definition Model (PDM) & Architectural Style \\   \hline
Platform Specific Model (PSM) & application- and style-specific Architecture \\ \hline

\end{tabular}
\caption{\label{fig:correspondence}Correspondence between terms of $Architectural maps$ and the formal structure of MDD}
\end{center}
\end{figure}

The analogy between two approaches is clear in the equations. Architectural styles plays the same role as platform descriptions in a MDD approach~\cite{schmidt_MDE_2006}. I believe that this observation opens a path to MDD tools that support architectural style as a separate variable in automated development of software architectures.
Figure~\ref{fig:correspondence} shows the correspondence between terms of $Architectural maps$ and the formal structure of MDD.
I introduce \emph{application types} as styles of application description that play the same role as application meta-models in MDD. Application models correspond to platform independent models in MDD;  architectural maps, to MDD transformations; architectural styles, to MDD platforms; and synthesized software architectural models, to platform-specific models in MBD.

Given the concepts of application {\em type} and architectural {\em style} I can now concisely describe the approach. A user selects an application type. This type selects a meta-model that parameterizes a model-based editing tool. 
Within such a tool one creates an application model as an instance of the selected application type. One then selects an architectural style. The combination of an application type and an architectural style selects the specification of a synthesis function: an \emph{architectural map} for that particular pair of input specifications. Each architectural map specifies the mapping of any application model (instance) of the given application type to an architectural model in the given style. Application types, architectural styles, and architectural maps are all formally specified in a notation that supports automated analysis and synthesis. The approach involves the synthesis of an architectural description that satisfies the constraints of both type and style specifications, and that does so in a particular manner described by architectural mapping constraints.

The next section shows that how these ideas can be reduced to practice.
There are many possible approaches to implementing tools that compute architectural maps. This work presents one approach, in which I use Alloy~\cite{daniel_jackson_alloy:lightweight_2002}, which is a specification language that has been optimized for automated analysis.


\section{Monarch: A Prototype Tool}
\label{framework}

\begin{figure}[tbh]
\includegraphics[width=\textwidth]{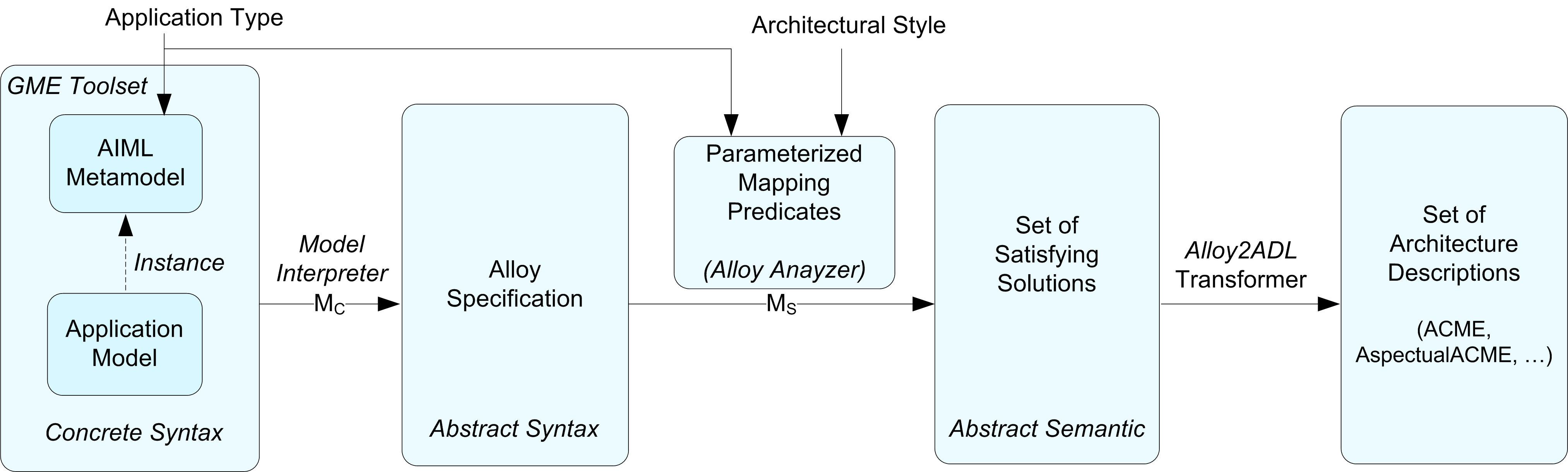}

\caption{\label{fig:ToolArchitecture}High-level overview of the Monarch framework.}
\end{figure}

Monarch is a framework for automated development of software architecture.
Figure~\ref{fig:ToolArchitecture} outlines the high-level overview of the Monarch framework. 
Formally, the definition of a modeling language~\cite{tony_clark_mmf_2001,kai_chen_semantic_2005} consists of: (1) the \emph{abstract syntax (A)} that defines the language concepts, and the relationships between them, (2) the \emph{concrete syntax (C)} as a human-centric notation, (3) the\emph{ syntactical mapping} $(M_C: C~\rightarrow A)$ that links the concrete syntax to the abstract syntax, (4) the \emph{semantic domain (S)} that specifies well-formedness rules for models as well as the meaning of the models, and finally (5) the \emph{semantic mapping} $(M_S: A~\rightarrow S)$ that links the abstract syntax to the semantic domain, giving a modeling language a meaning.

The Monarch framework comprises (1) an approach to architecture-independent application modeling using the Generic Modeling Environment (GME)~\cite{ldeczi_composing_2001}, with application types realized concretely as GME architecture-independent modeling language (AIML) meta-models; (2) interpreters that transform application models, viewed as \emph{concrete instances} of architecture-independent  modeling languages into an \emph{abstract syntax} based on Alloy~\cite{daniel_jackson_alloy:lightweight_2002}; (3) a mapping engine, based on the Alloy Analyzer, that takes such an application model and a formal specification of an architectural style and that finds architectural models that refine the application model in conformance with the given architectural style. The mapping employed is based on the combination of the selected application type and the selected architectural style. The engine uses the Alloy constraint solver to compute architectures, represented as satisfying Alloy solutions; (4) a final post-processing phase, \emph{Alloy2ADL}, translates the resulting Alloy instances into human-readable architectural description languages (ADLs).
The rest of this section describes each of the modules.

\subsection{AIML Meta-model}
The definition of a concrete syntax by a meta-model is supported thoroughly by many meta-modeling environments, e.g. Generic Eclipse Modeling System~\cite{jules_white_introduction_2007}, MetaEdit+~\cite{_metaedit+_2010} and  Generic Modeling Environment~\cite{ldeczi_composing_2001}. I have developed Architecture-Independent Modeling Languages (AIMLs) on top of the GME to support the specification of application content at the abstract modeling level. The reasons for choosing GME for this study include its straightforward mechanisms for developing extensions, and its availability and proven success for MBD.

\begin{figure}[tbh]
\includegraphics[width=\textwidth]{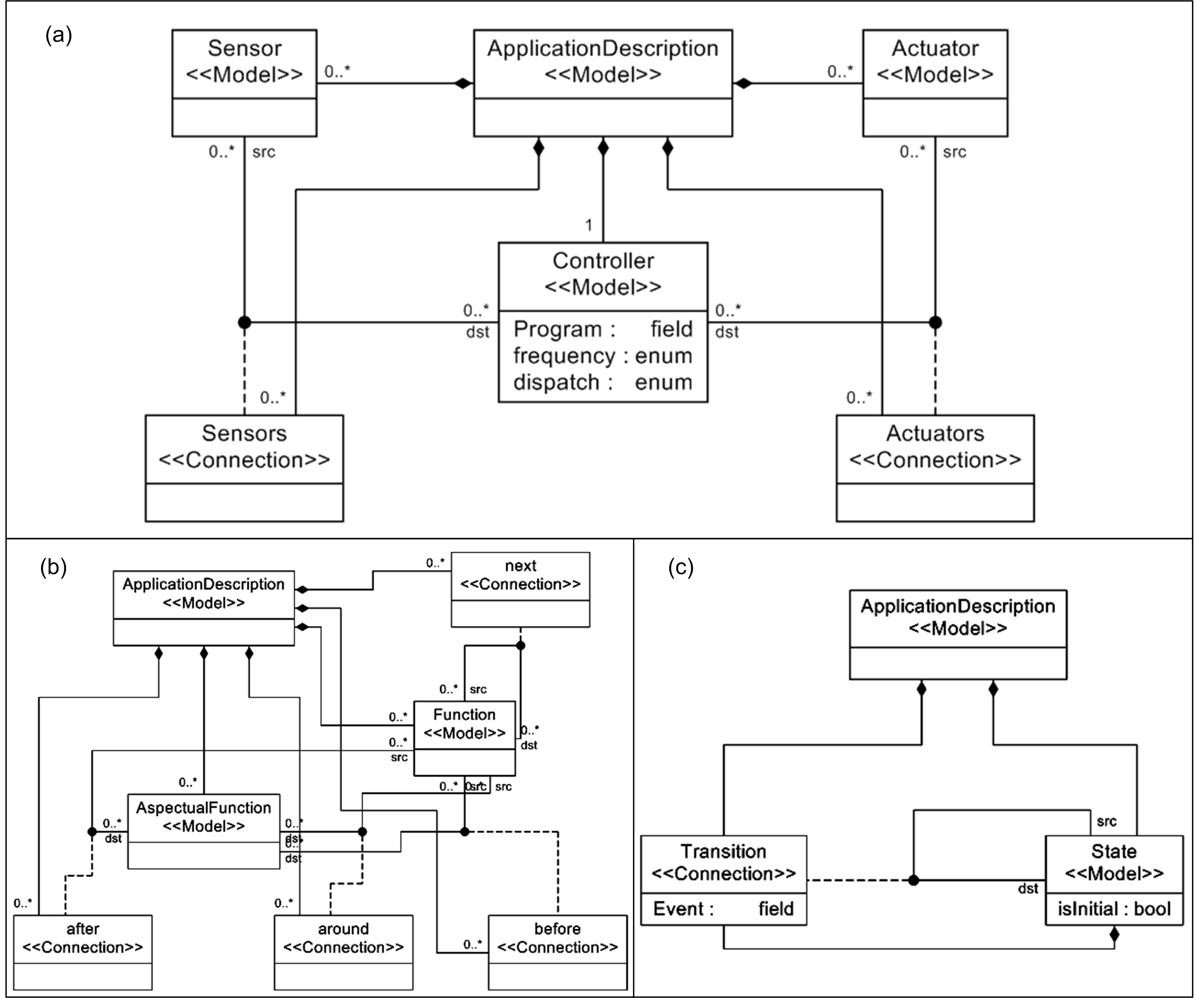}
\caption{\label{fig:Metamodels}Meta-models in GME for application types: a) Sense-Compute-Control b) Aspect-enabled Composition-of-Functions c) State-Driven behavior}
\end{figure}

A meta-model specification describes a particular form of model. In our earlier work~\cite{h._bagheri_architectural_2010}, we identified several possible forms of architecture-independent model, including {\em composition of functions}, {\em aspect-enabled composition-of-functions (ACF)}, {\em state-driven behavior (SD)} and {\em sense-compute-control (SCC)}. Our meta-model for the {\em sense-compute-control (SCC)} application type is shown in Figure~\ref{fig:Metamodels}a. I have developed GME meta-models for several previously identified but not well elaborated application types: {\em composition-of-functions (CF)}, {\em aspect-enabled composition-of-functions(ACF)} (Figure~\ref{fig:Metamodels}b), and {\em state-driven behavior (SD)} (Figure~\ref{fig:Metamodels}c). For brevity, and because it suffices to make our points, I describe only the \emph{SCC} meta-model in this work. Monarch supports the others as well.

Sense-compute-control (SCC) is an \emph{application type} for embedded control systems. The SCC application type is used to model applications in which a set of sensors and actuators are connected to controllers that cycle through the steps of fetching sensors values, executing a set of functions, and emitting outputs to the actuators~\cite{taylor_software_2009}. Figure~\ref{fig:Metamodels}a shows a UML class diagram for the SCC meta-model as represented in GME. The \emph{ApplicationDescription} class, which is common among application-type meta-models, denotes the application specification diagram, which contains elements of an application model. The \emph{Controller}, \emph{Sensor} and \emph{Actuator} classes represent three main elements of the SCC application type. The Controller\rq{}s frequency is abstracted into discrete ranges of \emph{slow} and \emph{fast}. The Controller can also be specified as using a \emph{periodic}, \emph{aperiodic} or \emph{sporadic} task.

Given such a meta-model specification of a modeling language for each architectural-style-independent application type, GME automatically creates an architectural-style-independent modeling environment. The designer of a system then specifies an application description as a model using the modeling environment. I believe this approach promises to allow domain experts to model their applications abstracted from details of software architectural styles.

\subsection{Transformation to Alloy Specification}
%

Monarch then translates an application model specified as a concrete instance of such an architecture-independent modeling language (AIML) for a given application type into an abstract syntax based on the Alloy language. The GME provides several ways to process data from the model automatically. I used the Java version of the Builder Object Network (BON), providing us access to the internal representation of the model through Java objects. I also developed a \emph{GME interpreter} for each meta-model as a syntax mapping, which elicits the model elements and transforms the constructs of the application model to formal specifications in the Alloy language.

Alloy is a lightweight set-theoretic specification language based on the first-order logic~\cite{daniel_jackson_alloy:lightweight_2002}. It has been applied by numerous researchers to formal work in software architecture. Kelsen and Ma~\cite{kelsen_lightweight_2008}, comparing the traditional methods of formal specifications for modeling languages with an approach based on the Alloy language, argue that because of both lower notational complexity and automatic analyzability,  Alloy provides more convenient facilities for defining the formal semantics of modeling languages.
I chose Alloy for this study for two reasons. First, its ability to compute solutions that satisfy complex sets of constraints is useful as an automation mechanism.
Second, and more importantly, it allows us to better validate our claims because we use, as inputs, architectural style specifications, in Alloy,  that others have published~\cite{jung_soo_kim_analyzing_Journal,stephen_wong_scalable_2008}. Reusing published models is important in that it shows our ideas and approach to be consistent with contemporary formal accounts of architectural style.

Essential constructs of the Alloy language include: \emph{Signatures}, \emph{Facts}, \emph{Predicates}, \emph{Functions} and \emph{Assertions}. Signatures represent the basic types of elements and the relationships between them. Alloy provides Facts to be used in defining constraints that any instance of a model must satisfy. Predicates are parameterized reusable constraints that are always evaluated to be either true or false. Functions are parameterized expressions. A function can be invoked by instantiating its parameter, but what it returns is either a true/false or a relational value instead. An assertion is a formula required to be proved. It can be used to check a certain property of a model.

With respect to the constraints in a given model, the Alloy Analyzer can be used either to find solutions satisfying them, or to generate counterexamples violating them. The Alloy Analyzer is a bounded checker, so a certain scope of instances needs to be specified. In the matter of architectural styles, the scope states the number of architectural elements of each type. To take advantage of partial models, its latest version uses \emph{KodKod}~\cite{emina_torlak_constraint_2009} as its constraint solver so that it can support incremental analysis of models as they are constructed. The generated instances are then visualized in different formats such as graph, tree representation or XML. I use the Alloy Analyzer to compute architectural models given the conjunction of an architecture-independent model represented using a particular meta-model, and a choice of formal specifications of an architectural style, also represented in Alloy. 

\subsection{Architectural Mapping}
\label{ArchitecturalMapping}

An architectural style description specifies the {\em co-domain} of an architectural map. To represent a map, itself, I extend the style description with mapping predicates. These predicates take types of applications as parameters, and define relationships required to hold between applications of given types and computed architectural models in the given style under consideration. Mapping predicates are responsible for ensuring that computed architectural models {\em refine} given application descriptions. Given an application description, and a map, Monarch using the Alloy Analyzer computes corresponding architectural models that {\em conform} to the given target architectural style.

\begin{figure}[tbh]
\centering
\includegraphics[width=\linewidth]{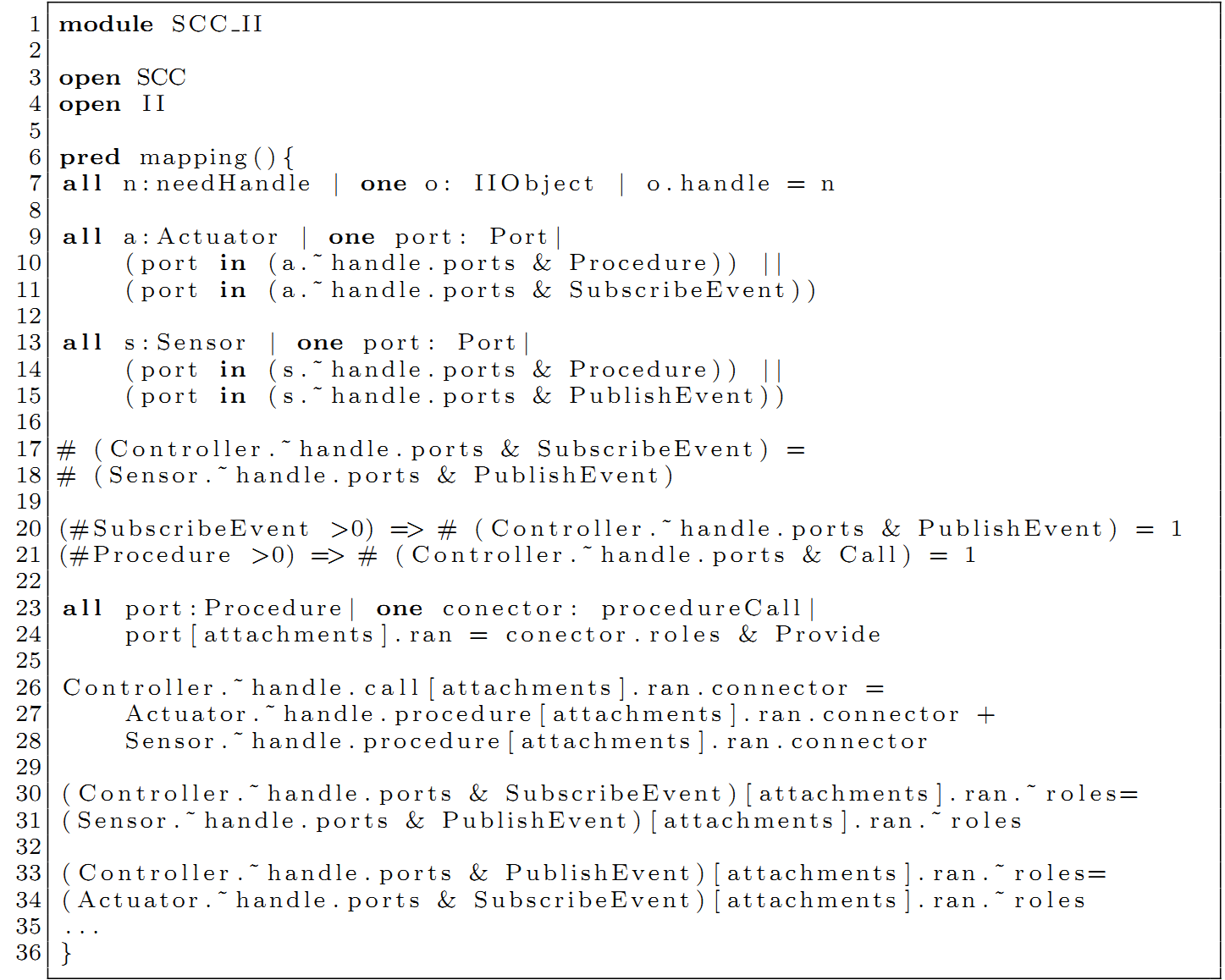}
\caption{\label{fig:AlloyCode}Part of the mapping predicate for the pair of SCC application type and II architectural style}
\end{figure}

As an example, 
Figure~\ref{fig:AlloyCode}
presents part of the predicate for mapping application models in the \emph{sense-compute-control} application type to architectures in the {\em implicit invocation} style. I describe it in more detail in Section~\ref{ArchitecturalMappingProcess}.

\subsection{Transformation to Architecture Description Language}
\label{AlloyToADLTransformer}

Having computed satisfying solutions, the \emph{Alloy2ADL transformer} component parses and transforms these solutions from low-level, XML formatted Alloy objects to high-level architecture descriptions in human-readable ADLs.
During the past several years, a considerable number of general and domain-specific ADLs have been proposed~\cite{nenad_medvidovic_classification_2000}. I use the Acme language in the prototype~\cite{david_garlan_acme:_2000}. Acme emerged as a generic language for describing software architectures, with particular support for architectural styles. It is also designed to work as an interchange format for mapping among other architecture description languages.
In an earlier work~\cite{bagheri_architecture_2010}, I briefly reported on the feasibility of treating architectural style as a separate variable in an aspect-oriented setting, with AspectualACME~\cite{alessandro_garcia_modular_2006}---an aspect-enabled extension of Acme---as a target ADL.

\section{Example}
\label{Case Study_1}

In this section, I use Monarch to  formally illustrate the process of mapping a \emph{SCC} description of the \emph{Lunar Lander} application~\cite{taylor_software_2009} to architectural models in the implicit-invocation style~\cite{implicit_invocation_1998}.




\subsection{Application Type: SCC}
In Section~\ref{framework}, I presented the concrete and human-centric  realization of $SCC$ application type as an AIML meta-model developed atop GME. Here I focus on its abstract syntax developed as an Alloy module.

\begin{figure}[tbh]
\centering
\includegraphics[width=\linewidth]{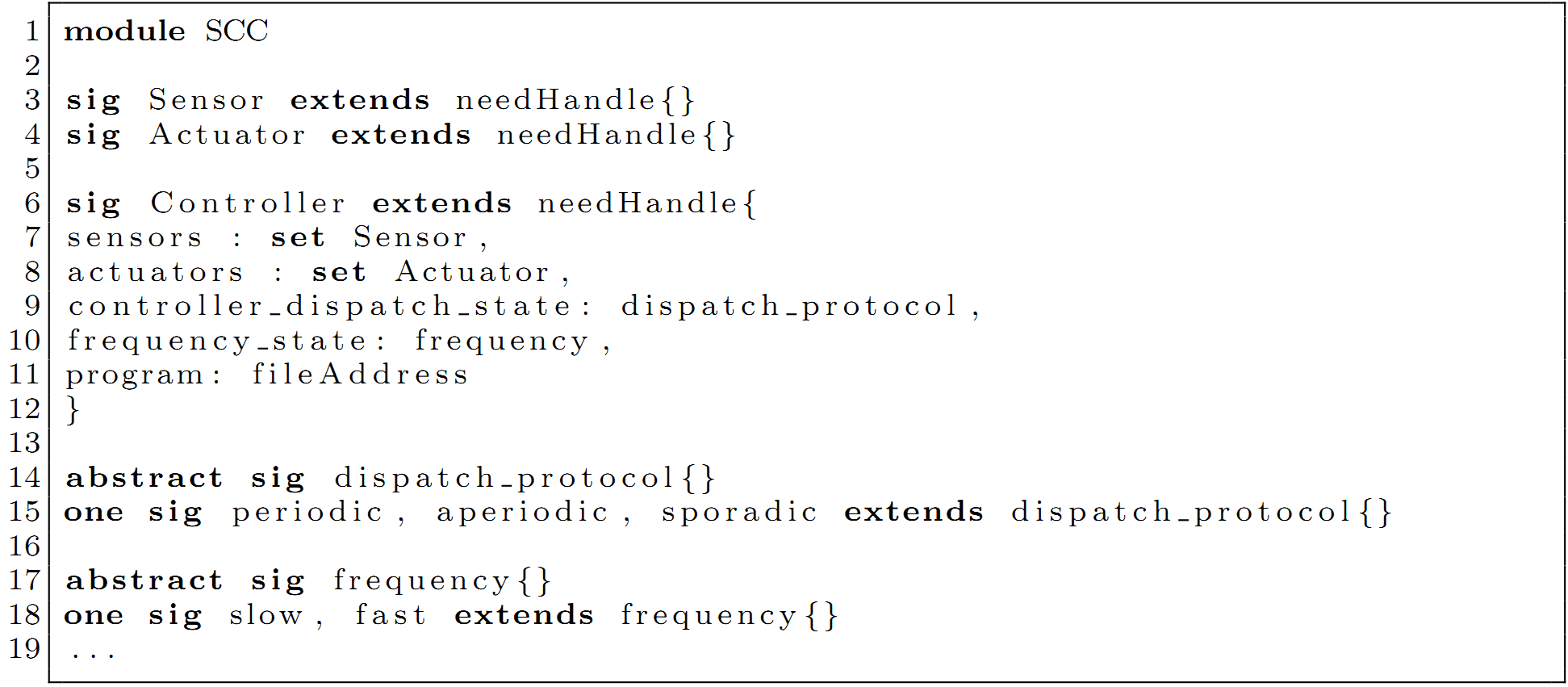}
\caption{\label{fig:SCCAlloy}Part of the sense-compute-control application type as an Alloy module}
\end{figure}

Figure~\ref{fig:SCCAlloy} partially outlines sense-compute-control application type represented in Alloy. Each meta-element in the SCC metamodel has a corresponding Alloy signature definition, except for \emph{ApplicationDescription} whose instances denoting specific application models are mapped to separate Alloy modules.
In particular, three Alloy signatures represent basic elements of SCC application type, i.e. \emph{Actuator}, \emph{Sensor} and \emph{Controller}.
The Alloy module further defines two abstract signatures of \emph{dispatch\_protocol} and \emph{frequency},
which are used in defining the specific properties of \emph{Controller} elements.

\subsection{Architectural style: II}
\begin{figure}[tbh]
\centering
\includegraphics[width=\linewidth]{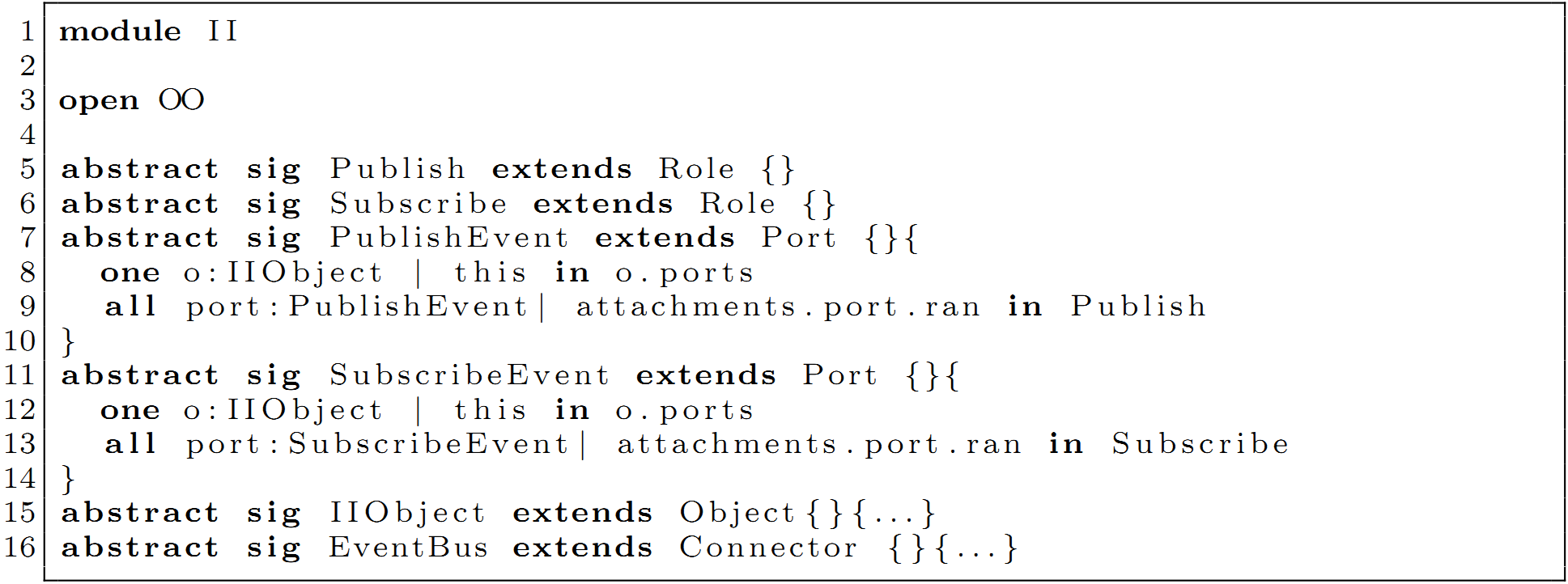}
\caption{\label{fig:IIAlloy}Part of II style described in Alloy}
\end{figure}

In some cases it is helpful to model one architectural style as inheriting rules from another. An implicit invocation object (IIObject) is thus an Object that provides both a collection of interfaces (as with Object) and a set of events. Procedures may also be called in the usual way. So, an IIObject extends the definition of an Object. It can, in addition, register some
of its procedures with events of the system; so those procedures will be invoked when the events are announced.
Figure~\ref{fig:IIAlloy}
(eliding details) makes these ideas precise in six signatures: Publish, Subscribe, PublishEvent,
SubscribeEvent, IIObject and EventBus. IIObject has PublishEvent and SubscribeEvent as its ports. EventBus is further a special kind of Connector and has two roles, i.e. Publish and Subscribe.

The Alloy dot operator denotes a relational join.
In expressions represented in lines 9 and 13, the $attachments$ is a relation of type $System~\times~Role~\times~Port$.
Therefore, the $attachments.port$ relation is from System to Role.
The function $ran$, defined in the Alloy module util/relation, returns the range of a binary relation. The ``in" operator furthermore declares the subset relation.
As such, the invariants under consideration specify that each $PublishEvent$ port of an $IIObject$ should be attached to a role of type $Publish$, and each $SubscribeEvent$ port of an $IIObject$, in a same way, should be connected to a $Subscribe$ role of an $EventBus$.

\subsection{Architectural Map: (SCC,II)}
\label{ArchitecturalMappingProcess}

The architectural mapping process takes as inputs the abstract application model (transformed directly from the concrete model to the Alloy module), an Alloy module specifying the application type (meta-model), an architectural style Alloy module that specifies the constraints to which the computed results conform, and the architectural mapping Alloy predicates that define relationships required to hold between the application model and computed architectural models. Each mapping predicate for the given application type and architectural style is responsible for confirming that the satisfying solutions refine the given application model in conformance with the given style. Overall, I have developed eight architectural mappings validating the approach (cf. Figure~\ref{fig:experiments-to-date}). 

Figure~\ref{fig:AlloyCode} shows such a predicate for the SCC application type and the implicit invocation architectural style.
At the top, the specification imports the Alloy modules for the SCC application type and implicit invocation architectural style.
The mapping predicate then, in line 7, states that for each sensor, actuator and controller, declared as subtype of the \emph{needHandle} abstract Signature, there is an \emph{IIObject} that handles it.
Expressions in lines 9--11, by using the Alloy inverse relation operator $\sim$, state that each Actuator's IIObject has a port of type \emph{SubscribeEvent} or \emph{Procedure} to be called implicitly or explicitly.
Likewise, each Sensor's IIObject has a port of type \emph{PublishEvent} or \emph{Procedure}. The number of SubscribeEvent ports of the Controller's IIObject equals to the number of PublishEvent ports of the Sensors' IIObjects, as mentioned in lines 17--18. So, each SubscribeEvent port of the Controller could be connected to a Sensor's PublishEvent port to be called implicitly. In addition, the specification, in lines 20--21, states that the Controller's IIObject has at most one PublishEvent port and one Call port so that the procedures of Actuators' IIObjects could be called explicitly or could register to be invoked when the PublishEvent port of the Controller's IIObject announces an event.

The II architectural style provides two ways of invoking methods: \emph{procedure call} and \emph{implicit invocation}. For the purpose of the former method, lines 23--28 state that for each Procedure port, there is a ProcedureCall connector connected to it, and the Call port of the Controller's IIObject is connected to the Procedure port of the IIObjects handling Sensors and Actuators via a connector of type ProcedureCall. For an implicit invocation, the SubscribeEvent ports of the controller's IIObject are connected to the PublishEvent ports of Sensors' IIObjects via an \emph{EventBus} connector, as mentioned in lines 30--31. In a similar way, the SubscribeEvent ports of the Actuators' IIObjects are connected to the PublishEvent port of the Controller's IIObject through an EventBus connector.

\begin{figure}[tbh]
\centering
\includegraphics[width=\textwidth]{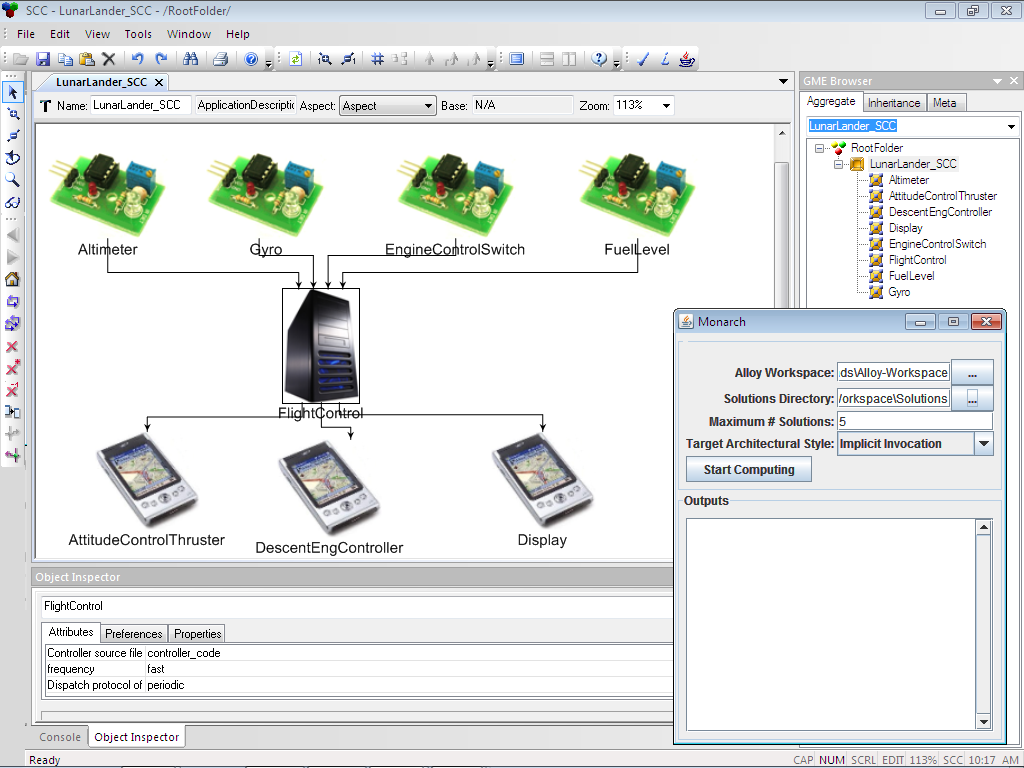}

\caption{\label{fig:LunarLander-GME}Lunar-Lander application model in GME modeling Environment}

\end{figure}

\subsection{Application Model: Lunar Lander in SCC application type}

In their textbook~\cite{taylor_software_2009}, Taylor et al. describe the informal mapping of a lunar lander application to architectures in a range of architectural styles. In this application, \emph{FlightControl} maintains the state of a  spacecraft based on the information provided by various sensors: Altimeter, Gyroscope, Fuel level indicator and the engine control switch. After processing control laws and computing values, FlightControl provides them for various actuators: Descent engine controller, Attitude control thruster and Display. Taylor et al. describe the lunar lander as an instance of a $sense-compute-control$  application. The notion of {\em application type} is implicit in their account. I make it explicit and formal in our theory. Figure~\ref{fig:LunarLander-GME} shows the lunar lander's application description modeled within GME using the generated modeling environment for the SCC meta-model.
What is also shown in the figure is a screenshot of the Monarch architecture-synthesizer environment and how the synthesis process is started.

\begin{figure}[tbh]
\centering
\includegraphics[width=\linewidth]{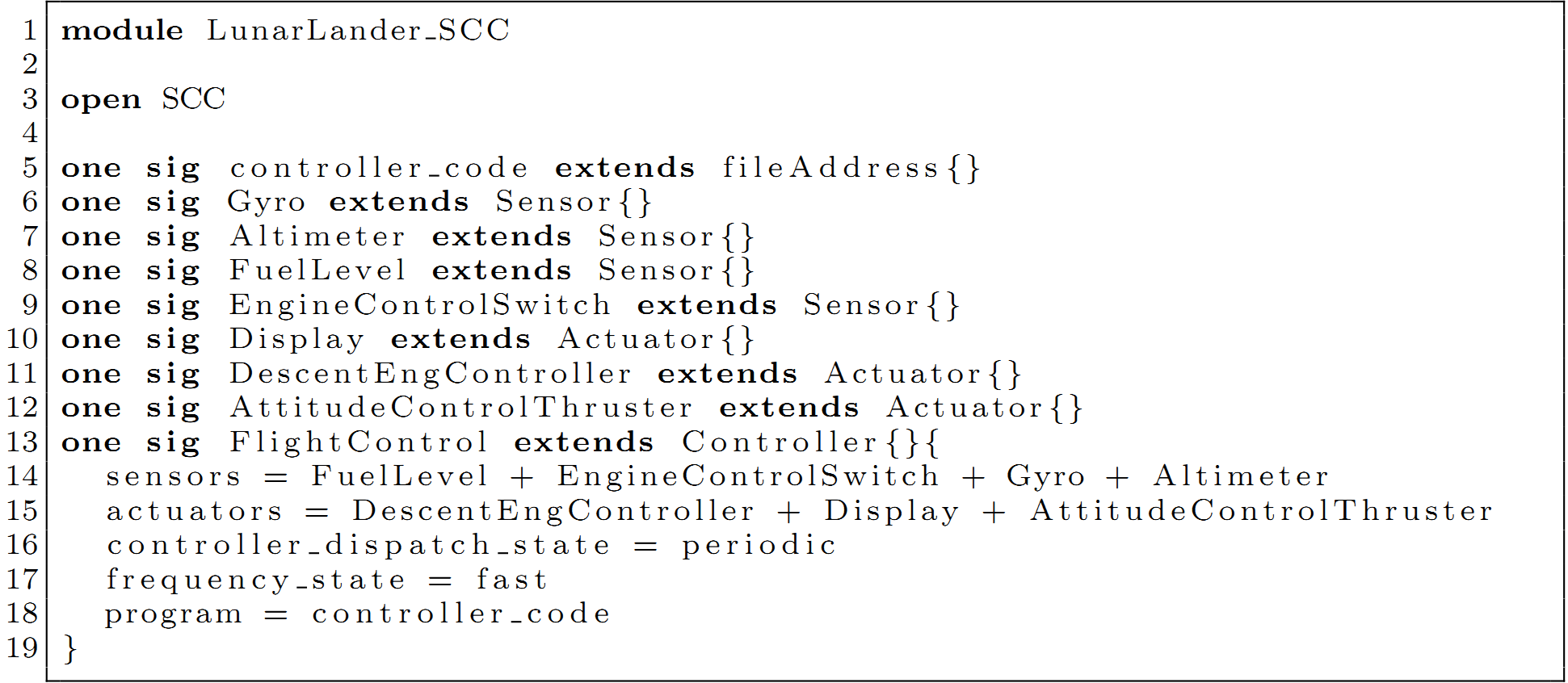}
\caption{\label{fig:LunarLander_Alloy}Lunar Lander application model represented in Alloy}
\end{figure}

Figure~\ref{fig:LunarLander_Alloy} illustrates the Alloy representation of the lunar lander application model generated directly from its concrete model by \emph{Monarch Interpreter} developed for the SCC meta-model.
A synthesized Alloy module contains a signature definition for each element in the concrete model as well as a set of facts corresponding to the properties of those elements.
More specifically, It starts by synthesizing the module name representing the name of the instance of the \emph{ApplicationDescription} class within the concrete model. It then imports the Alloy specification module(s) for application type(s). For each instance of \emph{Sensor}, \emph{Actuator}, and \emph{Controller} classes in a concrete model, it synthesizes a signature definition that represents the inheritance of a concrete element from its corresponding abstract class. The element's properties (if any) are also specified as Alloy facts for the corresponding signature of that element, e.g. \emph{FightControl} has a periodic task with high frequency.

\subsection{Satisfying Models}

Using the Alloy Analyzer, Monarch computes architectural models, represented as satisfying solutions to the constraints of a map applied to an application model. Alloy Analyzer guarantees that computed descriptions {\em conform} to the given architectural style. The mapping predicates are responsible for ensuring that computed architectural models {\em refine} given application models.

\begin{figure}
\centering
\includegraphics[height=\textheight]{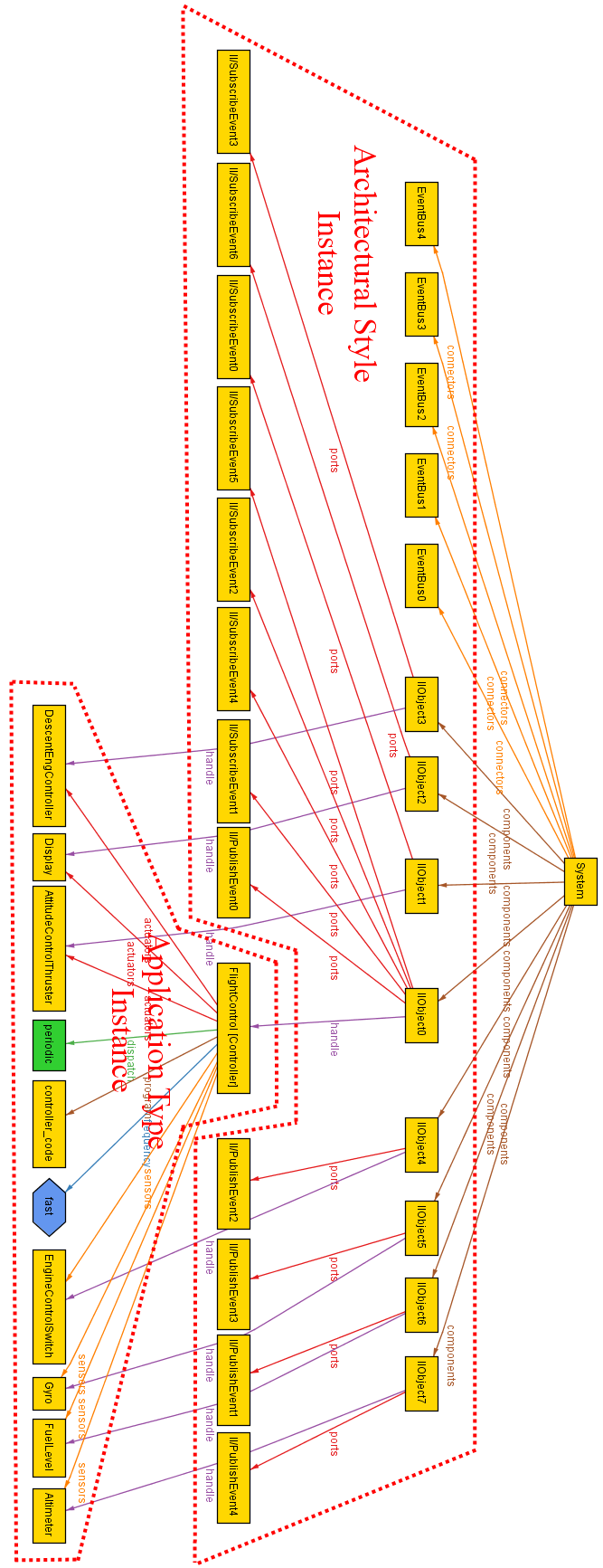}

\caption{\label{fig:SatisfyingSolution}The internal structure of a result of mapping Lunar Lander application into the II style }

\end{figure}


Mapping \emph{SCC} description of the Lunar Lander to the II architectural style yields a set of satisfying solutions, among them, for instance, Figure~\ref{fig:SatisfyingSolution} depicts the internal structure of a result for the lunar lander example.
In this diagram, the architectural description has eight \emph{IIObject}s. The \emph{FlightControl} element along with related sensors and actuators, inferred from the input specification, represents the Lunar Lander System.
Each IIObject handles an element. As a case in point, \emph{IIObject6} handles \emph{FuelLevel} sensor and publishes a notification of new value through \emph{PublishEvent1} that should be connected to an \emph{EventBus} (connections are omitted for the sake of readability).
On the other hand, \emph{IIObject0}, handles \emph{FlightControl}, subscribes to input events through \emph{SubscribeEvent}s ports, and will be implicitly invoked. This allows it to update the state of the spacecraft.
This in turn causes \emph{Display}, for example, to be invoked so that it refreshes its display based on new data.

\subsection{Architectural Models}

\begin{figure}[tbh]
\begin{center}
\setlength\fboxsep{0pt}
\setlength\fboxrule{0.5pt}
\fbox{
\includegraphics[height=2.15in]{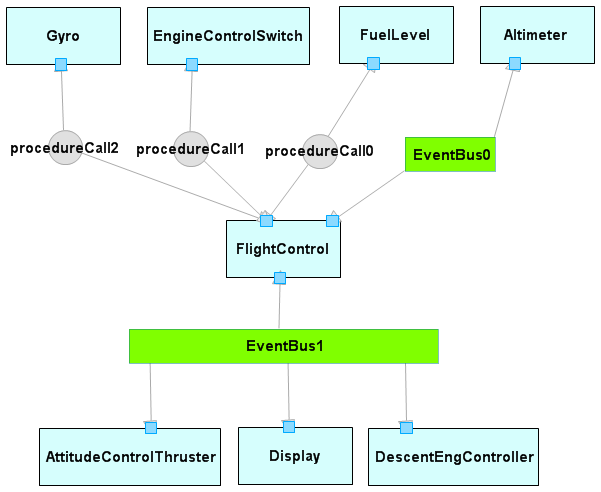}
}
\end{center}
\caption{\label{fig:LunarLander_SCC_II-Acme}One of the computed instances of mapping SCC description of the Lunar Lander into the implicit invocation architectural style in Acme}
\end{figure}

To make the outputs humanly readable and useful, the \emph{Alloy2ADL} transformer converts the Alloy-generated results, also available in an abstract XML-format, to a traditional architecture description language.
Figure~\ref{fig:LunarLander_SCC_II-Acme} represents one of the automatic computed instances of architecture description models in Acme. These models refine the Lunar Lander application description specified using the SCC application type, in conformance with the fully formal definition of the implicit invocation architectural style. The result is a set of formally derived architectural models for the given application in the selected architectural style.

According to the diagram, in this particular, arbitrarily selected case, three of the four top components, for handling sensors, are connected to the \emph{FightControl} through \emph{ProcedureCall} connectors, i.e., using explicit invocation, except for the Altimeter component, which is called implicitly via the \emph{EventBus0} connector. The actuators' components are also connected to the \emph{FlightControl} through the \emph{EventBus1} connector.

The example illustrates the point that architectural styles, viewed as mappings to platforms, can be one-to-N. In general, there are many possible architectures, consistent with a given style, for a given application. The architecture in the diagram should thus be viewed as just one instance (and not necessarily the best) of a family of architectures, given the specification of the II style. In general, mapping to families of architectures could be useful in enabling optimizing search for properties influenced by architecture but not constrained in the modeling stage. Alternatively, one could strengthen the definition of the style or the application type and instance to limit the family of conforming instances.

\section{Evaluation \& Discussion}
\label{Evaluation_1}

The claim that I make in this work is that it is feasible to formally separate application models from architectural styles, and to synthesize architectural models in given styles from given application models, and that the proposed separation of concerns enables automation and the production of model-based tools for architecture synthesis.
In support of the feasibility of these claims, I summarize a set of case studies formally conducted using \emph{Monarch} tool-suite. Specifically, I offer support in the form of a set of implemented architectural maps. I have developed maps for four application types and three architectural styles, as mentioned in Figure~\ref{fig:experiments-to-date}. The types are {\em composition-of-functions (CF)}, {\em state-driven behavior (SD)}, {\em sense-compute-control (SCC)} and {\em aspect-enabled composition-of-functions(ACF)}. The styles are {\em pipe-and-filter (PF), object-oriented (OO),} and {\em implicit invocation (II)}. 

\begin{figure}[bth]
\centering
\begin{tabular}{|l||c|c|c|c|}
  \hline
  & {\bf Comp. Fun.} & {\bf State-Driven}
  & {\bf SCC} & {\bf ACF} \\
  \hline
  {\bf Pipe-And-Filter} & KWIC, LL & & & KWIC\\
  \hline
  {\bf Object-Oriented} & KWIC, LL & KWIC & MIDAS, LL & \\
  \hline
  {\bf Implicit Invocation} & KWIC, LL & & MIDAS, LL & \\
  \hline
\end{tabular}
\caption{\label{fig:experiments-to-date}Maps defined and experiments performed. (Rows represent architectural styles; Columns represent application types.)}
\end{figure}


Each non-empty cell in Figure~\ref{fig:experiments-to-date} indicates an architectural map that I have implemented for the given type-style pair, and a corresponding experiment using the map.
The entries in the table indicate the case studies from the literature to which I have applied developed maps, to test the consistency of the automatically synthesized results with the informal, manually derived results in the literature. I have applied the approach to three case studies: KWIC, long used in studying architectural styles and their properties~\cite{parnas_criteria_1972}; the Lunar Lander (LL) case study of Taylor et al \cite{taylor_software_2009}; and the case study of an alarm system of type sense-compute-control, called MIDAS~\cite{malek_reconceptualizingfamily_2007}.
The last case study is inspired by Edwards et al.~\cite{george_edwards_model_2008}. They illustrated the structuring of the MIDAS application~\cite{malek_reconceptualizingfamily_2007} from a family of embedded applications at Bosch, in different architectural styles, to assess the influence of architectural style on quality attributes.

Most of the experiments attempt to replicate previously reported informal architectural mappings.
For example, \emph{(CF, OO, KWIC)} experiment focuses on reproducing previous informal studies by Parnas \cite{parnas_criteria_1972} and later studies by Shaw and Garlan \cite{shaw_software_1996}. I map a $CF$ description of KWIC to an architectural description in the OO style, or \emph{(SD, OO, KWIC)} experiment addresses the work of Garlan, Kaiser \& Notkin~\cite{garlan_using_1992}, who explored, among other things, how changing the KWIC application from batch-sequential to interactive might demand corresponding changes in the architectural style. I note that change can be seen as involving, at a more abstract level, a change in the {\em type} of application description, and that this change is what really drives the need for a new {\em architectural} style. I employ {\em state-driven behavior} as a type for interactive application description.
Details of these studies can be found in my previous technical reports~\cite{hamid_bagheri_architecture_2009,bagheri_architecture_2010}.
A complete list of these case studies, including complete versions of Alloy models, are also available at \emph{Monarch} website~\cite{_monarch_2010}.




The experiments show that architectural maps can be formalized and implemented as executable specifications.
The results of formal and automated computations are consistent with the informally and manually produced results documented in the literature.
This work also suggests that the concept of \emph{application type} is important.
The experiments further show that the concept of application type leads naturally to an abstract, user-friendly approach to application modeling. I have demonstrated an approach taking fully formal specifications of application models and architectural styles as inputs and producing software architectures as outputs within the framework of MDD.

In an earlier work~\cite{bagheri_architecture_2011}, we showed that architectural maps can also incorporate architectural tactics~\cite{bass_software_2003}, in a formal and reusable form.
More specifically, this work shows that our approach supports formal specifications of architectural tactics with respect
to the architectural maps developed for each pair of an application type and an architectural style.
This in turn enables automatic synthesis of architectural instances which conform to the rules implied by both the application model and the target architectural style as well as supporting the given architectural tactic.

The prototype implementation tool suite supports reasonable extension for new types and mappings. To that end, one specifies an application type and its corresponding GME meta-model, as well as the architectural mapping predicates for relevant styles, so that
by swapping between implementations of architectural maps, one can produce architectures in a range of styles for a given system from a high-level application description. This work appears to support the idea that being able to treat architectural style as a separate variable is a plausible aspiration.
With automated architectural mapping, the software architect may also readily examine the feasibility of various architectural alternatives.
The ease of examining more architectural alternatives will also increase the quality of software architecture.
The more various alternatives are studied, the more likely it is that the most satisfying option will be found.

It is also worth mentioning that the current tool mainly considers structural refinements with respect to the target architectural styles.
It would be of significant importance to represent and handle other style-related aspects, such as architectural behavior and quality attributes, which is considered as an interesting avenue for future work.
Moreover, our experiments to date were conducted on the accounts of architectural styles published in the literature. Targeting more complicated 
architectural styles, such as those induced by widely-used industrial platforms~\cite{nitto_exploiting_1999} or multi-dimensional architectural spaces is of tremendous value. I believe our synthesis approach can be naturally extended to support them, which remains an active area for future work.

Overall, the intellectual contribution of this work is the insight that software architectural styles can serve as analogs to choices of platforms in model-based development. This idea then leads naturally to a new kind of tool: one that allows for modeling of applications independent of subsequent architectural style choices, and for the automated mapping of models to architectures once such style choices are made. Whereas traditional MDD work seeks to replace the programmer, this work points the way to a future in which MDD replaces at least some of the work of the software architect.

\section{Summary}
\label{conclusion_1}

This work makes several contributions. First, 
I developed a theoretical framework for the notion of architectural map to separating application models from architectural styles and automating the associated mappings.
Second, I developed the concepts of application type and architectural map as key constructs needed, in addition to that of architectural style, to achieve such a separation of concerns. 
This work uses application types, architectural styles and architectural maps as source languages, target languages, and mapping rules, respectively.
Third, I showed that this separation of concerns gives rise to a natural form of model-driven development. Fourth, I presented experimental data and a prototype tool that support the feasibility claims and the proposition that these ideas are worth pursuing.


There are many opportunities for future work. First, architectural transformations are generally one-to-many. The more complex a system is, and the less tightly constrained the architectural style, the larger the set of architectures is. The need to choose from among candidates provides an opportunity to optimize in dimensions not already constrained by the application description. I further note that automated search could occur not only within one style but across styles---perhaps leading to tools that can not only synthesize in user-selected styles, but that can actually select appropriate styles for given applications.

Second, to date this work remains rather academic in that the target architectural styles are quite simple compared to those used in actual practice. 
An interesting avenue for future work thus would be to investigate the following issues among others:
synthesis of multi-style architectures; targeting of styles induced by middleware platforms and production frameworks~\cite{nitto_exploiting_1999}; application models with sub-models of heterogeneous types; 
and replacing Alloy with more scalable synthesis technology.

Finally, a shortcoming of current approaches to \emph{code generation }from architectural models is in the lack of flexibility with respect to architectural style~\cite{sam_malek_effective_2008}. The architect is forced to develop models in the particular architectural style supported by a given approach, rather than a suitable style chosen by the architect. In the next chapter, I present my work in this direction to include subsequent mappings from architectural descriptions to code level implementations. 


\newpage
\thispagestyle{empty}
\mbox{}

\chapter{Synthesizing Code Frameworks from Application Architectures}

In this chapter, I present Pol.
This is the second experiment the I have conducted in the context of this dissertation.
This work provides evidence in support of reducing the costs of model-driven development by demonstrating a specification-driven synthesis of platform-specific code frameworks without the need for traditional hand coding of translators.
Specifically, Pol shows that a combination of formal application architecture and platform models, implementation mappings, and platform-usage code fragments suffice.
In support of the claim of broadening the applicability of MDD in various stages of the software development lifecycle, this work 
targets synthesis of object-oriented application frameworks that use a range of widely-used industrial software platforms.
Finally, this work incorporates benefits of formal specifications of architectural styles and platform models to reliably synthesize platform-specific implementation models, from which code framework is then directly generated in a straightforward manner.
This chapter is organized as follows.
Section~\ref{problem_2} presents motivation and research problems for this aspect of my research.
Section~\ref{approach_2} details the specification-driven synthesis approach, and describes how it fits into an evolutionary development process.
Section~\ref{cyberhealth} introduces the system that I took as a subject for the case study.
Section~\ref{example_2} explains the approach in more detail with concrete examples drawn from this system.
Sections~\ref{experience_2} and \ref{results_2} report and interpret the data I measured during experimental evaluation of the approach.
Section~\ref{discussion_2} presents an overall evaluation of the ideas, experimental approach, and results, addressing some possible objections to this work.
Finally, Section~\ref{conclusion_2} concludes this chapter.


\section{Motivation and Research Problem}
\label{problem_2}
Software developers have to deal with many concerns when designing a system.
Examples include architectural style, method for data persistence, component interaction protocols, and authentication.
In each such dimension of concern there is generally a range of possible choices.
For example, data persistence might be handled using an SQL or a NoSQL database;
interactive web services might be handled using RESTful~\cite{roy_t._fielding_principled_2000} or WSDL/SOAP-based protocols~\cite{pautasso_restful_2008};
authentication might be handled using OAuth~\cite{OAuth} or basic HTTP authentication.
These examples point to a central point in this chapter: many such decisions are in turn supported by complex platforms. In this chapter I use this term to refer to middleware and frameworks.
For example, the Restlet~\cite{restlet} framework supports implementing RESTful web applications,
the HornetQ~\cite{hornetq} middleware supports reliable distributed messaging, and CouchDB~\cite{couchdb} is a document-oriented database supporting NoSQL for data persistence.
Software-intensive systems often rest on a complex set of platforms.

Such platforms provide developers with the capability of reusing not only code, but whole platform designs. However, developing applications on such platforms is costly, time-consuming and error prone~\cite{Fairbanks_DesignFragments_2006,garlan_evolution_styles_2009,Hou_SCL_2006}. Their complexity and poor documentations often make it hard to learn and use such platforms~\cite{Hou_SCL_2006}.
One approach to addressing these problems is the use of model-driven development methods to synthesize code for given platforms. Such approaches can ease development by expressing key features in models and refining them into code \cite{Santos_JSS_2010}. MDD holds out the promise of improved software productivity, timeliness, quality and costs.


The problem is that developing the code generators that MDD approaches require is difficult, costly and error-prone~\cite{Karsai_et_al_2003,white_improving_2009}.
Karsai et al. recognized that developing such code generation model interpreters is a very difficult problem~\cite{Karsai_et_al_2003}.
It is hard to develop and evolve modeling languages and code generation model interpreters for diverse, evolving applications~\cite{white_improving_2009}.

\section{Synthesis Approach}
\label{approach_2}
\begin{figure*}[tbh]
\centering
\includegraphics[width=6.0in]{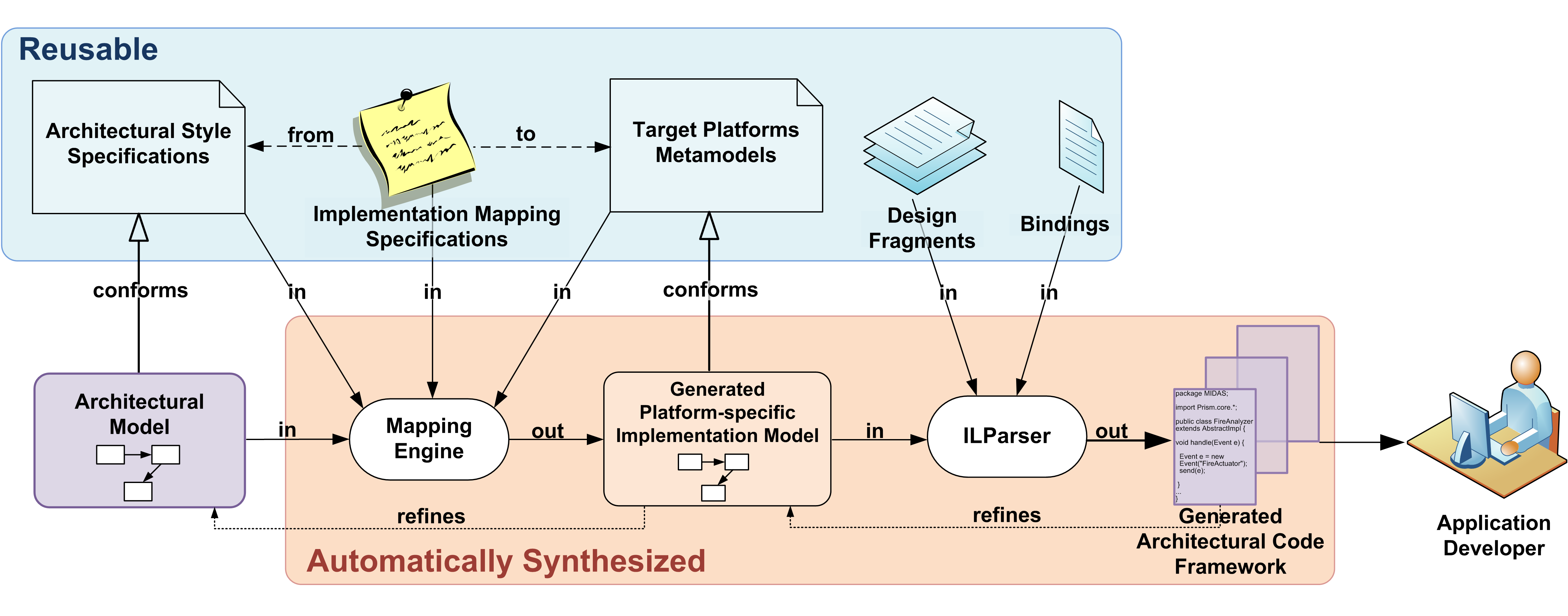}
\caption{\label{fig:Framework}High-level overview of the Pol framework}
\end{figure*}

In this section I discuss the specification-driven synthesis approach and briefly how it fits into an evolutionary development process. Figure~\ref{fig:Framework} illustrates the overall data flow structure of the approach and its two main transformational elements: a \emph{mapping engine} and an intermediate language (IL) parser, \emph{ILParser.}  This section presents these constructs in more detail.

I have developed a prototype tool implementing the approach: {\em Pol}, a Persian word for bridge. I chose to use Alloy~\cite{daniel_jackson_alloy:lightweight_2002} as a specification language and constraint solver. Alloy is helpful for formalizing modeling languages \cite{kelsen_lightweight_2008}. Its simple set theoretic language was sufficiently expressive for prototyping and evaluation purposes.

\subsection{Mapping Engine}

The function of the {\em mapping engine} is to transform an  {\em application architecture} in a given {\em architecture style} to a target-platform-specific,  {\em intermediate implementation model (IIM)}, from which the ILParser generates code. The engine uses a general constraint solver to synthesize the IIM, with no hand written, platform- or application-specific code. Rather, it works from specifications of (1) architectural style, (2) application architecture in that style, (3) target platforms, and (4) mapping predicates that define relationships between architectural style and target platform constructs.

An architectural style is formalized in the manner of Kim and Garlan~\cite{jung_soo_kim_analyzing_Journal}. I define application-specific architectural styles as extensions of more general component and connector styles. An application architecture is then given as an Alloy instance of such a style.  A {\em platform model (PM)} is a partial formal model of a target platform, tailored to the ways in which the platform is to be used in a given application. A PM models platform classes, methods, and constraints as Alloy signatures, relations, and invariants.  An implementation mapping (IM) is a set of Alloy predicates that explain how architectural style elements map to intermediate implementation (PM) constructs. IMs address the problem of expressing architecture-to-implementation mappings noted by Shaw~\cite{shaw_abstractions_1995}. All of these specifications are partial, lightweight models. They do not attempt to capture the complete structure or semantics of a platform or application, but are designed to enable a desired level of partial code synthesis. Given these inputs, Pol computes an IIM as a solution to the set of constraints. From an IIM, the ILParser then generates code in a manner that is also driven by specifications.

\subsection{ILParser}

The function of the ILParser is to transform an intermediate implementation model (IIM) into an object-oriented \emph{architectural code framework}. Such a framework comprises a set of classes that refine an architectural model into platform-specific code that developers extend by subclassing. The ILParser is parameterized by (1) platform-specific code usage patterns documented as \emph{design fragments}~\cite{Fairbanks_DesignFragments_2006}, and (2) \emph{binding} elements that relate fragments to elements of the IIM. No custom coding (beyond provision of code-level design fragments) is required. The ILParser implementation is independent of architectural style and platform models.

The ILParser generates two sets of framework classes from a given IM: platform-independent, and platform-specific. Focusing first on the {\em platform-independent} part, for each IIM signature that extends the \emph{Component} signature of the underlying component-and-connector style, the ILParser produces an interface declaration. Port entities in an architectural model expose the interfaces of components, and for each port in the IIM, the ILParser generates an abstract method declaration. Certain other abstract classes are synthesized to represent other general IIM constructs.


Second, the ILParser maps {\em platform-specific} implementation model elements to code based in part on our variant of the \emph{design fragments} method~\cite{Fairbanks_DesignFragments_2006} that Fairbanks et al. developed to capture platform usage patterns. A design fragment specifies how an application uses a platform to achieve some goal. Fairbanks showed that each platform exposes a limited number of fragments. Three principal pieces of a fragment are (1) a \emph{framework-provided} element that describes a relevant construct of the platform that an application needs to interact with, (2) a \emph{programmer-required} element that specifies both dependencies between design fragments and the constructs that application developers must provide to realize the design fragment under consideration, and (3) a~\emph{code snippet} element that defines a pattern of platform use in the form of a parameterized code template. Template parameters, identified by enclosure within \$ symbols, bind IIM constructs into these code snippets.

Fairbanks et al. proposed to use Java annotations to bind design fragments into code. We bind fragments to code using a separate \emph{bindings} file. Each binding element in a ``\emph{bindings}'' file welds \emph{template parameters} of a design fragment to the constructs in the implementation model (IIM). I specify binding elements declaratively. The binding process is supported by the Alloy query analyzer, with elements of a IIM accessed through query definitions. Constructs of a platform instance are declared in terms of intermediate language variables (\emph{IL-Variables}). Each \emph{IL-Variable} is defined as a query over the IM, and the values of these variables are dynamically realized by \emph{ILParser} while parsing the implementation model. They are then assigned to the \emph{template parameters} to generate code.

\section{Case Study Subject}
\label{cyberhealth}

In this section, I introduce the system that I took as a subject for the case study: {\em CyberHealth.}  In the next section I explain the approach in more detail with concrete examples drawn from this system.

CyberHealth is part of an ongoing effort to develop requirements and architectures for national-scale health information systems, akin to those suggested by the U.S. Department of Health and Human Services~\cite{nhin-exchange} and the President's Council of Advisors on Science and Technology (PCAST)~\cite{john_p._holdren_realizing_2010}. CyberHealth is an evolving, laboratory-scale, operational model of a national-scale health information ecosystem. In our model of a {\em nation-wide health information network} connecting diverse producers, stakeholders, and consumers of health data. Its development has involved the progressive introduction of widely used platforms, making it a reasonable subject for an initial evaluation of our approach.

The system models corporate and individual entities in the healthcare system, and healthcare data production, transmission, and use by and among entities. Inter-corporate communications involving clinical data are mediated by a centralized mechanism, which I refer to as our {\em exchange.} In a running simulation, each corporate entity is observable through a web site. A special web site is run by which the user of the model can control it, e.g., to simulate patient arrivals at hospitals, the production and flow of clinical data, and the distribution and use of such data. Web page views are kept up to date dynamically using standard server push technology (currently Comet, soon the WebSocket protocol).

A key health system requirement reflected in this model is {\em principals}---entities {\em including patients} that have rights to access and use data---should be empowered to establish automated data-flow rules that configure the exchange and its connected entities for transmission of such data. For example, a patient being seen at a hospital and several clinics should be able to exercise her legal rights, provided by HIPAA regulations, to access, cause transmission of, and authorize the use of her own data: by configuring the exchange to create data subscriptions that cause relevant past and future record updates to flow across channels and on to destinations she designates.

Our health system vision thus includes what we call a {\em principal control system (PCS)}: a web portal by which principals (patients, agencies, etc) manage data and control flows consistent with their rights. Each principal has an account in the PCS, and can use it to register with clinical and non-clinical service providers and to establish connections among them. Connection sources support both query/response and publish/subscribe mechanisms. A patient can register a subscription with a hospital, connect it to a specified channel causing updates to her records to be sent to that channel, and then forward them to her account in a personal health record or research study in which she is enrolled.

Each corporate entity and the exchange itself runs as a RESTful web service~\cite{roy_t._fielding_principled_2000} with a web browser view.
(Representational State Transfer, or REST, is an architectural style for systems on the internet~\cite{roy_t._fielding_principled_2000}; and systems in this style are called \emph{RESTful}.) Entities include hospitals, non-clinical institutions with authorization to use clinical data, and citizen-patients themselves. Patients are first-class entities in this concept of a future health information ecosystem. Other entities include personal health record data banks, research institutions, public health agencies, and data search services of the kind envisioned by the PCAST. The actual data processed by our system is synthetic and not yet reflective of real clinical data.

\section{Example }
\label{example_2}

This section now illustrates the approach with snippets of the artifacts from each stage of architectural code synthesis for the CyberHealth system. I discuss the following data elements: (1) architectural style specification; (2) application architecture model; (3) platform model; (4) implementation mapping; (5) intermediate implementation model; (6) design fragments; (7) bindings; (8) and synthesized architectural code framework.

\subsection{Architectural Style}
CyberHealth is designed in an application-specific architectural style that combines and extends several other styles, including REST and implicit-invocation. I used existing Alloy formalizations of such styles~\cite{stephen_wong_scalable_2008} when available. Not finding one for REST, I crafted one~\cite{bagheri_formal_2011}. Listing~\ref{ArchModel} presents an excerpt of the Alloy definition of the CyberHealth application architecture, which starts by importing the style we defined: \emph{CH\_style.} This style provides the language in terms of which the application architecture model is developed.

\subsection{Application Architecture}
Continuing with Listing~\ref{ArchModel} the Hospital1 signature, lines 5-7, extend the \emph{Publisher} component definition to model a Hospital. Each component has ports describing interaction points with its environment. Here, \emph{Hospital1Encounter} is defined as a port of Hospital1. Each port has a set of processes. A Process represents a port activity. \emph{Hospital1Encounter} extends the \emph{PublishEvent} port and names \emph{AnnounceEncounter} as its process (lines 8 to 11). The \emph{AnnounceEncounter} is further specified as being of the \emph{Announce} Process type with \emph{Encounter} and \emph{Topic} as corresponding data elements.

\begin{figure}
\lstset{ %
basicstyle=\scriptsize,       
numbers=left,                   
numberstyle=\tiny,      
stepnumber=1,                   
numbersep=7pt,
backgroundcolor=\color{white},  
showspaces=false,               
showstringspaces=false,         
showtabs=false,                 
frame=bottomline, 
tabsize=2,                    
captionpos=b,                   
breaklines=true,                
breakatwhitespace=false,        
numberbychapter=false,
xleftmargin=1.5em,
morekeywords={module,sig,abstract,extends,one,some,set,open,pred,all,in}
}
\lstinputlisting[caption=CyberHealth application architecture (elided).,numberbychapter=false,label=ArchModel]
{Arch_model_Alloy_short1.als}
\vspace{-0.6cm}
\end{figure}


\subsection{Platform Model}

The goal of approach is to produce platform-specific code without custom synthesizers. Rather, it requires partial formal platform models. I developed models for each key CyberHealth platform: \emph{HornetQ}~\cite{hornetq}, \emph{Cometd}~\cite{cometD}, \emph{OAuth}~\cite{OAuth}, and \emph{Restlet}~\cite{restlet}. HornetQ is a Java-based message-oriented middleware system, popular for scalability and performance. Cometd is a HTTP-based event routing framework for server push messaging. OAuth 2.0 enables clients to authorize third party applications to access protected web resources without sharing of passwords.

Consider the Restlet framework for RESTful web services. There are several popular platforms that support the REST style. Among them is the Java {\em Restlet} platform.  Listing~\ref{Restlet-Alloy} presents part of the Restlet platform model. The \emph{Restlet} Alloy signature models the core \emph{Restlet} class of the platform, which exposes the uniform REST interface. \emph{Resource} represents the target of a hypertext reference, including HTTP methods such as GET, POST, PUT and DELETE. A \emph{Handler} provides thread-safe processing of calls. \emph{Resource} extends \emph{Handler.} A \emph{Connector} enables communication between components, encapsulating the activities of accessing resources and transferring resource representations. Clients and Servers are \emph{Connector}s. \emph{VirtualHost} is a router of calls from Server connectors to Restlet instances, such as Applications. \emph{Component} is a \emph{Restlet} that manages a set of \emph{Connector}s, \emph{VirtualHost}s and \emph{Application}s. An {Application} is in charge of coordinating deployment of functionally connected Restlet instances and is directly attached to \emph{VirtualHost}.

\begin{figure}
\lstset{ %
basicstyle=\scriptsize, 
numbers=left,                   
numberstyle=\tiny ,      
stepnumber=1,                   
numbersep=7pt,
backgroundcolor=\color{white},  
showspaces=false,               
showstringspaces=false,         
showtabs=false,                 
frame=bottomline, 
tabsize=2,                    
captionpos=b,                   
breaklines=true,                
breakatwhitespace=false,        
numberbychapter=false,
xleftmargin=1.5em,
morekeywords={module,sig,abstract,extends,one,some,set,open,pred,all,and,in}
}
\begin{lstlisting}[caption=Restlet platform model in Alloy.,numberbychapter=false,label=Restlet-Alloy]
module Restlet

abstract sig Restlet {...}
abstract sig Handler extends Restlet{}
sig Resource extends Handler{
	reps: set Representation
}
abstract sig Connector extends Restlet{}
sig Server extends Connector{}
sig Router extends Restlet{
	attachedResources: some Restlet 
}{
	attachedResources in Application + Resource
}
sig VirtualHost extends Router {}
sig Application extends Restlet{
	router: one Router,
	reources: some Resource
}{
	router.attachedResources in reources
}
\end{lstlisting}
\vspace{-0.6cm}
\end{figure}


\subsection{Implementation Mapping}

An implementation mapping associates architectural style constructs to platform model constructs and is the critical element in the synthesis of intermediate implementation models from application architectures.  Listing~\ref{implMapping} shows an extract of the implementation mapping for refining architectural models in the CyberHealth style to the target platforms. This mapping predicate states that for each architectural \emph{EventBus} connector (defined by the imported implicit-invocation architectural style) there is a corresponding HornetQ platform {\em Topic}.  The next statement says that for each port process of type \emph{Announce}, there is an instance of the {\em TopicPublisher} class handling the port process, such that its destination, {\em Topic}, is in charge of the connector to which the given port is connected. (The dot operator denotes relational join, and $\sim,$ relational inverse.) Next, the specification states that for each \emph{Port} of a component and each \emph{ServerPushConnector}, there is an instance of a Restlet Resource and a \emph{CometdService} handling them. In this way the mapping specification maps architectural elements to abstract platform-specific elements (not yet code).

\begin{figure}
\lstset{ %
basicstyle=\scriptsize,       
numbers=none,                   
numberstyle=\tiny,      
stepnumber=1,                   
numbersep=7pt,
backgroundcolor=\color{white},  
showspaces=false,               
showstringspaces=false,         
showtabs=false,                 
frame=bottomline, 
tabsize=2,                    
captionpos=b, 
breaklines=true,                
breakatwhitespace=false,        
numberbychapter=false,
xleftmargin=0em,
morekeywords={module,sig,abstract,extends,one,set,open,pred,all,and,in}
}
\lstinputlisting[caption=Implementation mapping predicate.,numberbychapter=false,label=implMapping] 
{implementation_mapping_short2.als}
\vspace{-0.6cm}
\end{figure}


\subsection{Intermediate Implementation Model}

\begin{figure}[tbh]
\centering
\includegraphics[width=6in]{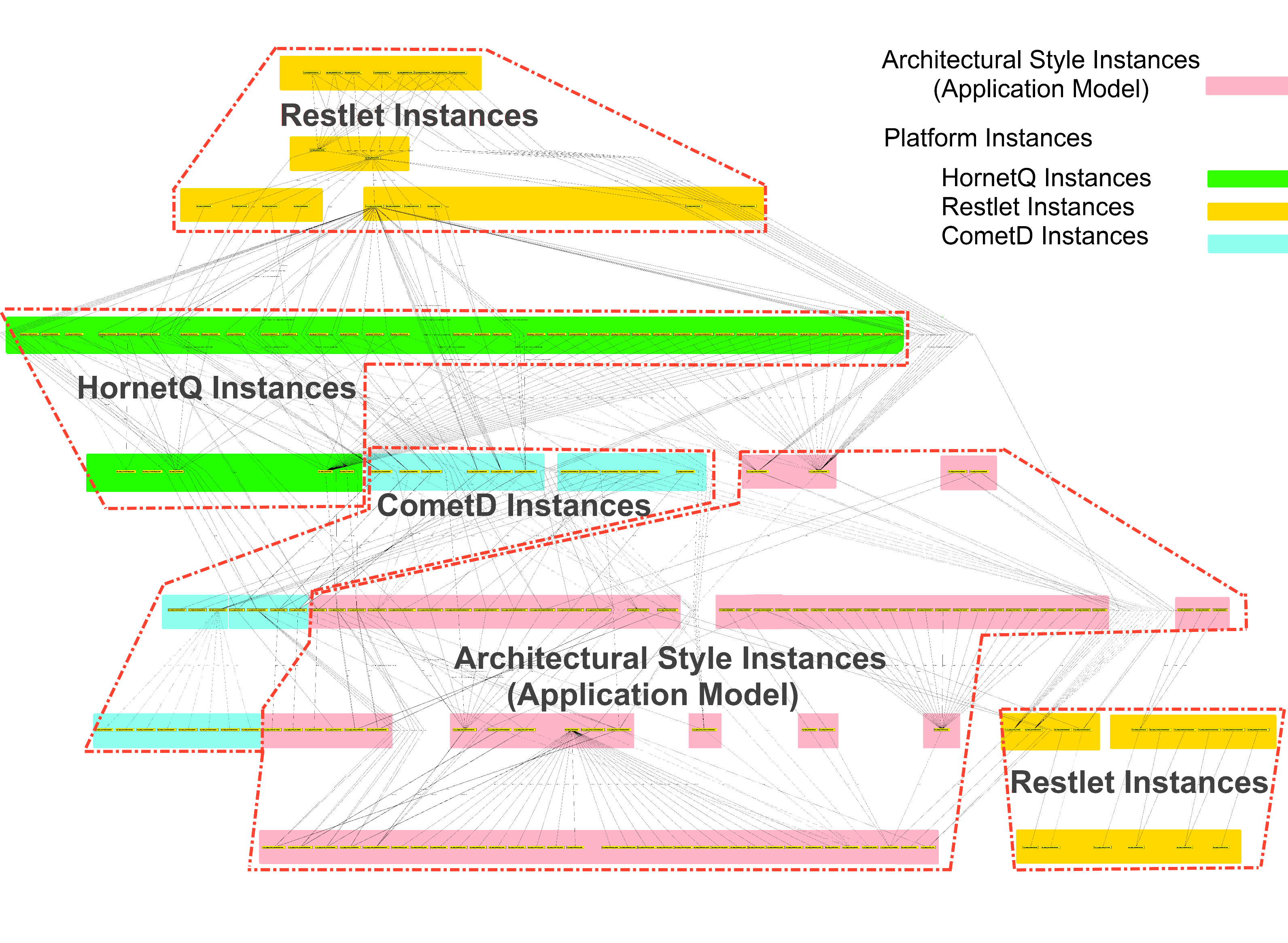}
\caption{\label{fig:impl_model}
Synthesized, platform-specific implementation model for the CyberHealth architecture.
}
\end{figure}

Figure~\ref{fig:impl_model} shows the result of applying the implementation mapping to the CyberHealth architecture. The model contains hundreds of elements including architectural and platforms constructs and the relationships among them. I have highlighted architectural style instances and the corresponding constructs of each platform. Having such a detailed, implementation-level architectural model serves as a basis for defining a clear programming model and promotes traceability from implementation artifacts to architecture.

\subsection{Design Fragment}

\begin{figure}[tbh]
\centering
\includegraphics[width=4.25in]{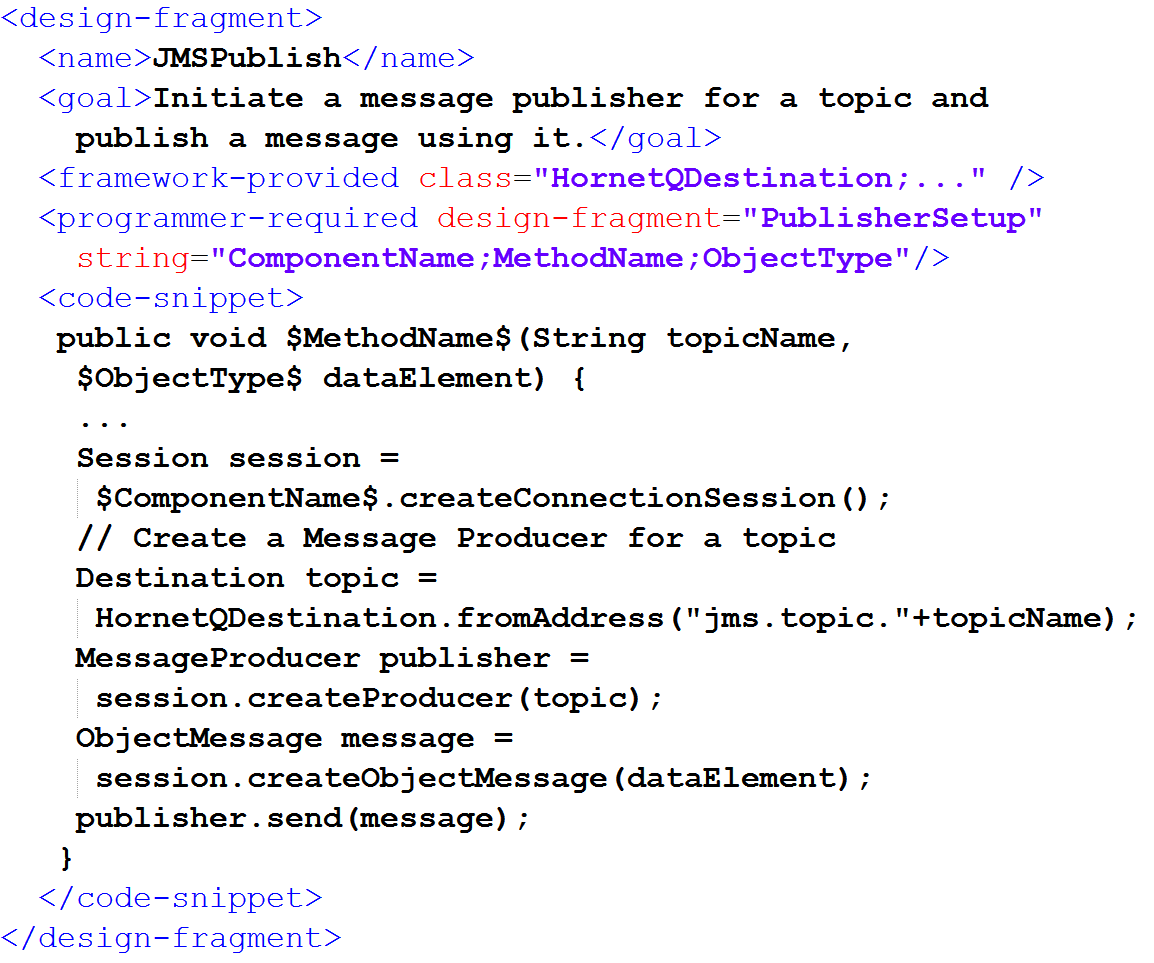}
\caption{\label{fig:DesignFragment}A design fragment.}
\end{figure}

With an intermediate implementation model synthesized, the ILParser now operates to produce code. Code is based on the matching of design fragments to elements of the implementation model. Figure~\ref{fig:DesignFragment} partially represents the \emph{JMSPublish} design fragment for the HornetQ platform. The intent of this design fragment is to initiate a message publisher for a topic and to publish a message using it.

\subsection{Bindings}
\begin{figure}[tbh]
\centering
\includegraphics[width=4.5in]{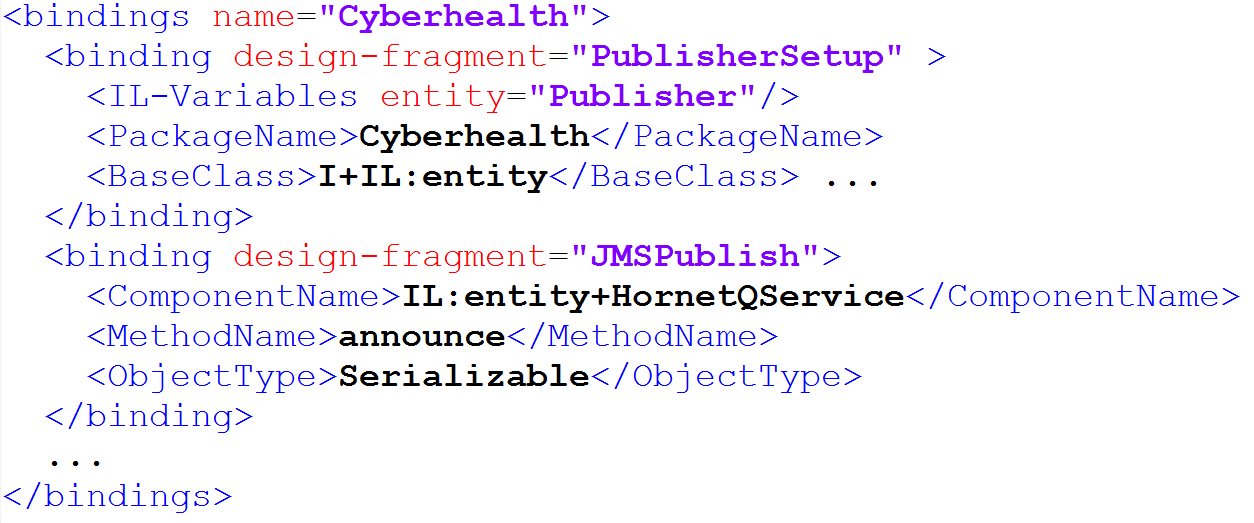}
\caption{\label{fig:binding}
A binding specification.
}
\end{figure}

Bindings state how design fragments are bound to and parameterized by intermediate implementation model elements. Figure~\ref{fig:binding} presents a snippet of a bindings file. In the first binding, declared for the \emph{PublisherSetup} design fragment, the \emph{entity} variable is specified as a query over the implementation model. It represents application elements extending the Publisher signature. Other binding definitions then can refer to this variable using the \emph{IL:} prefix. Given the bindings declaration and the synthesized implementation model as inputs, the \emph{ILParser} synthesizes the architectural code framework for the target platforms.

\subsection{Architectural Code Framework}

The final result of synthesis is a set of platform-independent and platform-dependent interfaces and classes constituting an object-oriented framework. Listing~\ref{AbstractHospital} presents an example of a synthesized interface for a Hospital component  in which \emph{AnnounceEncounter} and \emph{RegisterForAnnouncement} are defined as port processes. Listing~\ref{Databank} shows an extract of a generated class providing a platform-specific implementation based on a mapping definition. I assign each port process of type \emph{announce} to a \emph{TopicPublisher} element from the HornetQ platform (cf. Listing~\ref{implMapping}). That element is then realized using the \emph{JMSPublish} design fragment.

\begin{figure}
\lstset{ %
basicstyle=\scriptsize, 
numbers=left, numberstyle=\tiny, stepnumber=1, numbersep=5pt, backgroundcolor=\color{white}, showspaces=false, showstringspaces=false, showtabs=false,
frame=bottomline, 
tabsize=2, captionpos=b,                   
breaklines=true,                
breakatwhitespace=false,        
numberbychapter=false,
language=Java,
xleftmargin=1.4em,
}
\begin{lstlisting}[caption=Synthesized platform-independent interface.,numberbychapter=false,label=AbstractHospital]
public interface IHospital1 {
  public abstract void AnnounceEncounter(Encounter e, Topic t);
  public abstract void RegisterForAnnouncement(Topic t);
  ...
}
\end{lstlisting}
\vspace{-0.6cm}
\end{figure}

\begin{figure}
\lstset{ %
basicstyle=\scriptsize,
numbers=none,
numberstyle=\tiny, stepnumber=1, numbersep=5pt, backgroundcolor=\color{white}, showspaces=false, showstringspaces=false, showtabs=false,
frame=bottomline, 
tabsize=1, captionpos=b,                   
breaklines=true,                
breakatwhitespace=false,        
numberbychapter=false,
language=Java,
xleftmargin=0em,
}
\lstinputlisting[caption=Synthesized platform-specific method.,numberbychapter=false,label=Databank] 
{generatedMethod-2.java}
\vspace{-0.6cm}
\end{figure}

\subsection{Hand-Written Extensions}
Referring to Listing~\ref{Hospital1}, I find code implementing the \emph{IHospital1} interface. Based on the formal specification of the Hospital1 component in the application architecture, the interface declares \emph{AnnounceEncounter} as an abstract method. The concrete implementation of this method relies on the platform-specific framework: On line 7, the developer calls the announce method provided by the synthesized framework. The synthesized framework helps the developer to implement an application that uses complex platforms by synthesizing required APIs and calls automatically.

\begin{figure}
\lstset{ %
basicstyle=\scriptsize, numbers=left, numberstyle=\tiny,
stepnumber=1, numbersep=5pt, backgroundcolor=\color{white}, showspaces=false, showstringspaces=false, showtabs=false,
frame=bottomline, 
tabsize=1, captionpos=b,                   
breaklines=true,                
breakatwhitespace=false,        
numberbychapter=false,
language=Java,
xleftmargin=0em,
}
\begin{lstlisting}[caption=Extract of a manually developed class.,numberbychapter=false,label=Hospital1]
public class Hospital1 implements IHospital1 {

  @Override
  public void AnnounceEncounter(Encounter e, Topic t) {
    EncounterImpl encounter = (EncounterImpl) e;
    Hospital1HornetQService service = new Hospital1HornetQService();
    service.announce(t.topicName, encounter);
    ...
  }
  ...	// defines other methods
}
	
\end{lstlisting}
\vspace{-0.6cm}
\end{figure}

\section{Evaluation Methodology}
\label{experience_2}
In this section, I briefly report and interpret the data I measured during experimental evaluation of the approach.
I have conducted evaluation of the approach using the case study method of Kitchenham, Pickard and Pfleeger~\cite{kitchenham_case_1995}. The steps in this method relevant to this work are: (1) defining the hypotheses to be tested; (2) selecting a pilot project; (3) planning (and executing) the case study; (4) analyzing and reporting the results. The following subsections address each of these elements.

\subsection{Hypotheses}
I frame three hypotheses for this approach. First, it is feasible to synthesize architectural code frameworks without custom code generators for software-intensive systems based on platforms in wide use today. Second, the choice of application-specific object-oriented frameworks as synthesis outputs helps limit the impacts on hand-crafted code of evolutionary changes in architectural and platform models. Third, the performance of the synthesis approach using constraint-solving techniques is adequate for modest applications.

\subsection{Project Selection}
To evaluate the aforementioned hypotheses, I have run the approach implementation on several case studies, including two examples inspired from the literature, namely MIDAS~\cite{george_edwards_model_2008} and Lunar Lander~\cite{taylor_software_2009}.
More usefully, I applied the proposed technology against the needs of a representative web-based, distributed software system. To this end, I have adopted the CyberHealth system as a challenging problem for the proposed technology. This system is meant as an operational model for a possible future national cyber-infrastructure for healthcare data liquidity and service integration. As discussed in section~\ref{example_2}, its development has involved the progressive introduction of widely used platforms, making it a reasonable subject for evaluation of the proposed approach.

\subsection{Case Study Planning and Execution}
The experimental procedure was to migrate the initial version of CyberHealth to a substantially synthesized version, then to evolve the result through additional architectural enhancements, while measuring key parameters.  To test the first hypothesis---technical feasibility---I attempted to develop a working tool. I also assessed the range and industrial importance of the platforms used in development of the system, the size of platforms and architectural specifications, and the difficulty and cost to develop these models. To test the second hypothesis---using object-oriented frameworks controlled the impacts of changes in synthesized code---for each of three significant changes to the system, I measured lines of code changed in the architectural model, lines of code changed in mapping specifications, lines of code changed in synthesized architectural code framework, and lines of code that had to be changed by hand. To test the third hypothesis---adequacy of synthesizer performance---I measured the computational time required for deriving architectural code frameworks.

The execution of this case study involved development of specifications for several platforms, as well as application-specific architectural models. I studied how platforms were used in the hand-crafted version and extracted usage patterns as design fragments. I developed an initial version of the architectural model for the system by reverse engineering of the hand-crafted version, abstracting key entities and connections. I evolved the architectural model through several stages as we sought to satisfy increasingly demanding requirement, such as the added support for multi-party authentication. After each change to the  architectural model I used Pol to synthesize a corresponding code framework, and I adapted the hand-crafted application code as required by changes to the underlying code framework.


\section{Results and Interpretation}
\label{results_2}

Executing the case study produced measurement data that I report and interpret in this section. I address each hypothesis in turn.

\paragraph*{Synthesis without Custom Generators}
This work demonstrates the technical feasibility of specification-driven architectural code synthesis without the need for custom code generators tailored to particular architectural styles, architectures, platforms, or framework usage patterns. A combination of formal application architecture and platform models, implementation mappings, and design fragments suffice. We interpret our evidence as providing fairly strong support for our hypothesis of technical feasibility.

The architectural model for the application under consideration, ended up at about 300 lines of Alloy code and 50 signatures. The Restlet, OAuth, CometD and HornetQ models are about 90, 80, 50, and 80 lines, respectively. The specifications defining our architectural style are about 400 lines of Alloy. The CyberHealth implementation is over 12 thousand lines of code with more than 34\% synthesized. This code implements key crosscutting decisions in a consistent, trustworthy manner, modulo possible errors in hand-crafted code fragments. I also measured the execution coverage of synthesized code. More than 80\% of synthesized code is executed during routine runs. Speaking subjectively, I found the ability to evolve the architecture and to re-synthesize a complex and substantial part of the code base to be useful. Regarding the effort required to support a new platform, it typically took a few days to develop a new model and mappings. The challenge was not so much in writing Alloy as in understanding the platform and how it will be used.

I interpret these data as suggesting that synthesis of architectural code skeletons within a hybrid automated-manual, evolutionary development process is an approach worth exploring further. Consolidating the prototype technology and assessing its use in field studies would be a natural next step for this research.

\paragraph*{Manageability of System Evolution}
Architectural change in complex software is common if not inevitable~\cite{garlan_evolution_styles_2009}. One of the goals was to limit the amount of hand-written code modifications required when an architectural code framework is resynthesized. The approach was to have synthesized code present an understandable interface to hand-written code in the form of an object-oriented framework. To assess how well this works, I conducted three small experiments. Figure~\ref{fig:experiments} summarizes the results. Columns represent lines of code changed in the architectural specification, mapping specifications, the synthesized code framework, and hand-written code, respectively. The set of implementation mapping specifications and binding definitions are collectively referred to as Mapping.

\emph{Experiment 1: Adding Support for Server-push technology.}
In an early version of CyberHealth, components pulled resource representations from servers using a pure REST style. Now push-based web messaging technologies are important. I changed the architectural style and model to use such technology for eager updating of client views. Components now publish events when their states change. Subscribers express interests in classes of events, and receive notifications for matching events. To extend the synthesis technique to such functions, I modeled the server-push web messaging middleware as a connector type ~\cite{nitto_exploiting_1999,nenad_medvidovic_role_2003} and added the \emph{ServerPushConnector} signature as an extension of the \emph{Connector} signature in the \emph{CH\_style} module.

Next I developed a model of the CometD platform as a widely used implementation of this technology, and refined the implementation mapping and the bindings file by adding a set of new constraints to specify the relationships required to hold between elements of the architectural model and elements of the CometD framework. These changed produced 450 new lines of synthesized code, declaring and initiating CometD channels, a \emph{CometdApplication} resource that manages those channels, and a servlet that initiates the \emph{CometdApplication} resource. I had to modify about 100 lines of hand written code to push specific messages to CometD service channels. The system continued to support pull-based communication through REST APIs, so no code had to change in that area of the system.

\begin{figure}[bt]
\centering
\begin{tabular}{|l||c|c|c|c|}
  \hline
  & {\bf Arch.}& {\bf Mapping}
  & {\bf Syn.} & {\bf Dev.}\\
  \hline
  {\bf Exp. 1} & 40& 52 & 450& 96\\
  \hline
  {\bf Exp. 2} & 12& 16& 108& 0\\
  \hline
  {\bf Exp. 3} & 84& 49& 704& 570\\
  \hline
\end{tabular}
\caption{\label{fig:experiments}Experiments statistics: each cell corresponds to the LOC modifications for the element and experiment given on the axes.}
\end{figure}

\emph{Experiment 2: Adding Support for Logging.}
Next I modified the CyberHealth system so that hospital components log requests received by their ports. I added a \emph{Log} field to the \emph{Process} signature, to specify the logging level for a particular process, and a corresponding binding definition to the \emph{bindings} declaration file. The regenerated code supports logging with no changes in hand-written code. Here connections to aspect-oriented programming become evident.

\emph{Experiment 3: Modifying Dataflow Protocol.}
An early version of the system supported a channel per patient, to which medical record updates for a patient would be pushed. When a patient visited a hospital, the hospital would publish an encounter record on that patient's channel. Entities such as the personal health record service would receive notifications of these updates. Later we decided to change the system so that channels are created by user request through the principal control system (PCS). Components publish/receive to/from a given channel only if a principal establishes the appropriate connections through the PCS.

Starting from the architectural model, I added a protected port to the component description of providers, such as hospitals, through which the PCS requests the provider to publish record updates to a channel. This is a protected port because as unauthorized clients do not have access to it. I also added a protected port to the component description of data receivers through which the PCS requests the data receiver component to subscribe to a channel. In the mappings, I added new rules and bindings for mapping the architecture to the constructs provided by the OAuth platform. Resynthesis of the architectural code framework generated about 700 LOC scattered in several framework classes, providing support for the protected dataflow protocol based on the OAuth framework. The developer needed to develop details of the application logic, which in this case required about 570 lines of hand-written code.

Overall, the experiments and our experience have provided support for the hypothesis that the use of an object-oriented framework approach to separating synthesized and hand-written code was consistent with a manageable change process. Architectural changes do {\em not} result in any overwriting of the hand-written code for a system.  This approach is different from approaches that generate code skeletons that are to be filled by developers, in many of which, regeneration overwrites hand-crafted code.  The technique certainly does not completely insulate hand-crafted code from changes in the architecture and platforms. In general, developers will have to adapt their code to the new framework code. Of course, they would have to do this if new underlying code were produced by hand, too. When the architectural code framework is regenerated, the compiler was often helpful in revealing mismatches between the hand-crafted code and regenerated code. A good topic for future work would be to tell developers more precisely which code might have been invalidated.

\paragraph*{Performance}
To measure performance, I recorded the computational time required to synthesize architectural code frameworks within an Alloy scope large enough to admit non-empty solutions to given constraints. The scope states the maximum number of architectural elements or framework constructs of each type in the case of architectural models or framework instances. I used a PC with an Intel Core i5 2.67 Ghz processor and 4 GB of main memory, with SAT4J as the SAT solver. I conducted measurements over an increasing scope, from at most 10 instances for each type to at most 30. The slowest synthesis time (for a bound of 30 elements) was about 100 seconds, confirming that the proposed technology based on a bounded model checker is reasonable at the scale of our experiments. Moreover, given the apparent trend of advancements in SAT solvers we were witnessing in recent years, similar improvements in the performance of formal application synthesis seem to be expectable.


\label{evaluation}
\section{Discussion}
\label{discussion_2}

In this section, I present an overall evaluation of the ideas, experimental approach, and results, addressing some possible objections to this work and conclusions.

\emph{Partial Synthesis.}
This is not an attempt to raise the level of abstraction entirely to the model level. Rather, it is a pragmatic attempt to automate a substantial part of code production, balanced against the need for simple modeling languages, targeted at capturing essential application structures, and to avoid the burdens of having to hand-craft synthesizers customized to particular source and target environments. The more detailed the input models, the more detailed a mapping specification a user can develop, and the more complete the code that can be generated. Architectural models generally exclude application details that are not important at the architectural level~\cite{malek_architecture-driven_2010}. A key supposition in this work is that one can  generate significant parts of an application in this style, and that, all things being considered, it can be valuable to do so.

\emph{Tractable specification languages vs. General-purpose modeling languages.}
This work suggests that tractable but lightweight formal languages might have a useful role to play in program synthesis. Limited but clean formalisms can perhaps replace broad-spectrum but complex and semantically messy notations, with general-purpose analyzers replacing hand-crafted transformers, at the cost of incomplete synthesis. I trade away completeness and richness of synthesis for simplicity of modeling languages and tools, avoiding the need for custom synthesizers.
Moreover, this approach involves more than synthesis. It also focuses on ways that {\em partial} synthesis can be embedded in an architecturally-focused, evolutionary development process. In other work~\cite{bagheri_spacemaker,h._bagheri_architectural_2010}, I have also addressed synthesis of {\em spaces} in the presence of under-constrained mappings and automated trade space tradeoff analysis. Merging these lines of work would be a valuable avenue for future work.

\emph{Benefits of using architectural styles.} This work started with the use of architectural styles to support code synthesis. The main benefit we have seen so far is that we can be assured that the intermediate architectures that we generate are absolutely consistent with the rules of given styles. Synthesized code is too, modulo possible errors in our design (code) fragments. In the future, I imagine to prove additional properties about architectural styles, so that architects are relieved of the burden of having to prove these properties for each implemented program. I believe that the use of fully formal and analyzable modeling notations creates valuable opportunities in this dimension. Earlier work by other researchers supports this idea.

\emph{Readability of formal specifications.}
The choice of identifier names, indentation and comments are significant for readability~\cite{ray_buse_metric_2008}. I specify architecture and platform models and mapping specifications in Alloy using the exact name of platforms classes and methods for their corresponding Alloy entities. In specifying mappings between architectural elements and platforms constructs, I employ concepts from the architectural style literature. All in all, I have found them readable and producible: more readable and easier to modify than code.


\emph{Architecture-to-platform mappings.}
By being able to mix code generation capabilities with declarative specifications, the developer can easily express otherwise tacit knowledge of architecture-to-platform mappings; architectural decisions and those regarding implementation technologies are decoupled, making both easier to evolve.
To the extent that such mappings can be matured into richer, general forms, they would enable capturing of important software design knowledge.

\emph{Traceability Support.}
Having architecture as an intermediate representation promotes the traceability of implementation artifacts to architectural artifacts and to abstract application elements. Using Pol, traceability links can be obtained by a query over formally synthesized models. For example, the query, \emph{TopicPublisher.destination.handle,} returns a set of \emph{EventBus} connectors handled by \emph{Topic} constructs from the HornetQ platform, where \emph{TopicPublisher} message producers have those Topics as destinations.


\emph{Tool support.}
Using the current tool, a developer must specify bindings manually. Other work suggests that known good patterns for using frameworks can be detected automatically~\cite{Supporting-Framework_2009}. Although I have not yet experimented with such techniques, they seem worth investigating in future work, to relieve developers from having to capture design fragments by hand.

\section{Summary}
\label{conclusion_2}

This chapter contributes an approach to using program synthesis within an evolutionary software development process, based on trading away completeness of synthesis for simplicity and rigor in modeling notations and use of general-purpose solvers in place of costly transformation engines. 
The experimental data suggests that this is a line of work worth exploring further, with potential for industrial impact.



\newpage
\thispagestyle{empty}
\mbox{}

\chapter{Formal Synthesis of Tradeoff Spaces for Object-Relational Mapping}

In this chapter, I present Spacemaker.
This is the third experiment the I have conducted in the context of this dissertation.
The specific aim of this system is to reduce the time to develop relational schemas and associated OR mappings, while improving quality of mappings (in conformance to formal \emph{correctness} constraints), and to enable designers to make quantitative tradeoffs in areas in which they have previously made decisions based on intuition or training.
To this end, I develop a formally precise approach for synthesis of large spaces of such mappings, and classifying individual mappings in these spaces into multidimensional quality equivalence classes.
This work, among other things, provides evidence in support of reducing the costs of development by
reducing the need for traditional hand coding of translators.
Specifically, formal specifications of OR mapping strategies along with the application object model enables automating the process of synthesizing the tradeoff spaces of corresponding relational schemas.
In support of the claim of broadening the applicability of MDD in various stages of the software development lifecycle, this work
targets development of effective persistence layers for object-oriented applications.
Finally, this aspect of my research shows the ability of the overall approach to formally validate correctness of mapping rules. I express essential properties expected to hold as analyzable specifications. I then use automated analysis to check them.
The rest of this chapter is organized as follows.
Section 5.2 presents the approach. 
Sections 5.3 and 5.4 reports and discusses data from the experimental testing of the approach. 
Finally, Section 5.5 concludes with an outline of future work in line this research component. 

\section{Motivation and Research Problem}
Relational database and object-oriented software development are both playing central roles in their respective areas.  Object-oriented applications often need to use relational databases for persistent storage. Transformations between instance models in these two paradigms encounter the so-called {\em impedance mismatch} problem~\cite{ireland_understanding_2009}. Object-relational mapping (ORM) systems are now widely used to bridge the gap between object-oriented application models and relational database management systems (DBMS), based on application-specific mapping definitions on how object models are to be mapped to database structures.

Developing object-relational (OR) mappings that achieve desirable quality attributes for object oriented applications is difficult, tedious, and costly.
Today one has to choose between automatic generation of mappings using \emph{pure} mapping patterns~\cite{luca_cabibbo_managing_2005,ireland_understanding_2009}, or the manual design of mixed mappings, in which different mapping strategies are applied to individual classes rather than to entire inheritance hierarchies. Producing pure mappings automatically is easy, but it often leads to sub-optimal results. Developing mixed mappings by hand can achieve much higher quality, but it is hard and error-prone. Among other things, it requires a thorough understanding of both object and relational paradigms, of large spaces of possible mappings, and of the tradeoffs involved in making choices in these spaces.

\section{Approach}
\label{Approach}

\begin{figure}[tbh]
\centering
\includegraphics[width=6in]{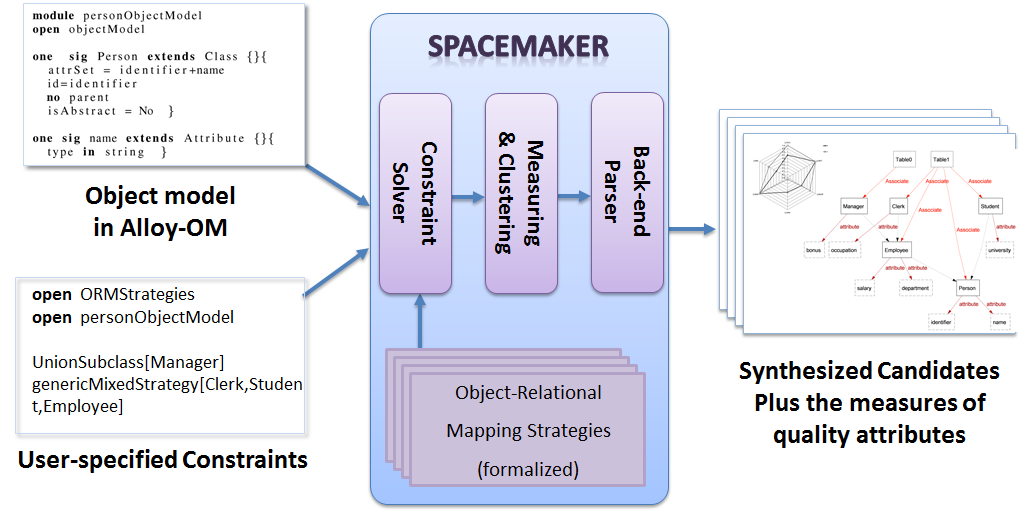}
\caption{\label{fig:spacemaker}High-level view of the spacemaker tool-suite.
}
\end{figure}

I present the approach in four parts.
I first introduce \emph{Alloy-OM}, a domain-specific language that I developed within the Alloy language to let developers specify object models in Alloy. Next, I present formalizations of object-relational mapping strategies, and then use these formalizations to automate synthesis of quality equivalence classes of OR mappings.
Finally, I describe algorithms that are important for the scalability of the approach.
In a nutshell, they serve to decompose large object models into smaller components for which mapping problems can be solved independently. This element of the work helps avoid combinatorial explosion in constraint solving.
Figure~\ref{fig:spacemaker} shows the high-level view of the spacemaker tool-suite implementing the approach.

\subsection{Alloy-OM}
The Alloy object model (Alloy-OM) is a domain specific language (DSL), I developed in Alloy. The user of Spacemaker then can describe object models in the Alloy-OM DSL. The Alloy-OM meta-model has three main constructs: Class, Attribute and Association. For each class in the object model, there is a corresponding ``Class'' signature in the Alloy-OM model. Similarly, for each attribute of a given class in the object model, the corresponding Alloy-OM Class signature contains a corresponding Alloy-OM Attribute as a member of its attribute set. Finally, each association in the object model has a corresponding ``Association'' signature in the Alloy-OM model. In the following, I will describe details of each of these elements.

\subsubsection{Class Definition}

Each Class signature in the Alloy-OM model has a set of fields: attrSet, id, parent and isAbstract. The ``attrSet'' field in each Class signature specifies the set of attributes as defined for the corresponding class in the object model. The ``id'' field in the Class signature represents the identifier of the corresponding class. The inheritance relationship in the object model is represented by the ``parent'' relation in the Alloy-OM model. Specifically, the inheritance relationship between classes c and p, where c inherits from p, will represent by the expression of ``\emph{parent = p}'' specified within the c class signature definition. Finally, ``isAbstract'' field of each Alloy-OM class signature denotes whether the class under consideration is abstract or not.

\begin{wrapfigure}{r}{0.5\textwidth}
\vspace{-25pt}
  \begin{center}
    \includegraphics[width=0.53\textwidth]{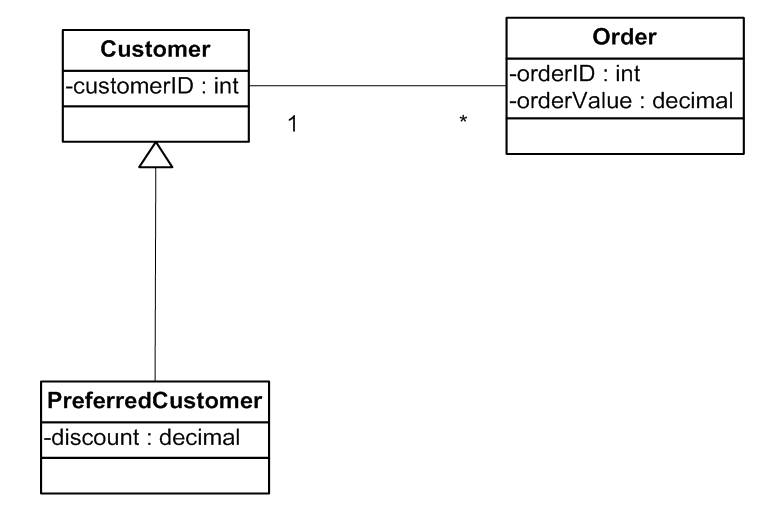}
  \end{center}
  \vspace{-25pt}
  \caption{\label{fig:object_model_diagram}customer-order object model.
  }
\end{wrapfigure}

To make the idea concrete, Figure~\ref{fig:object_model_diagram} shows an object model diagram for a simple customer-order example. The object model contains three classes of Order, Customer, and PreferredCustomer. There is a one-to-many association between the Customer and Order classes, and PreferredCustomer inherits from the Customer class.

To describe each class in the Alloy-OM, I define a signature that extends ``Class'' signature defined within the Alloy-OM meta-model. For example, Listing~\ref{orderClass} represents ``Order'' class in Alloy-OM DSL.

\begin{figure}
\lstset{ %
basicstyle=\scriptsize,       
numbers=left,                   
numberstyle=\tiny,      
stepnumber=1,                   
numbersep=7pt,
backgroundcolor=\color{white},  
showspaces=false,               
showstringspaces=false,         
showtabs=false,                 
frame=bottomline, 
tabsize=2,                    
captionpos=b,                   
breaklines=true,                
breakatwhitespace=false,        
numberbychapter=false,
xleftmargin=1.5em,
morekeywords={module,sig,abstract,extends,one,some,set,open,pred,all,in,no}
}
\lstinputlisting[caption=Order class in Alloy-OM.,numberbychapter=false,label=orderClass]
{Pics/ch4/Order-class.als}
\vspace{-0.6cm}
\end{figure}

The order class has two attributes of ``orderID'' and ``orderValue'', which are assigned to the ``attrSet'' field of the Order class. The id field specifies the orderID as the identifier of this class. The last two lines of the Order signature specification denote that Order is not an abstract class and has no parent.
Similarly, the following code snippet represents ``PreferredCustomer'' signature definition. According to Fig.~\ref{fig:object_model_diagram}, this class inherits from the Customer class. The expression on line 3, thus, specifies Customer as the parent of the PreferredCustomer class.

\begin{figure}
\lstset{ %
basicstyle=\scriptsize,       
numbers=left,                   
numberstyle=\tiny,      
stepnumber=1,                   
numbersep=7pt,
backgroundcolor=\color{white},  
showspaces=false,               
showstringspaces=false,         
showtabs=false,                 
frame=bottomline, 
tabsize=2,                    
captionpos=b,                   
breaklines=true,                
breakatwhitespace=false,        
numberbychapter=false,
xleftmargin=1.5em,
morekeywords={module,sig,abstract,extends,one,some,set,open,pred,all,in,no}
}
\lstinputlisting[caption=PreferredCustomer class in Alloy-OM.,numberbychapter=false,label=PreferredCustomer]
{Pics/ch4/PreferredCustomer.als}
\vspace{-0.6cm}
\end{figure}

\subsubsection{Attribute Definition}

To describe each attribute in the Alloy-OM, I define a signature that extends the corresponding data type signature. Alloy-OM DSL contains a set of predefined data types, namely Integer, Real, string and Bool. As a concrete example, orderValue is an attribute of type Real for the Order Class. The following code snippet represents its specification in Alloy-OM:
\emph{one sig orderValue extends Real{}}

The user can also specify a new data type by defining an abstract signature that extends the ''Type'' signature. The following expression defines a data type named NewDataType.
\emph{abstract sig NewDataType extends Type{}}

\subsubsection{Association Definition}

Each Association signature in the Alloy-OM model has four fields: src, dst, $src\_multiplicity$ and $dst\_multiplicity$. The src and dst fields of each Association signature specify the source and destination classes for that association, respectively. Association multiplicity defines the number of object instances that can be at each end of the association. The $src\_multiplicity$ and $dst\_multiplicity$ fields represent Association multiplicities of source and destination classes respectively, and can have values of either ONE or MANY.

\begin{figure}[bh]
\lstset{ %
basicstyle=\scriptsize,       
numbers=left,                   
numberstyle=\tiny,      
stepnumber=1,                   
numbersep=7pt,
backgroundcolor=\color{white},  
showspaces=false,               
showstringspaces=false,         
showtabs=false,                 
frame=bottomline, 
tabsize=2,                    
captionpos=b,                   
breaklines=true,                
breakatwhitespace=false,        
numberbychapter=false,
xleftmargin=1.5em,
morekeywords={module,sig,abstract,extends,one,some,set,open,pred,all,in,no}
}
\lstinputlisting[caption=CustomerOrder Association in Alloy-OM.,numberbychapter=false,label=CustomerOrderAssociation]
{Pics/ch4/CustomerOrderAssociation.als}
\vspace{-0.6cm}
\end{figure}

As a concrete example, to describe the association between customer and order in the Alloy-OM model, I define a corresponding signature that extends ``Association'' signature. The code snippet of Listing~\ref{CustomerOrderAssociation} represents CustomerOrderAssociation specification in the Alloy-OM.
The above specification states Customer and Order classes as source and destination of the CustomerOrder association, respectively. The CustomerOrderAssociation is a ``one-to-many'' association, which is specified using $src\_multiplicity$ and $dst\_multiplicity$ fields.

\begin{figure}[tbh]
\centering
\includegraphics[width=6in]{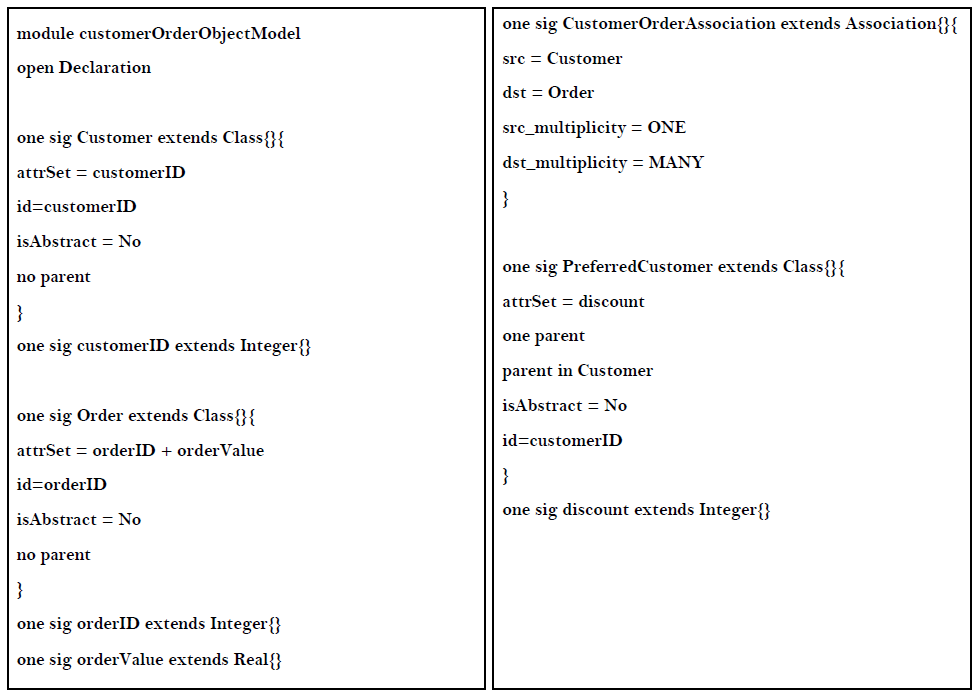}
\caption{\label{fig:customerOrderAlloyOM}
customer-order Alloy-OM model.
}
\end{figure}

Now, we have all the elements to specify the customer-order object model diagram in the Alloy-OM DSL. Putting all together, the specification of the Alloy-OM model for the customer-order example is represented in Fig.~\ref{fig:customerOrderAlloyOM}. The alloy module starts by specifying the name of the model, here module customerOrderObjectModel. We then import the Alloy-OM declaration module by the following expression: open Declaration. The rest of the module specifies three classes, their attributes and the association between Customer and Order classes.

\subsection{Formalization}

The issue of mapping an object model to a set of relations is described thoroughly in the research literature~\cite{luca_cabibbo_managing_2005,ireland_understanding_2009,keller_mapping_1997,philippi_model_2005}.
To provide a basis for precise modeling of the space of mapping alternatives, I have formalized OR mapping strategies in an appropriate level of granularity.
As an enabling technology, I chose Alloy~\cite{daniel_jackson_alloy:lightweight_2002} as a specification language and satisfaction engine for three reasons.
First, its logical and relational operators makes Alloy an appropriate language for specifying object-relational mapping strategies. Second, its ability to compute solutions that satisfy complex constraints is useful as an automation mechanism. Third, Alloy has a rigorously defined semantics closely related to those of relational databases, thereby providing a sound formal basis for our approach.

The principal mapping strategies are explained in terms of the notations suggested by Philippi~\cite{philippi_model_2005}, and Cabibbo and Carosi~\cite{luca_cabibbo_managing_2005}.
To manage association relationships, I have formally specified three ORM strategies of {\em own association table}, {\em foreign key embedding} and {\em merging into single table}. I have also defined three more ORM strategies for inheritance relationships: {\em class relation inheritance (CR)}, {\em concrete class relation inheritance (CCR)} and {\em single relation inheritance (SR)}. Furthermore, as the aforementioned ORM strategies for inheritance relationships are just applicable to the whole inheritance hierarchies, I have defined extra predicates for more fine-grained strategies: {\em Union Superclass}, {\em Joined Subclass} and {\em Union Subclass}, suitable to be applied to part of inheritance hierarchies, letting the developer design a detailed mapping specification using the combination of various ORM strategies.
To make the idea concrete, I illustrate the semantics of one of these strategies in the following.

\begin{figure}
\lstset{ %
basicstyle=\scriptsize,       
numbers=left,                   
numberstyle=\scriptsize ,      
stepnumber=1,                   
numbersep=5pt,                  
backgroundcolor=\color{white},  
showspaces=false,               
showstringspaces=false,         
showtabs=false,                 
frame=bottomline,                    
tabsize=2,                    
captionpos=b,                   
breaklines=true,                
breakatwhitespace=false,        
numberbychapter=false,
xleftmargin=1.8em,
morekeywords={module,sig,abstract,extends,one,set,open,pred,all,in,none}
}
\begin{lstlisting}[caption=Part of the Alloy predicate for the $UnionSubclass$ strategy,numberbychapter=false,label=CCR]
pred UnionSubclass[c: Class]{
  c in (isAbstract.No) =>{
    one Table<:c.~tAssociate	
  }
  (c.isAbstract=No) =>{ 
    all a:Attribute|a in c.attrSet =>{
      one f:Field|f.fAssociate=a 
      && f in (c.~tAssociate.fields) } 
  }
  (c.isAbstract=No)&&(c.^parent != none) =>{
    all a:Attribute | a in c.^parent.attrSet =>{
      one f:Field|f.fAssociate=a && 
      f in (c.~tAssociate.fields) }	
  }
  (c.~tAssociate).foreignKey = none
  ...
}
\end{lstlisting}
\vspace{-0.6cm}
\end{figure}

Listing~\ref{CCR} partially outlines the Alloy predicate for the {\em Union Subclass} strategy, where each concrete class within the hierarchy is represented by a separate table. The strategy predicate then states, in lines 5--14, 
that each table encompasses relational fields corresponding to both attributes of the associated class and its inherited attributes. As such, to retrieve an individual object, only one table needs to be accessed.
Finally, this strategy implies no referential 
constraint over the mapped relations.

\subsection{Design Space Exploration}
\label{DSE}
In the previous section, I showed how analyzable specifications can be used to formalize OR mapping strategies.
In this section, I tackle the other aspect that needs to be clarified:
how one can apply a design space exploration approach to generate \emph{quality equivalence classes} of OR mappings based on those specifications.

A design space is a set of possible design alternatives, and design space exploration (DSE) is the process of traversing the design space to determine particular design alternatives that not only satisfy various design constraints, but are also optimized in the presence of a set of objectives~\cite{saxena_mde-based_2010}.
The process can be broken down into three key steps:
(1) Modeling the space of mapping alternatives;
(2) Evaluating each alternative by means of a set of metrics;
(3) Traversing the space of alternatives to cluster it into equivalence classes.

\subsection*{Modeling the Space of Mapping Alternatives}
For each application object model, due to a large number of mapping options available for each class, its attributes and associations, and its position in the inheritance hierarchy, there are several valid variants.
To model the space of all mapping alternatives, I develop a generic mixed mapping specification based on fine-grained strategies formalized in previous section.
This generic mixed mapping specification lets the automatic model finder choose for each element of the object model any of the relevant strategies, e.g. any of the fine-grained generalization mapping strategies for a given class within an inheritance hierarchy.

Applying such a loosely constrained mixed mapping strategy into the object model leads to a set of ORM specifications constituting the design space.
While they all represent the same object model and are consistent with the rules implied by a given mixed mapping strategy, they exhibit totally different quality attributes. For example, how inheritance hierarchies are being mapped to relational models affects the required space for data storage and the required time for query execution.

I called this mapping strategy loosely constrained because it does not concretely specify the details of the mapping, such as applying, for example the \emph{UnionSubclass} strategy to a specific class.
An expert user, though, is able to define a more tightly constrained mixed mapping by means of the parameterized predicates \emph{Spacemaker} provides, as I demonstrate in the next section.
The more detailed the mapping specifications, the narrower the outcome design space, and the less the required postprocessing search.

\subsection*{Measuring Impacts of OR mappings}

Mapping strategies have various kinds of impacts in terms of quality attributes of applications.
There are several approaches proposed in the literature dealing with the challenge of defining metrics for OR mapping impacts on non-functional characteristics.
It has been shown that 
maintainability, storage space and performance, among the set of all quality attributes defined by the ISO/IEC 9126-1 standard, are characteristics significantly influenced by OR mappings~\cite{stefan_holder_towardsmetrics_2008}.
For each of those attributes, I use a set of metrics suggested by Holder et al.~\cite{stefan_holder_towardsmetrics_2008} and Baroni et al.~\cite{baroni_formal_2005}. The metrics are \emph{Table Access for Type Identification} (\emph{TATI}), \emph{Number of Corresponding Table} (\emph{NCT}), \emph{Number of Corresponding Relational Fields} (\emph{NCRF}), \emph{Additional Null Value} (\emph{ANV}), \emph{Number of Involved Classes} (\emph{NIC}) and \emph{Referential Integrity Metric} (\emph{RIM}).

To measure these metrics, I developed a set of queries to execute over synthesized alternatives.
For brevity, and because it suffices to make the point, I concisely describe one of these metrics and the corresponding query in the following. \emph{Spacemaker} supports the others as well.

The \emph{Number of Corresponding Relational Fields} (NCRF) metric specifies the extent of change propagation for a given OR mapping. Specifically, the NCRF metric manifests the effort required to adapt a relational schema after applying a change, such as inserting or deleting an attribute, over a class. According to the definition, given a class \emph{C}, \emph{NCRF(C)} specifies the number of relational fields in all tables that correspond to each non-inherited, non-key attribute of \emph{C}.
The specification of a query we designated to measure the NCRF metric over synthesized alternatives is given below:

\emph{NCRF(C) = \#(C.attrSet - C.id).$\sim$fAssociate.$\sim$fields}

The Alloy dot operator denotes a relational join.
While \emph{attrSet} specifies a set of non-inherited attributes of a class, \emph{fAssociate} is a relation from a table field to its associated class attribute. The query expressions then, by using the Alloy set cardinality operator \emph{(\#)}, defines the NCRF metric.

\subsection*{Exploring, Evaluating and Choosing}

The next step is to explore and prune the space of mapping alternatives according to quality measures.
Spacemaker partitions the space of satisfactory mixed mapping specifications into equivalence classes and selects at most a single candidate from each equivalence class for presenting to the end-user.

To partition the space, Spacemaker evaluates each alternative with respect to previously described relevant metrics. So each equivalence class consists of all alternatives that exhibit the same characteristics.
Specifically, two alternatives $a_1$ and $a_2$ are equivalent if \emph{value($a_1$, $m_i$)} = \emph{value($a_2$, $m_i$)} for all metrics ($m_i$). Because equivalent alternatives all satisfy the mapping constraints, it suffices to select one alternative in each equivalence class to find a choice alternative.
Given that quality characteristics are usually conflicting, there is generally no single optimum solution but there are several pareto-optimal choices representing best trade-offs.

\subsection*{Back-end Parser}
\label{AlloyToSQLTransformer}

Having computed satisfying solutions, the \emph{back-end parser} component parses and transforms these solutions from low-level, XML formatted Alloy objects to SQL counterparts. This back-end parser enables integrating the formally precise object-relational mapping synthesis approach with industrial OR mapping tools.

\subsection{Model Splitting}
As with many formal techniques, the complexity of constraint satisfaction restricts the size of models that can actually
be analyzed~\cite{ethan_k._jackson_components_2010}. This approach also requires an explicit representation of the set of all quality equivalence classes of mapping alternatives, which in general grows exponentially in the number of elements in a model.

To address these scalability problems, I split the object model into sub-models.
The key idea is that since for association relationships with cardinality of many-to-many, there is just one applicable mapping strategy, i.e. \emph{own association table}, I make use of such relations to split the object model into sub-models.

I consider an object model as a graph, $G_{objModel}= <V,E>$, where nodes \emph{V} represent classes,
and there is an edge $<v_i, v_j>$ joining two nodes $v_i$ and $v_j$ if there is a direct relationship including association and generalization link between them.
I assume that $G_{objModel}$ is connected. Otherwise, I consider each sub-graph separately.
An edge joining two nodes $v_i$ and $v_j$ in a graph is a \emph{bridge} if removing the edge would cause $v_i$ and $v_j$ to lie in two separate sub-graphs~\cite{david_easley_networks_2010}.
A bridge is the only route between its endpoints.
In other words, bridges provide nodes with access to parts of the graph that are inaccessible by other means.
So, to decompose a graph $G_{objModel}$, I remove all bridges of type many-to-many association.

\begin{figure*}[tbh]
\centering
\includegraphics[width=\linewidth]{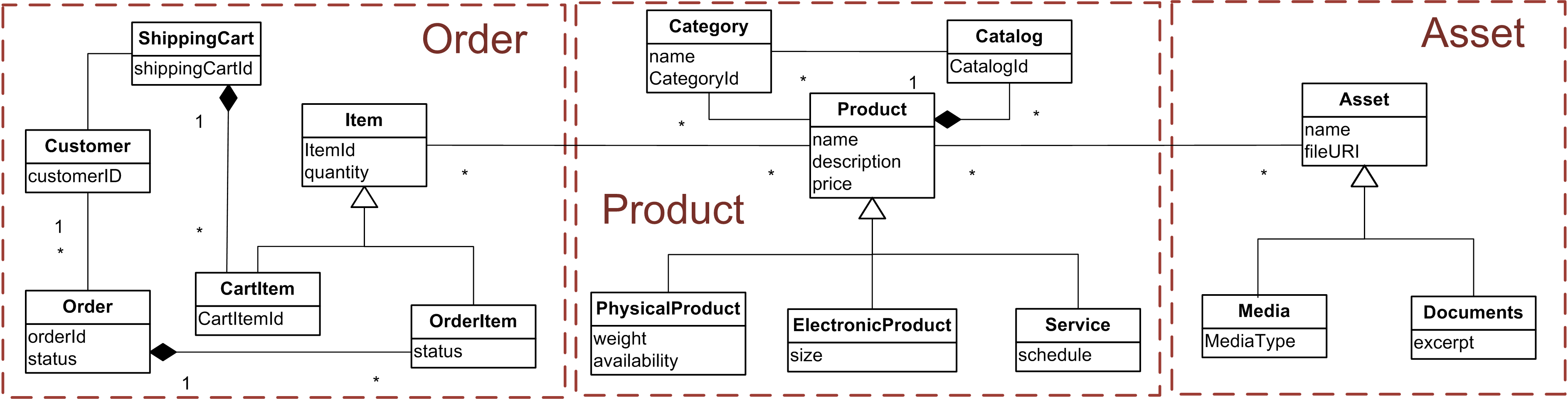}
\caption{\label{fig:ecommerce}The ecommerce object model.}
\end{figure*}

To make the idea concrete I consider the ecommerce domain model adopted from Lau and Czarnecki~\cite{sean_quan_lau_domain_2006}.
According to the diagram shown in Figure~\ref{fig:ecommerce}, there are two such bridges: $<Product,Asset>$ and $<Item,Product>$.
By removing those bridges we obtain three smaller sub-graphs.
The gain then comes from the reduction in the sizes of the constraint solving problems. That is, I replace a large constraint solving problem with smaller and more manageable problems that can in particular be addressed by the formally precise synthesis technique.

\subsection{Formal Validation of Mapping Rules}
Formalizing ORM strategies in an analyzable specification language not only enables automatic synthesis of mapping alternatives for each application object model, but also provides the basis to formally validate correctness of mapping rules. By expressing essential properties of object-relational mappings as analyzable specifications, we can use automated analysis to check them, albeit within limited scopes.
We specify such implications required to be checked as assertions. These assertions express properties expected to hold. In other words, assertions state a set of constraints intended to follow from the specifications of the model~\cite{daniel_jackson_alloy:lightweight_2002}.

Alloy is based on relational logic which is undecidable. As such, it is impossible to automatically prove whether an assertion holds for every possible case or not. Rather, the Alloy Analyzer checks the assertion against a huge set of model instances that can be considered as test cases. More precisely, the Alloy Analyzer is a bounded checker, guaranteeing the validity of assertions only within a bounded instance space. 
If the assertion does not hold for a certain instance, Alloy Analyzer reports it as a counterexample. Counterexample is a particular model instance that makes the assertion false. When the analyzer finds no counterexample it means that the assertion holds for all considered model instances. That means the assertion is valid within the specified scope. Spacemaker bounds execution of assertions with the ultimate scope of elements considered for the synthesis of application-specific OR mappings, we thus expect the validity of assertions for all generated mappings.

To make the idea concrete, I illustrate the contents of two assertions, represented in Listing~\ref{assertions}, in the following.

\begin{figure}
\lstset{ %
basicstyle=\scriptsize,       
numbers=left,                   
numberstyle=\scriptsize ,      
stepnumber=1,                   
numbersep=5pt,                  
backgroundcolor=\color{white},  
showspaces=false,               
showstringspaces=false,         
showtabs=false,                 
frame=bottomline,                    
tabsize=2,                    
captionpos=b,                   
breaklines=true,                
breakatwhitespace=false,        
numberbychapter=false,
xleftmargin=1.8em,
morekeywords={module,sig,abstract,extends,one,set,open,pred,all,in,none}
}
\begin{lstlisting}[caption=Two examples of assertions,numberbychapter=false,label=assertions]
assert noTableForAbstractClasses{
	all c:Class| c.isAbstract=Yes => no c.~tAssociate
}

assert tableFields{
	all c:Class| c.~tAssociate.fields in
	c.~tAssociate.tAssociate.attrSet.~fAssociate + c.~tAssociate.foreignKey + DType.~fAssociate
}
\end{lstlisting}
\vspace{-0.6cm}
\end{figure}

The first assertion states that no table should be associated to abstract classes. The next assertion is about the relational fields of each table. The first part specifies that each table can encompass relational fields corresponding to attributes of all relevant classes. I specify that expression as ``c.~tAssociate.tAssociate.attrSet.$\sim$fAssociate" in Alloy, rather than simply defining it as ``c.attrSet.$\sim$fAssociate". This is because in case of applying the UnionSuperclass strategy, one table is assigned to a set of classes within an inheritance hierarchy, and the associated class contains fields for attributes of all those classes. The ``tAssociate" is a relation from a table to a corresponding class. Using the reverse join operator, $\sim$, ``c.$\sim$tAssociate" states  the table associated to the class c, and then another join, ``.tAssociate", returns a set of all classes handled by that table. In some cases, the associated table also contains a separate field to indicate the most specific class for the object represented by each tuple. This field is indicated as DType in the assertion under consideration.

To check assertions, we issue the \emph{check} command to the analyzer, that instructs the analyzer to search for situations where the assertion set is violated. In conducted experiments, the Alloy Analyzer reports no counterexamples to the assertions, after checking all possible model instances up to the scope within which OR mappings are synthesized, which confirms the validity of assertions in generated models.


\section{Evaluation}
\label{evaluation}

The claim I make for this work is twofold:
(1) It is feasible to formalize the \emph{correctness constraints} for object-relational mapping strategies, thereby to automate the synthesis of an exhaustive set of mixed object-relational mapping candidates, and that it is possible to statistically analyze each of the candidates in dimensions of six major mapping quality metrics, and thereby to cluster them into quality equivalence classes;
(2) the performance of the technology implementation based on a bounded model checker is adequate (on the order of minutes) to support synthesis of modest applications (with up to 40 tables).


To test the feasibility hypothesis, I develop a prototype tool that implements it, called Spacemaker~\cite{_spacemaker}. 
I show that the ideas are practical by applying Spacemaker to several case studies from the object-relational mapping literature.
I then compare the discrepancies between our formally derived OR mappings and the manual mappings published in the literature.
The differences revealed problems with their mappings, suggesting again that manual development of OR mappings can be error-prone.

\begin{figure*}[tbh]
\centering
\includegraphics[width=\linewidth]{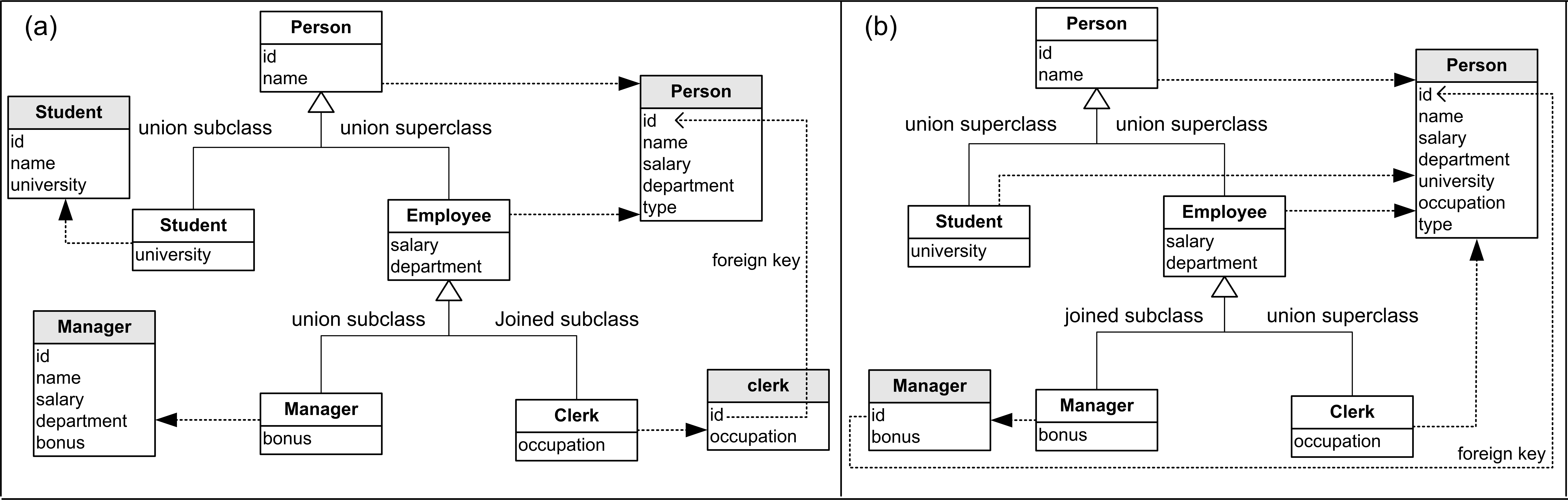}
\caption{\label{fig:examples}Examples of mixed mapping strategies}
\end{figure*}

Figure~\ref{fig:examples} shows two applications of mixed ORM strategies, adopted from Holder et al.~\cite{stefan_holder_towardsmetrics_2008}.
White boxes represent classes, while boxes having grey titles represent corresponding mapped tables. Black and white arrows represent mapping and inheritance relationships, respectively. Finally, foreign keys as well as the applied mapping strategies are also mentioned in the diagrams.

Listing~\ref{appModel} formally describes part of the {\em personObjectModel} according to the diagram. At the top, it imports the declaration of {\em objectModel}, and then defines {\em Person} and its attribute, {\em name}, using signature extension as a subtype of {\em Class} and {\em Attribute} types. The other characteristics of the class are also specified.

\begin{figure}
\lstset{ %
basicstyle=\scriptsize,       
numbers=left,                   
numberstyle=\scriptsize ,      
stepnumber=1,                   
numbersep=5pt,                  
backgroundcolor=\color{white},  
showspaces=false,               
showstringspaces=false,         
showtabs=false,                 
frame=bottomline,                    
tabsize=2,                    
captionpos=b,                   
breaklines=true,                
breakatwhitespace=false,        
numberbychapter=false,
morekeywords={module,sig,abstract,extends,one,no,set,open,pred,all,in,none}
}
\begin{lstlisting}[caption=Person object model in Alloy-OM,numberbychapter=false,label=appModel]
module personObjectModel
open objectModel

one  sig Person extends Class {}{
	attrSet = identifier+name
	id=identifier
	no parent 
	isAbstract = No  }

one sig name extends Attribute {}{
	type in string  }
\end{lstlisting}
\vspace{-0.6cm}
\end{figure}

\lstset{ %
basicstyle=\scriptsize,       
numbers=none,                   
backgroundcolor=\color{white},  
showspaces=false,               
showstringspaces=false,         
showtabs=false,                 
frame=none,                    
tabsize=2,                    
captionpos=none,                   
breaklines=true,                
breakatwhitespace=false,        
numberbychapter=false,
}

To specify a mixed OR mapping, the developer can call fine-grained ORM strategies, given as inputs those classes to be mapped in a specific manner.
{\em Spacemaker} then automatically generates the corresponding mapping specifications, should they exist.
The followings outline the high-level definition of mapping specifications for Figure~\ref{fig:examples}a.

\begin{figure}[tbh]
\lstset{ %
basicstyle=\scriptsize,       
numbers=none,                   
backgroundcolor=\color{white},  
showspaces=false,               
showstringspaces=false,         
showtabs=false,                 
frame=none,                    
tabsize=2,                    
captionpos=none,                   
breaklines=true,                
breakatwhitespace=false,        
numberbychapter=false,
}
\begin{lstlisting}[caption=High-level definition of mapping specifications for figures~\ref{fig:examples}a and~\ref{fig:examples}b using our approach,numberbychapter=false,label=AlloyCode]
open ORMStrategies
open personObjectModel

UnionSubclass[Manager]
JoinedSubclass[Clerk]
UnionSuperclass[Employee]
UnionSubclass[Student]
\end{lstlisting}
\vspace{-0.6cm}
\end{figure}

\begin{figure}[tbh]
\centering
\includegraphics[height=3.4in]{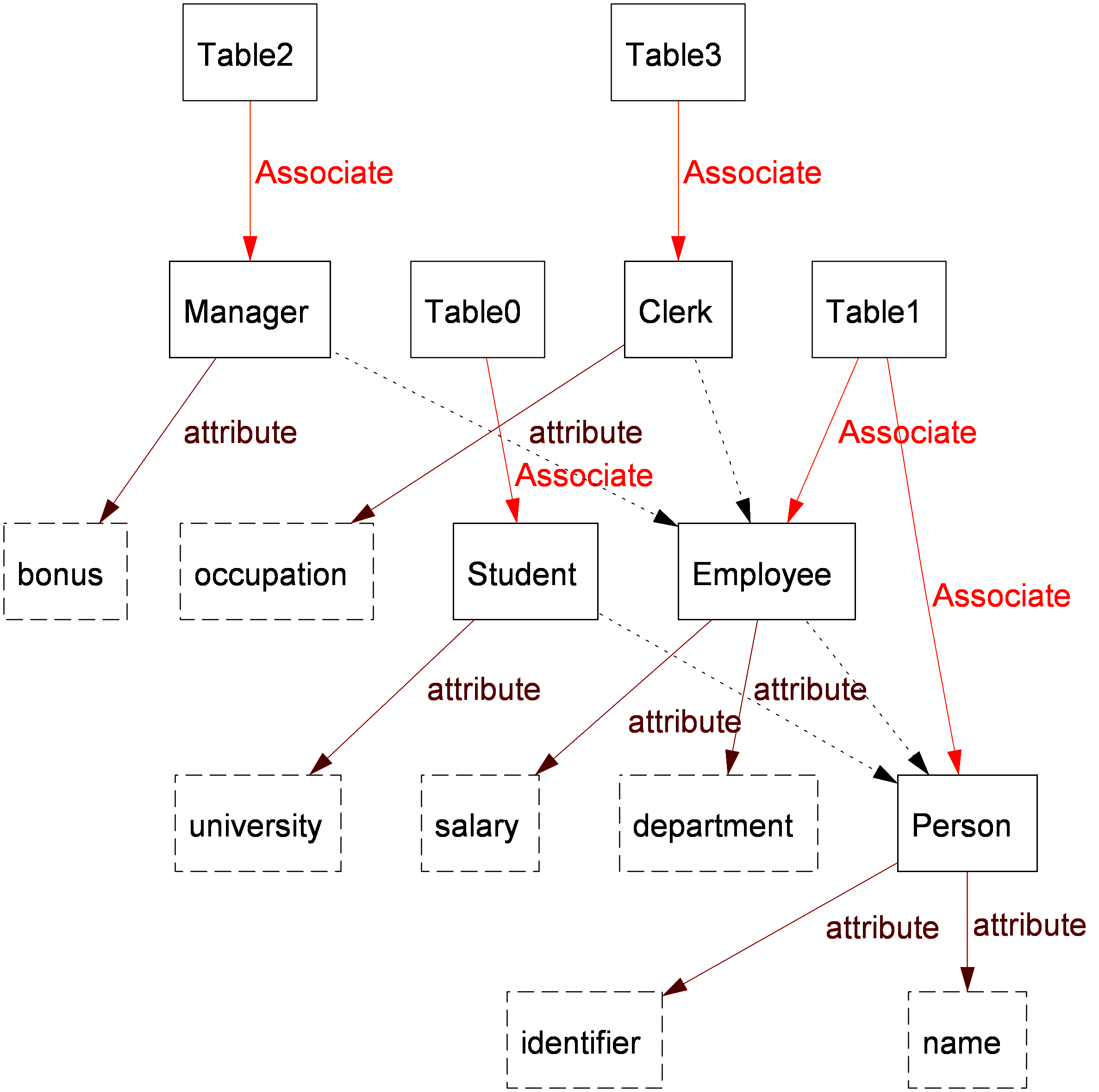}
\caption{\label{fig:AlloyModel}Mapping diagram for Figure~\ref{fig:examples}a derived automatically based on mixed mapping strategies of {\em union superclass}, {\em joined subclass} and  {\em union subclass}}
\end{figure}

Figure~\ref{fig:AlloyModel} illustrates the computed result for the example of Figure~\ref{fig:examples}a.
The diagram is accurate for the result automatically computed, but I have edited it to omit some details for readability (fields of tables and primary key relationships, for example). In this diagram, Table 1 is associated to {\em Person} and {\em Employee} classes, which are being mapped by the {\em union superclass} strategy. Separate tables are associated to both {\em Student} and {\em Manager} classes, according to the {\em union subclass} strategy. Finally, application of the {\em joined subclass} strategy leads to a separate table for {\em Clerk} with a foreign key, omitted in the diagram, to its superclass corresponding table.

To enumerate the space of mappings for the given object model, I use the \emph{genericMixedStrategy}, with the set of classes within the hierarchy as inputs. This generic strategy lets the automatic model finder, here Alloy analyzer, choose for each class any of the fine-grained strategies and to see whether their combinations applied to classes within the hierarchy is satisfiable or not.
Alloy guarantees that all computed mapping candidates conform to the rules implied by mapping predicates formalizing correctness constraints.

I used a PC with an Intel Core i5 2.67 Ghz processor and 4 GB of main memory, and leveraged \emph{SAT4J} as the SAT solver during the experiments.
Given all the specifications and mapping constraints, Spacemaker using the Alloy analyzer then generate 760,000 mapping candidates, assess them, and reduce them to 40 equivalence classes, in less than 10 seconds.

\begin{wrapfigure}{r}{0.5\textwidth}
\vspace{-25pt}
  \begin{center}
    \includegraphics[width=0.53\textwidth]{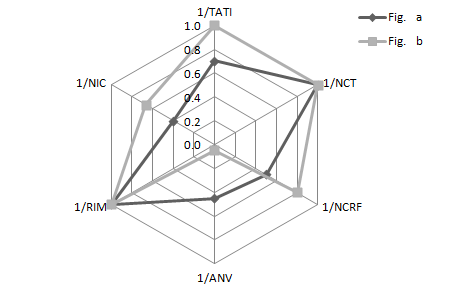}
  \end{center}
  \caption{\label{fig:spider_diagram}Multi-dimensional \emph{quality measures} for two mapping candidates 
  }
\end{wrapfigure}

The spider diagram, shown in Figure~\ref{fig:spider_diagram}, illustrates the 6-dimensional ``quality measures'' for two mapping candidates represented in Figure~\ref{fig:examples}. To display quality measures in one diagram, I normalized the values.

According to the diagram, if the designer opts for the resource utilization, the mapping depicted in Figure~\ref{fig:examples}a would be a better option. More specifically, with respect to the ANV metric, representing additional storage space in terms of null values, the mapping of Figure~\ref{fig:spider_diagram}b requires more \emph{wasted space}. This is because instances of four different classes, namely Person, Student, Employee and Clerk, are stored together in a shared table. Thus, each row in the shared table that represents an instance of the Student class, for example, contains a null value at each relational field corresponding to the other classes.
On the other hand, if the designer opts for maintainability and performance, the mapping depicted in Figure~\ref{fig:examples}b would be a better choice.
More precisely, the mapping of Figure~\ref{fig:examples}a negatively affects the NCRF metric reflecting the effort required to adapt the relational schema. This is partly because applying the \emph{UnionSubclass} strategy results in duplication of relational fields. With respect to the TATI metric which is a performance indicator of polymorphic database queries, this mapping also poses performance problems.

Focusing on the second hypothesis, to test that Spacemaker is able to handle also non-trivial OR mappings, I select an object model of a real ecommerce system~\cite{sean_quan_lau_domain_2006}. This object model, shown in Figure~\ref{fig:ecommerce}, represents a common architecture for the kind of open source and commercial ecommerce systems. It includes 15 classes connected by 9 associations and consists of 7 inheritance relationships.

Without decomposition, the Alloy analyzer ran out of memory before synthesizing the whole space of mapping alternatives. Given the splitting algorithm, we decompose the object model to three sub-models and feed them into the Spacemaker.

Interpretation of data shows that similar to the former experiment, the synthesis time for the \emph{Asset} sub-model is in the order of seconds, but for the other two sub-models is in the order of minutes.
This is mainly because in the former case, the analyzer just considers inheritance mapping strategies as there are no associations in those models, while in the other models the constraints of both inheritance and association mapping strategies are involved. So it takes more time for the model finder to generate satisfying solutions for them.

As Spacemaker solves sub-models separately, the constraint solving bottleneck depends on the largest sub-model to solve.
Although the number of valid solutions is high, i.e. hundreds of thousands of satisfiable solutions, Spacemaker is able to generate quality equivalence classes of mappings in an acceptable amount of time, which confirms that the proposed synthesis technology is feasible.

\section{Discussion}
\label{discussion}

This work shows that ORM strategies can be formalized and implemented as executable specifications, and that Spacemaker can automatically synthesize and prune the space of mapping alternatives in an effective manner.
The formal recapitulations of previous studies also reveals some problems. For example, the referential integrity constraint in Figure~\ref{fig:examples}b, is not mentioned in the source reference~\cite{stefan_holder_towardsmetrics_2008}, but exists in the mapping specifications automatically derived using {\em Spacemaker}.
Discovery of such inconsistencies provides an example of how formal synthesis technique can help designers in an error-prone task of developing OR mappings.


Overall, this work appears to support the idea that shifting the responsibility of finding an optimized mapping specification from technicians---who better understand mapping strategies, their implications, and techniques for mapping object models to relational models---to the domain experts, more aware of requirements and specifications
is a plausible aspiration.

\section{Summary}
\label{conclusion}

While a wealth of research has been performed on bridging application models and databases to address the {\em impedance mismatch} problem, little has been done on automated support for the derivation of mapping specifications for ORM frameworks.
This chapter presents a novel approach that substantially supports automatic generation of such mapping specifications to deliver the quality of expert-hand-crafted mappings and the productivity benefits of fully automated techniques.
This approach ultimately promises to reduce the engineering personnel costs involved in producing high-quality modern software systems.
The new mapping approach exposes many interesting research challenges.
These challenges include exploring symmetry breaking techniques customized for the specific domain of OR mappings to reduce the size of the solution space
and integrating the mapping compiler with industrial object-relational mapping tools.

\newpage
\thispagestyle{empty}
\mbox{}

\chapter{Evaluation of this Research}

In this chapter, I first summarize how well the thesis of this dissertation is supported by the evidence and analyses presented. After that, I evaluate the novelty, potential, as well as the shortcomings and remaining problems in the proposed approach. 

\section{Thesis and Evidence}

The overall goal of this research is to enable rapid and reliable model-driven development of software applications, while broadening its applicability in various development phases. To this end, I proposed a novel approach to substantially automating the synthesis process without domain-specific languages or hand-crafted translators. The main claim at the heart of this thesis is that it is feasible to use analyzable formal specification languages for encoding general MDD abstractions and to use associated formal analyzers for synthesis purposes.
I evaluated the thesis in three dimensions of software synthesis: synthesizing architectural models from abstract application models, synthesizing partial code frameworks from application architectures, and synthesizing object-relational mapping tradeoff spaces for object-oriented application architectures.



Chapter 2 
presents  the first experiment with my approach focusing on formal synthesis of software architectures.
This work shows that the vision of formally precise refinement of application models into architectural models in various styles 
is feasible. It also supports a new kind of model-based engineering tool: one that is parameterized by input specifications of architectural styles and application types and that is then capable of supporting the editing of models of that type and automatically mapping them to architectures in the given style.

The evaluation strategy for the first claim includes two parts: First, I evaluated the feasibility of the proposed method for architecture synthesis by conducting a set of experiments, replicating prominent earlier architectural studies from the literature using our formal approach. I offered support in the form of a set of implemented architectural maps. 
The experiments have shown that architectural maps can be formalized and implemented as analyzable specifications, and automatically generated results are consistent with the informally produced results documented in the literature.
Second, I developed a quasi-formal description of the isomorphism between the separation of application and architecture concerns and that which is central to MDD. I demonstrated the utility of this bridge in a form of a model-based prototype tool. This tool combines major achievements in MDD, including meta-modeling tools like the Generic Modeling Environment (GME), with major results in software architecture, including formalization of architectural styles, and with new synthesis concepts introduced in this work, to produce a novel and promising kind of application modeling and transformation tool.


Chapter 3 has presented the evaluation of the second element of my thesis. The evaluation approach was to test the proposed technology against the needs of a representative web-based, distributed software system. To this end, I have adopted the \emph{CyberHealth} system as a challenging problem for the proposed technology.
This system is meant as an operational model for a possible future national cyber-infrastructure for healthcare data liquidity and service integration. Its development has involved the progressive introduction of widely used platforms, making it a reasonable subject for evaluation of the synthesis approach.
The experiments and our experience have provided support for the hypothesis that practical synthesis of platform-specific frameworks from formal architectural specifications without hand-crafted translators is feasible, and a hybrid automated-manual, evolutionary development process is an approach worth exploring further.


Chapter 4 has presented the evaluation of the third element of my thesis.
The evaluation strategy includes two parts: First, to evaluate its feasibility I developed a prototype tool, Spacemaker~\cite{_spacemaker}, that implements the approach; by implementing the technique for OR mapping tradeoff space synthesis and analysis, I demonstrated that the proposed idea is tractable in computational terms~\cite{bagheri_spacemaker}.
I have then conducted formal validation of mapping rules in terms of developing a set of assertions.
I applied the technology implementation to several case studies from the object-relational mapping literature~\cite{stefan_holder_towardsmetrics_2008,drago_quality_2011}, among others an object model of an ecommerce system developed by Lau and Czarnecki~\cite{sean_quan_lau_domain_2006}, that represents a common architecture for the kind of open source and commercial ecommerce systems. During the experiments, I measured performance of the approach implementation based on a bounded model finder by recording the computational time required for deriving the ORM space.
The experimental data shows that formalizing ORM strategies as well as application object models as analyzable specifications enables us to automatically synthesize and prune the space of mapping alternatives in an effective manner.

In summary, the specific goals set forth have been achieved and the evidence and analyses presented have supported my thesis. However, there are still some limitations and shortcomings which will be discussed in section 6.3.

\section{Novelty and Potential}
The work reported in this dissertation is novel in several important dimensions.
Most fundamentally, this work contributes a novel approach to automates software synthesis, balanced against the need for custom-built modeling languages, and to avoid the burdens of having to hand-craft synthesizers customized to particular source and target environments.

Second, the work on ``effective separation of application essence from architectural style"---discussed in Chapter 3---does not pursue incremental elaboration of established research directions. Rather, it develops a novel formal account of mappings from application models to architectures in given architectural styles. Such a rigorous account of mappings support automating the process of refining application models to architectures. 
An important insight arising from this work is that 
while previous work formalized architectural styles, we need a new, parallel concept of \emph{application type}. 
Application types serve as source languages, and architectural styles as target in this work.
Just as an architecture in a given style can be seen as an {\em instance} of that style, and just as an architectural style can be seen as a specification of the family of architectures in that style, we need to view an application model as an instance of an application type, and to develop this concept of application type as a formal specification of a family of application models. The idea that there are {\em types} of applications, just as there are styles of architecture is not new, but the formal development of this idea in this work is a novel contribution that can help to advance work on taxonomies of application types.

Third, the work on specification-driven synthesis of architectural code frameworks---discussed in Chapter 4---
is novel in its formal underpinnings: that abstracting from application details and focusing on essential aspects of the system, such as architectural aspects, relieves the synthesizer of responsibility for full application synthesis, and in turn enables the use of formal methods for modeling and synthesis. This work trades away completeness and richness of synthesis for manageability of modeling languages and tools, avoiding the need for custom synthesizers.
This approach also makes otherwise tacit knowledge of \emph{architecture-to-platform mappings} explicit in the form of mapping predicates. To the extent that such mappings can be matured into richer, general forms, they would enable capturing of important software design knowledge.


This work has contributed potentials in several areas.  One is that the emphasis of this work on the partial formal synthesis points to the possibility of a re-conception of model-driven software engineering~\cite{bagheri_nier}. The next chapter introduces and discusses a bottom-up approach to model-driven development based on the key notion of partial synthesis.

Finally, regarding the work on synthesis of tradeoff space for OR mapping, the formal account of ORM strategies 
enables the rigorous and automatic generation of OR mappings for each application object model. It promises to deliver the quality of expert-hand-crafted mappings and the productivity benefits of automated techniques.
This work appears to have a significant potential to contribute a novel formal approach to search-based software engineering (SBSE)~\cite{harman_search_2010}. 
Success in applying such a formally-specified tradeoff synthesis and analysis can open a new path to applying \emph{formal SBSE} in other disciplines. 

\section{Limitations and Remaining Problems}
In this section, I discuss limitations and shortcomings of this work.


First, to date the work has mainly considered structural refinements, and system behavior is addressed only to the extent that constraints on behavior are implicit or explicit in the specifications of the employed architectural styles. 
As Alloy's emphasis is on specification and automatic analysis of structural properties of systems, it may not be the best option to specify behavioral aspects of systems. I envisage that in an ultimate implementation of this technology, one uses several specification languages and corresponding synthesis technologies handling different aspects of the system.


Second, the architectural styles explored in the Monarch work for architecture synthesis are canonical styles of academic interest. This work did not address more complicated and multi-dimensional architectural styles developed based on elaborations of prominent architectural styles.
Targeting complicated styles, especially those which are particularly relevant to practice is of tremendous value.
I believe the synthesis approach can be naturally extended to support them, subject to the scalability issue discussed in the following.

Third, as with many formal techniques, the complexity of constraint satisfaction restricts the size of models that can actually
be analyzed and synthesized using the proposed approach. Although modularization techniques and the decomposition approach, especially the one presented in Chapter 4, has alleviated the problems encountered in the case studies, the scales of the models considered in these case studies are still relatively small. 
There are several possible approaches to deal with this issue. The first one is to explore symmetry breaking techniques customized for the specific domain of software architecture to reduce the size of the solution space. The other, possibly more pragmatic, approach to limit the size of the solution space is to use partial model instances defining already known aspects of the architecture solution. The third possible approach to avoid combinatorial explosion in constraint solving is to decompose large models into smaller components for which synthesis and analysis problems can be solved independently.


Another important question concerns the potential usability of the tools presented here. Clearly the production of the required specification inputs will be a specialized skill requiring knowledge of formal methods.
Conducting human studies with subjects selected for their knowledge of formal methods, and engaging them in selected modeling and synthesis tasks, would be the natural next step for evaluating this work.
We have recently begun to collaborate with external partners to further evaluate the approach against their external needs. 

Ultimately, the values of a technology such as the one presented in this work is demonstrated in its actual application to solve previously unsolved problems in engineering practice. This work is not yet at a stage where such an application has been possible. Rather, the main goal of this work has been to show that the technique has enough potential to address fundamental issues of cost, quality and breadth of applicability in MDD that they are worth pursuing. More aggressive empirical validation in industrial application areas will be important in future work of this approach.

\newpage
\thispagestyle{empty}
\mbox{}

\chapter{Future Work}

The synthesis approach contributed in this dissertation exposes 
a range of new research opportunities. These research directions include: (1) integrating forward synthesis technique with back-mappings, i.e., abstraction from code back to architectural models and hence to application models; (2) integrating quality-attribute analysis into the presented method and tool; (3) dealing not only with system structure but also with behavior at the abstract modeling level; (4) providing support for automated checking of substitutability properties to guarantee that the evolved architecture is compatible with the previous one; (5) integrating the front- and back-end synthesis phases---from abstract model to architecture, and from architecture to code---to provide end-to-end transformations; (6) conducting human subjects studies for evaluating ease of use the approach; and (7) integrating the tradeoff space decision-support proposed in this dissertation with the formally precise synthesis technique.

In the rest of this section, I discuss in more detail a few of these potential directions for future work.

\section{Formal approach to search-based software engineering}
A very interesting area of research would be to explore a possibility of a novel formal technique to Search-based Software Engineering (SBSE)~\cite{harman_search_2010}. The notion of SBSE is first introduced by March Herman in 2006, where a software engineering task is formulated as a search problem over the space of candidate solutions as well as a fitness function to distinguish solution candidates~\cite{harman_search_2010}. It provides an automated approach to address hard and highly constrained problems that involve conflicting objectives.

What constitutes a promising avenue for future work is to explore a formally precise approach to synthesizing the solution space. This kind of synthesizing the solution space lets us to formally explore such spaces. By formal exploration, I  mean to iteratively restrict boundaries of the solution space, rather than just applying meta-heuristic search techniques which generally lead to local optimum solutions. I envision applying such a formally-precise SBSE technique to a broad set of problems in architectural decision making and development of embedded systems, which are involved making tradeoff decisions for various constraints.



\section{Bottom-up Model-driven Development}
The technologies of model-driven development are generally coupled with a {\em rationalist stance} on software development. Such a stance holds that {\em abstract} models should be the principal artifacts developed by human hands, and {\em concrete} code, derived by top-down refinement, should be made incidental and is best hidden from view. Software evolution, in this paradigm, occurs through model evolution and replaying of evolving, automated refinement procedures.

Important voices in the software engineering research community have questioned the legitimacy of the rationalist stance.
Among practitioners, Ambler says, ``I'm concerned about the viability of the [Model Driven Architecture] MDA.\ldots  Although the MDA is a very wonderful idea I suspect that it will succeed in only a very small percentage of organizations~\cite{scott_ambler_examining_MDA}.'' He argues that current modeling languages do not support the real-world needs of most projects (e.g., the user interface and database components needed in many systems); developers lack adequate modeling skills; and tooling is inadequate. Fowler says, ``Although I've been involved, to some extent, in \ldots [model driven development] for most of my career, I'm rather skeptical of its future. Most fans \ldots base their enthusiasm on the basis that models are ipso facto a higher level abstraction than programming languages. I don't agree with that argument - sometimes graphical notations can be a better abstraction, but not always\ldots~\cite{martin_fowler_modeldrivensoftwaredevelopment_2008}.'' McNeily says that unless executability and translatability can be brought to the kinds of models that business, as opposed to real time embedded systems, developers use, that modeling will remain subject to criticism as ``tool-centric busy work of dubious value, and that we should go back to using a whiteboard~\cite{ashley_mcneile_mda:_2003}.''

Among researchers, Finkelstein is a notable critic. In a recent blog post on the {\em Bottom 10 Challenges in Software Engineering Research}~\cite{anthony_finkelstein_bottom_2012}, he stated that ``...the idea that changes could be made in a high-level specification and then somehow `replayed' is appealing but ignores the ways in which such learning arises, in the context of specific representations and through verification or testing tied to that representation.''
In this formulation, he takes an aggressively {\em empirical} stance on the nature of software development. The empirical stance holds that concrete representations must remain as the principal subject of human effort, 
because it is only by exploring the design space at that level one can {\em learn} what must be learned for projects to succeed.

The question is whether the technologies of MDD can be rescued from the empirical critique?
I envision a synthesis arising out of the juxtaposition of the rational and empirical stances. My stance is {\em scientific-empirical}\footnote{What I mean by the word ``empirical" in the context of this chapter is different from what it entails generally in software engineering literature. Indeed, the focus here is on empirical validation of models within a particular application, rather than empirical testing of a new development method and tool.} (SE).
It is empirical, holding that developers must work at the concrete level, in general.
Yet, it recognizes that just as scientists seek to formulate and test {\em abstract} theories about selected aspects of the concrete, empirical world,
so can software developers profitably develop useful, partial, formal models from experience with concrete artifacts. Moreover, just as theories can have generative power (e.g., in supporting analysis and synthesis of engineered systems), so software models can also support system analysis and synthesis.

The technical key to the viability of this position is found in the
recent developments in partial MDD, including my recent work~\cite{Pol_2012}, that provides the crucial enabling technology for an {\em empirical approach to formal synthesis of software}, by supporting formal modeling of {\em selected} aspects of systems, top-down formal refinement from such models to support {\em partial synthesis}, and a clean separation and integration of model-based and hand-crafted code artifacts.

The rest of this section introduces the working scheme of this idea, and puts it in context with related efforts.

\subsection*{Bottom-up Partial Formal Synthesis}
\label{Approach}

The essence of my approach is based on a set of principles different from those typically held in traditional model-driven development (MDD) approaches.
In this section, I present the approach based on those principles. 

\emph{- Bottom-up.} The approach is bottom-up. In this view, models follow from code and other concrete artifacts, rather than the other way around. Specifically, I posit that software engineers, having worked diligently in the concrete, empirical world of code, can find it profitable to derive and validate abstract models of {\em selected aspects} of code, which then support analysis and synthesis.
The benefits are not in hiding the code behind abstract models, but rather in leveraging the technologies of MDD for improved abstraction, productivity and reliability going forward.

\emph{- Partial Models.} Software development, in traditional MDD, is centered around model specifications of the system, and everything is then derived from those model specifications. 
The second principle, by contrast, states that it is often not practical to develop abstract models for an entire system. Rather, it is often better to extract models for certain stabilized aspects of the system.
Indeed, the code-base for a system is divided into two parts: (1) a part that is synthesized from partial models, and (2) a part that continues to be developed manually. The artifacts that software engineers develop thus include both code and models from which additional code is synthesized.

A criterion for making a decision to lift some idiomatic aspect of the code to the model level is that the aspect has become sufficiently well understood and stabilized. Once we have learned what kind of code is needed for a particular kind of concern in a system, then we can profitably mechanize the development of future instances.
In some sense, for those concerns we expect that the learning is over. As a concrete example, architectural styles are among such aspects of the system, representing recurrent architectural situations~\cite{taylor_software_2009}. They are expressive abstractions for software understanding. We thus learn architectural styles of systems and use them in documenting application architecture. Technically speaking, we consider architectural styles as metamodels to which abstracted architectural models conform.


\emph{- Partial Synthesis.}
The abstracted models are partial with respect to the underlying empirical domain. They thus support only selective system analysis and synthesis. I use the term \emph{partial synthesis} to refer to an approach in which part of the code base is synthesized from partial models~\cite{Pol_2012}. In fact, partial synthesis technologies give us the ability to decide which aspects of the system are dealt with formally in terms of model representations---from which code will be synthesized---and which aspects of the system are dealt with in terms of code. Partial synthesis techniques required for this approach thus should provide a basis for separation of generated and non-generated code with support for merging that limits the impacts on hand-written code modifications required when the synthesized parts are regenerated.
I envisage that in an ultimate implementation, one uses several specification languages and corresponding synthesis technologies handling different aspects of the system. 

Different from traditional MDD, where developers produce domain specific languages for use by non-programmer domain experts, in this approach developers both produce and consume concern-specific modeling notations within the scope of the application under development. 
Specifically, given that a new requirement can be modeled in terms of the already identified and captured model elements, development starts from the model-level. Otherwise, it starts at the code-level.

\subsection*{Related Work}
\label{related}

This work on bottom-up MDD is related to many other efforts, including work on inferring partial specifications from code, model-driven modernization, and partial code synthesis.

\emph{Inferring Partial Specifications.}
A large body of research focuses on inferring partial specifications from code, albeit more for property checking than for synthesis. 
Among others, \emph{Daikon}~\cite{ernst_daikon_2001} discovers likely program invariants by detecting patterns and relationships among values taken by variables during program executions.
\emph{SLAM}~\cite{ball_SLAM_2001_short} automatically and incrementally abstracts a given program based on a set of user-provided predicates. The abstraction is captured in terms of a boolean program, which exhibits an identical control-flow structure to the program, but contains only boolean variables, each of which represents a given predicate.
The extracted boolean program is partial in that program parts irrelevant to the predicates remain abstract, and are captured as \emph{``do nothing"} operators.
Taghdiri and Jackson~\cite{taghdiri_inferring_2007} also proposed an iterative approach based on the \emph{CEGAR} scheme~\cite{e._clarke_counterexample-guided_2003} to check a procedure code against a given assertion. The procedure code is incrementally replaced by a set of partial specifications. These specifications are partial in that they are only as complete as required to verify the given assertion.

The important concept these research efforts have in common with mine is the emphasis on \emph{selective specification recovery} rather than extracting complete specifications. In my work, developers progressively select aspects of the system to be captured by formal specifications to enable, among other things, model-driven synthesis and formal analysis.

\emph{Model-driven Modernization.}
Model-driven modernization~\cite{ADM} is about migrating from heterogenous implementation technologies to the homogenous world of models, from which everything is generated. The initial step though would be to obtain representative models of the legacy systems.
Reus et. al~\cite{reus_harvesting_2006} proposed to reconstruct UML models from existing code. Their approach is based on several model transformations to automate parts of the migration.  Mansurov and Campara~\cite{mansurov_managed_2005} also proposed to extract a specific type of architectural models, called \emph{Container Models}, from existing source code as a first step in migration towards using model-driven architecture (MDA).
Along the same line of research in migrating legacy systems to model-based systems, van Deursen et al.~\cite{van_deursen_symphony:_2004} proposed \emph{Symphony}, a reverse engineering process designed for reconstructing architectural models in terms of appropriate viewpoints. The process consists of two steps of reconstruction design and reconstruction execution. During the former phase, relevant viewpoints are identified, and mapping rules from source to target views are designed. The latter step extracts the source views, and generates the target views through applying the mapping rules to the source views.
More recently, Bruneliere et al.~\cite{bruneliere_modisco:_ASE_2010} proposed an extensible framework (\emph{MoDisco}) for model-driven reverse engineering. They suggested two consecutive steps of model discovery (extracting high-level models from the legacy system) and model understanding (transforming extracted models into required target formats). MoDisco is aimed at facilitating the development of model discoverers.

While my approach is built upon reverse engineering techniques used in this area, 
it is different in several ways. First, they rely on after-the-fact model extraction from an already developed application. In contrast,
my work is geared towards the application of an iterative model abstraction during the software development lifecycle, as opposed to a one-time reverse engineering for software \emph{modernization} of legacy systems. Second, model is the only first class citizen in theirs, while in my approach code base is the main place for learning and modeling is a means to capture obtained knowledge from code. 

To conclude discussion of this future research direction, some researchers and leading practitioners are concerned about the future of model-driven development (MDD).
They argue that MDD has a dim future because of its core {\em rational} assumption, that one can develop software effectively in an iterated top-down manner from abstract specifications, is untenable. Rather, the empiricist argues one must work directly with concrete artifacts (designs, code, etc) to learn what needs to be built and how to build it.
This section presents a novel idea towards a pragmatic, bottom-up approach to model-driven development based on the key notion of \emph{partial formal synthesis}. 
Early validation through experience of applying these ideas to a healthcare-related experimental system in our lab supports the claim that it promises many benefits of MDD, in intellectual control, reliability, and productivity,
while escaping the rationalist trap.
This change in software development perspective, however, imposes a set research challenges which will be discussed in the next section.

\section{Exploring Impacts of bottom-up MDD}
The other interesting area of research would be to explore the impacts of bottom-up model-driven development.
The emphasis of this work on synthesis of partial code frameworks from partial formal systems specifications
points to the possibility of a re-conception of model-driven software engineering, from a top-down approach based on abstract models and hidden code, to a bottom-up approach based on incremental abstraction from code to partial models and subsequent synthesis of visible \emph{parts} of a code base from such models. 
This work can lead to the notion that future developers might work with hybrid code bases, comprising both traditional imperative source code as well as formal models and code that is synthesized from them, evolving in ways that include ongoing refactoring between imperative code and declarative specifications.
This change in software development perspective imposes a set of interesting research impacts.
In the following I discuss some of those consequences that form important areas for future work.

\subsubsection*{Modularity mechanisms}
A complex system has a range of concerns, and its development requires multiple modeling languages and multiple synthesizers appropriate for different aspects of the system. While the approach can improve separation of concerns in two dimensions: between model-driven and manually-developed code artifacts and between different aspects of the system, expressed in different modeling notations, at the same time, utilizing heterogeneous modeling notations imposes development complexity.
This calls for a modularization mechanism so that different models can be changed independently within certain constraints without breaking the whole system.
Important steps in this direction can be found in some recent efforts in aspect-oriented software development (AOSD), specifically on weaving aspect mechanisms in multi-language aspect-oriented frameworks~\cite{kojarski_lorenz_2007}.

\subsubsection*{Property Preservation}
Replacing a concern by a formal representation and introducing model-driven code generation for that concern, requires validation to ensure the property preservation.
Specifically, the application's behavior is not altered once we improve its structure using MDD technologies.
One approach to deal with behavior preservation would be to conduct the so-called regression testing. A related issue is that some test cases, such as those relying on the application structure might be invalidated, even if the application behavior does not alter. The other more pragmatic approach is based on the concept of \emph{call preservation}~\cite{mens_formalising_2002} ensuring that all method calls are still preserved after modification.
Property preservation is an active area of research.

\subsubsection*{Refactoring mechanisms}
The abstraction process in the presented approach in some sense is a refactoring operation.
Refactoring is the process of modifying a software system in such a way that it does not change the external behavior of the system while improving its code structure~\cite{mens_survey_2004}. Similarly, we are refactoring the original
code base into a modified version of the code and extracted high-level specifications from which additional code would be derived.
This calls for new kinds of rafactoring tools that support poly-lingual programming at different level of abstractions.

\newpage
\thispagestyle{empty}
\mbox{}

\chapter{Conclusions}

The overall contribution of this work is a novel formal, specification-driven approach to addressing fundamental problems in the current state of the art in model driven development, including its cost, reliability, and breadth of applicability.
I have evaluated this approach with experiments in three key dimensions of software synthesis:
synthesizing architectural models from abstract application models, synthesizing partial code frameworks from application architectures, and synthesizing object-relational mapping tradeoff spaces from object-oriented application
architectures.

Focusing on the first research component---targeting software architecture synthesis---this work shows that it can be used to separate decisions about abstract application structure and about specific architectural styles to be employed in detailed architectural descriptions. This separation, in turn, enables a formally precise, automated synthesis of architectural models from application models and choice architectural styles, which supports a model-based development to architecture synthesis with style as a separate design variable. The second research component shows that with modest and principled development of code fragments capturing idiosyncratic use of given platforms in given applications it can map architectural descriptions to object-oriented application frameworks that use a range of modern software platforms and standards. Finally, in the context of the object-relational mapping problem, I showed that it creates valuable opportunities for novel forms of trade-space analysis.
In the following, I summarize contributions of this dissertation in the context of each research component.

The initial phase of this research, focused on software architecture synthesis, identifies the treatment of architectural style as a separate variable as a key problem area and goal for software engineering.
This work contributes a theoretical framework to make the notion of \emph{architectural maps} precise. This in effect supports a model-based development to architecture synthesis with style as a separate design variable, which is realized in a prototype tool, called Monarch~\cite{_monarch_2010}.
The results of experiments indicating the viability of the idea in supporting automated synthesis of architectural models from formal specification of applications and choices of architectural style, suggest that this research direction is worth pursuing.
Advances in the science in this area have the potential to catalyze a range of other technologies, among others, in product line architectures from  support for variation within architectures to variation of architectures, and in design space synthesis and analysis.

The second phase of this research contributes a novel approach for architecture-based, model-driven development of systems on multiple software platforms. This work addresses two of the significant problems in software engineering: dependable automation of software production, and facilitating evolution of application architectures. Architecture evolution is often problematical as it has impacts across a whole system. This approach is novel in its formal underpinnings: that abstracting from application details and focusing on certain aspects of the system relieves the synthesizer of responsibility for full application synthesis, while facilitating the use of formal methods for modeling and synthesis.


The third research component contributes a novel, formally precise approach that substantially supports automatic generation of application-specific object-relational mapping specifications to deliver the quality of expert-hand-crafted mappings and the productivity benefits of fully automated techniques. This approach ultimately promises to reduce the engineering personnel costs involved in producing high-quality modern software systems.
This work also offers a novel formal technique to tradeoff space synthesis and analysis.

Having laid a solid foundation, this research project reveals a range of new research opportunities. These research directions include
(1) exploring the possibility of a novel formally precise technique to search-based software engineering~\cite{harman_search_2010};
(2) composing the front- and back-end phases of the work that I have done (from abstract model to architecture, and from architecture to code) to provide an end-to-end transformation approach; (3) integrating tradeoff space decision-support with the formally precise end-to-end synthesis approach; (4) dealing not only with system structure but also with behavior at the abstract modeling level by bringing in new formalisms and synthesis technologies handling different aspects of the system; and (5) exploring the possibility of an agile bottom-up approach to formal model-driven software development. This change in software development perspective have a set of research impacts, among others, in modularity mechanisms, testing, and refactoring tools.

\newpage
\thispagestyle{empty}
\mbox{}



\backmatter

\ifwww
\else
\addcontentsline{toc}{chapter}{Bibliography}
\fi

\bibliographystyle{plain}
\bibliography{Bibliography}


\end{document}